\documentclass[longauth]{aa}
\usepackage[T1]{fontenc}
\usepackage{amstext}
\usepackage{amsmath}
\usepackage{graphicx}
\usepackage[varg]{txfonts}  
\bibpunct{(}{)}{;}{a}{}{,}  
\usepackage[colorlinks, citecolor=blue, linkcolor=blue,draft]{hyperref}
\usepackage[T1]{fontenc}
\usepackage[utf8]{inputenc}

\newcommand{\toprule}{
     \hline\hline
     \noalign{\smallskip}
}
\newcommand{\midrule}{
      \noalign{\smallskip}
      \hline
      \noalign{\smallskip}
}
\newcommand{\bottomrule}{
        \noalign{\smallskip}
        \hline
}

\makeatletter

\iftrue
    \usepackage{twoopt}
    \usepackage{xstring}
    \newcommand{\adshref}[2]{\StrRight{#1}{19}[\adsid]\href{http://adsabs.harvard.edu/abs/\adsid}{#2}}
    
    \let\orgcitep\citep
    \let\orgcitet\citet
    \let\orgcitealt\citealt
    \let\orgcitealp\citealp
    \let\orgciteauthor\citeauthor

    \renewcommandtwoopt{\cite}[3][][]{\adshref{#3}
        {\def\hyper@linkstart##1##2{}
        \let\hyper@linkend\@empty\orgcitet[#1][#2]{#3}}}
    \renewcommandtwoopt{\citep}[3][][]{\adshref{#3}
        {\def\hyper@linkstart##1##2{}
        \let\hyper@linkend\@empty\orgcitep[#1][#2]{#3}}}
    \renewcommandtwoopt{\citet}[3][][]{\adshref{#3}
        {\def\hyper@linkstart##1##2{}
        \let\hyper@linkend\@empty\orgcitet[#1][#2]{#3}}}
    \renewcommandtwoopt{\citealt}[3][][]{\adshref{#3}
        {\def\hyper@linkstart##1##2{}
        \let\hyper@linkend\@empty\orgcitealt[#1][#2]{#3}}}
    \renewcommandtwoopt{\citealp}[3][][]{\adshref{#3}
        {\def\hyper@linkstart##1##2{}
        \let\hyper@linkend\@empty\orgcitealp[#1][#2]{#3}}}
    \renewcommandtwoopt{\citeauthor}[3][][]{\adshref{#3}
        {\def\hyper@linkstart##1##2{}
        \let\hyper@linkend\@empty\orgciteauthor[#1][#2]{#3}}}
    \newcommandtwoopt{\citeyearads}[3][][]
        {\href{http://adsabs.harvard.edu/abs/#3}
        {\def\hyper@linkstart##1##2{}
        \let\hyper@linkend\@empty\citeyear[#1][#2]{#3}}}
   \iftrue
        \renewcommandtwoopt{\cite}[3][][]{\adshref{#3}{\orgcitet[#1][#2]{#3}}}
        \renewcommandtwoopt{\citep}[3][][]{\adshref{#3}{\orgcitep[#1][#2]{#3}}}
        \renewcommandtwoopt{\citet}[3][][]{\adshref{#3}{\orgcitet[#1][#2]{#3}}}
        \renewcommandtwoopt{\citealt}[3][][]{\adshref{#3}{\orgcitealt[#1][#2]{#3}}}
        \renewcommandtwoopt{\citealp}[3][][]{\adshref{#3}{\orgcitealp[#1][#2]{#3}}}
    \fi
\fi

\providecommand{\tabularnewline}{\\}

\def\instrefs#1{{\def\scsep{\def\scsep{,}}\@for\w:=#1\do{\scsep\ref{inst:\w}}}}
\renewcommand{\inst}[1]{\unskip$^{\instrefs{#1}}$}

\usepackage{xspace}

\renewcommand*\aa@pageof{, page \thepage{} of \pageref*{LastPage}} 

\usepackage{xcolor}
\newcommand{\orcid}[1]{\href{https://orcid.org/#1}{\textcolor[HTML]{A6CE39}{\aiOrcid}}}

\usepackage{scalerel}
\usepackage{tikz}
\usetikzlibrary{svg.path}
\definecolor{orcidlogocol}{HTML}{A6CE39}
\tikzset{
  orcidlogo/.pic={
    \fill[orcidlogocol] svg{M256,128c0,70.7-57.3,128-128,128C57.3,256,0,198.7,0,128C0,57.3,57.3,0,128,0C198.7,0,256,57.3,256,128z};
    \fill[white] svg{M86.3,186.2H70.9V79.1h15.4v48.4V186.2z}
                 svg{M108.9,79.1h41.6c39.6,0,57,28.3,57,53.6c0,27.5-21.5,53.6-56.8,53.6h-41.8V79.1z M124.3,172.4h24.5c34.9,0,42.9-26.5,42.9-39.7c0-21.5-13.7-39.7-43.7-39.7h-23.7V172.4z}
                 svg{M88.7,56.8c0,5.5-4.5,10.1-10.1,10.1c-5.6,0-10.1-4.6-10.1-10.1c0-5.6,4.5-10.1,10.1-10.1C84.2,46.7,88.7,51.3,88.7,56.8z};
  }
}
\newcommand\orcidicon[1]{\href{https://orcid.org/#1}{\mbox{\scalerel*{
\begin{tikzpicture}[yscale=-1,transform shape]
\pic{orcidlogo};
\end{tikzpicture}
}{|}}}}

\onecolumn

\begin{document}

\lefthyphenmin=3

\title{The CARMENES search for exoplanets around M dwarfs}
\subtitle{LP\,714-47\,b (TOI\,442.01): Populating the Neptune desert\thanks{Based on observations carried out at the Centro Astronómico Hispano 
Alemán (CAHA) at Calar Alto, operated jointly by the Junta de Andalucía 
and the Instituto de Astrofísica de Andalucía (CSIC), on observations carried out at the European Southern Observatory under ESO programme 0103.C-0152(A), and data collected with the 6.5 meter Magellan Telescopes located at Las Campanas Observatory, Chile.}} 

\titlerunning{TOI\,442.01 = LP\,714-47\,b: populating the Neptune desert}
\authorrunning{S. Dreizler et al.}

\author{
    S.~Dreizler\inst{iag}
    \and I.\,J.\,M.~Crossfield\inst{ku}
    \and D.~Kossakowski\inst{mpia}
    \and P.~Plavchan\inst{gmu}
    \and S.\,V.~Jeffers\inst{iag}
    \and J.~Kemmer\inst{lsw}
    \and R.~Luque\inst{iac,ull}
    \and N.~Espinoza\inst{stsci}
    \and E.~Pall\'e\inst{iac,ull}
    \and K.~Stassun\inst{vanderbilt}
    \and E.~Matthews\inst{mit}
    \and B.~Cale\inst{gmu} 
    \and J.\,A.~Caballero\inst{cabesac}
    \and M.~Schlecker\inst{mpia}
    \and J.~Lillo-Box\inst{cabesac}
    \and M.~Zechmeister\inst{iag}
    \and S.~Lalitha\inst{iag} 
    \and A.~Reiners\inst{iag} 
    \and A.~Soubkiou\inst{umarr} 
    \and B.~Bitsch\inst{mpia}
    \and M.\,R.~Zapatero~Osorio\inst{cab}
    \and P.~Chaturvedi\inst{tls}
    \and A.\,P.~Hatzes\inst{tls}
    \and G.~Ricker\inst{mit}
    \and R.~Vanderspek\inst{mit}
    \and D.\,W.~Latham\inst{cfa}
    \and S.~Seager\inst{mit,DE-mit,DA-mit}
    \and J.~Winn\inst{princeton}
    \and J.~M.~Jenkins\inst{NASA-Ames}
\and J.~Aceituno\inst{CAHA}\inst{iaa}
\and P.\,J.~Amado\inst{iaa}
\and K.~Barkaoui\inst{uliege,umarr}
\and M.~Barbieri\inst{atacama}
\and N.\,M.~Batalha\inst{ucsc}
\and F.\,F.~Bauer\inst{iaa}
\and B.~Benneke\inst{umontreal}
\and Z.~Benkhaldoun\inst{umarr}
% ...
}





\institute{
    \label{inst:iag}Institut f\"ur Astrophysik, Georg-August-Universit\"at, Friedrich-Hund-Platz 1, 37077 G\"ottingen, Germany\\
    \email{dreizler@astro.physik.uni-goettingen.de}
    \and\label{inst:ku}Department of Physics and Astronomy, University of Kansas, Lawrence, KS, USA
    \and \label{inst:mpia}Max-Planck-Institut f\"ur Astronomie, K\"onigstuhl 17, 69117 Heidelberg, Germany
    \and\label{inst:gmu}Department of Physics and Astronomy, George Mason University, 4400 University Drive MS 3F3, Fairfax, VA 22030, USA
    \and \label{inst:lsw}Landessternwarte, Zentrum f\"ur Astronomie der Universit\"at Heidelberg, K\"onigstuhl 12, 69117 Heidelberg, Germany
    \and \label{inst:iac}Instituto de Astrof\'isica de Canarias (IAC), V\'ia L\'actea s/n, 38205 La Laguna, Tenerife, Spain
    \and \label{inst:ull}Departamento de Astrof\'isica, Universidad de La Laguna (ULL), 38206 La Laguna, Tenerife, Spain
    \and \label{inst:stsci}Space Telescope Science Institute, 3700 San Martin Drive, Baltimore, MD, USA
    \and \label{inst:vanderbilt}Physics \& Astronomy Department, Vanderbilt University, 6301 Stevenson Center Ln., Nashville, TN 37235, USA 
    \and\label{inst:mit}Department of Physics, and Kavli Institute for Astrophysics and Space Science, M.I.T., Cambridge, MA 02193, USA
    \and \label{inst:cabesac}Centro de Astrobiolog\'ia (CSIC-INTA), ESAC, Camino bajo del castillo s/n, 28692 Villanueva de la Ca\~nada, Madrid, Spain
    \and \label{inst:umarr}Oukaimeden Observatory, High Energy Physics and Astrophysics Laboratory, Cadi Ayyad University, Marrakech, Morocco
    \and \label{inst:cab}Centro de Astrobiolog\'ia (CSIC-INTA), Carretera de Ajalvir km 4, 28850 Torrej\'on de Ardoz, Madrid, Spain
    \and \label{inst:tls}Th\"uringer Landessternwarte Tautenburg, Sternwarte 5, 07778 Tautenburg, Germany
    \and\label{inst:cfa}Center for Astrophysics / Harvard \& Smithsonian, 60 Garden Street, Cambridge, MA 02138, USA
    \and\label{inst:DE-mit}Department of Earth, Atmospheric and Planetary Sciences, Massachusetts Institute of Technology, Cambridge, MA 02139, USA
    \and\label{inst:DA-mit}Department of Aeronautics and Astronautics, MIT, 77 Massachusetts Avenue, Cambridge, MA 02139, USA
    \and\label{inst:princeton}Department of Astrophysical Sciences, 4 Ivy Lane, Princeton University, Princeton, NJ 08544, USA
    \and \label{inst:NASA-Ames}NASA Ames Research Center, Moffett Field, CA 94035, USA
    \and \label{inst:CAHA}Centro Astron\'omico Hispano-Alem\'an (CSIC-Junta de Andaluc\'ia), Observatorio Astron\'omico de Calar Alto, Sierra de los Filabres, 04550 G\'ergal, Almer\'ia, Spain 
    \and \label{inst:iaa}Instituto de Astrof\'isica de Andaluc\'ia (IAA-CSIC), Glorieta de la Astronom\'ia s/n, 18008 Granada, Spain
    \and \label{inst:uliege}Astrobiology Research Unit, Université de Liège, 19C Allée du 6 Août, 4000 Liège, Belgium
    \and\label{inst:atacama}Universidad de Atacama, Instituto de Astronom\'ia y Ciencias Planetarias, Copiap\'o, Chile
    \and\label{inst:ucsc}Department of Astronomy and Astrophysics, University of California, Santa Cruz, CA, USA
    \and\label{inst:umontreal}Département de Physique, Université de Montréal, QC, Canada
% ...
}

\abstract{We report the discovery of a Neptune-like planet (LP\,714-47\,b, $P=4.05204\,{\rm d}$, $m_{\rm b}=30.8\pm 1.5$\,M$_\oplus$, $R_{\rm b}=4.7\pm 0.3$\,R$_\oplus$) located in the `hot Neptune desert'. Confirmation of the {\em TESS} Object of Interest (TOI\,442.01) was achieved with radial-velocity follow-up using {CARMENES}, {ESPRESSO}, {HIRES}, {iSHELL}, and {PFS}, as well as from photometric data using {\em TESS}, {\em Spitzer}, and ground-based photometry from {MuSCAT2}, {TRAPPIST-South}, {MONET-South}, the {George Mason University telescope}, the {Las Cumbres Observatory Global Telescope} network,  the {El Sauce} telescope, the {T\"UB\.{I}TAK National Observatory}, the {University of Louisville Manner Telescope}, and {WASP-South}. 
We also present high-spatial resolution adaptive optics imaging with the Gemini Near-Infrared Imager. 
The low uncertainties in the mass and radius determination place LP\,714-47\,b among physically well-characterised planets, allowing for a meaningful comparison with planet structure models.
The host star LP\,714-47 is a slowly rotating early M dwarf ($T_{\rm eff}=3950\pm51$\,K) with a mass of 0.59\,$\pm0.02$\,M$_\odot$ and a radius of 0.58\,$\pm0.02$\,R$_\odot$. 
From long-term photometric monitoring and spectroscopic activity indicators, we determine a stellar rotation period of about 33\,d. The stellar activity is also manifested as correlated noise in the radial-velocity data. 
In the power spectrum of the radial-velocity data, we detect a second signal with a period of 16\,days in addition to the four-day signal of the planet. This could be shown to be a harmonic of the stellar rotation period or the signal of a second planet. It may be possible to tell the difference once more {\em TESS} data and radial-velocity data are obtained.
}


\date{Received dd March 2020 / Accepted dd Month 2020}

\maketitle
\sloppy

\section{Introduction}

Currently, approximately 4000 planets have been detected using the transit method, primarily with the \emph{Kepler} space telescope. A statistical analysis of these detections indicates that at short orbital periods, there is a bimodal distribution of planets, characterised by a population of Earth-sized planets (including super-Earths)   and a population of larger Jupiter-sized planets \citep{SzaboKiss2011ApJ...727L..44S,Mazeh2016A&A...589A..75M}. While intermediate Neptune-mass planets have been detected at larger orbital periods, there is a distinct lack of Neptune-mass planets at short orbital periods.  The rarity of planets with masses of approximately 0.1 M$_{\rm Jup}$ and periods of less than about 4\,d is referred to as the `hot Neptune desert'.  

Deliberations on the cause of the hot Neptune desert have been addressed in several papers involving photo-evaporation \citep[e.g.][]{OweWu2013,LopezFortney2014}, the preferential formation of Jupiter-sized rather than Neptune-sized planets, as a consequence of core accretion \citep[e.g.][]{IdaLin2008,Mordasinietal2009}, the early migration of low-mass planets \citep[e.g.][]{Flock2019}, or core-powered mass loss \citep{2020MNRAS.tmp..303G}. In an investigation of possible formation mechanisms, and to explain the distinctive triangular shape of the desert in the mass-period diagram, \cite{Owen2018MNRAS.479.5012O} showed that photo-evaporation of Neptune-mass planets occurs very close to the host stars.

Photo-evaporation may not necessarily be required in this process. Instead, atmospheric mass-loss can be driven by the release of the primordial energy from the formation, which is comparable to the atmospheric binding energy \citep{2016ApJ...825...29G,2018MNRAS.476..759G,2019MNRAS.487...24G}. With this scenario, it is possible to explain the radius valley spanning between Earth-sized and sub-Neptune-sized planets. Properties of the radius valley are predicted to show a different dependence on age and metallicity between the photo-evaporation and the core-powered mass-loss scenario. Another possibility is that of scattering onto high-eccentric orbits followed by a circularisation (high-eccentricity migration). This could explain the lower and upper boundary of the Neptune desert \citep{2016ApJ...820L...8M}.

A few planets or planet candidates have been detected in the hot Neptune desert.  These include K2-100b \citep{Mann2017AJ....153...64M},
NGTS-4\,b \citep{West2019MNRAS.486.5094W}, 
TOI-132\,b \citep{Diaz2020}, 
TOI-824\,b \citep{2020AAS...23534909B},
LTT~9779\,b \citep{Jenkins2019ESS.....410307J}, 
 and TOI-849\,b \citep{2020Natur.583...39A}. The latter two are extreme cases with orbital periods shorter than one day. More planets close to or inside the Neptune desert with precise mass and radius determinations would help to shed more light on the distinction between the various explanations for the existence of the Neptune desert.

The {\em Transiting Exoplanet Survey Satellite} \citep[{\em TESS},][]{Ricker2014SPIE}, launched in April 2018, aims to detect close-in planets transiting bright and nearby stars.  
Such transiting systems are ideal for ground-based follow-up observations focused on the study of the atmospheres of exoplanets by using transmission or emission spectroscopy. Given its instrument characteristics, {\em TESS} is ideal for the detection of hot Neptunes. In this paper, the {\em TESS} Object of Interest TOI\,442.01 is confirmed to be a 30-Earth-mass planet orbiting its late-type host star \object{LP\,714-47} with an orbital period of 4.05 days. From the radial velocity measurements, as well as the {\em TESS}, {\em Spitzer}, and ground-based photometry, presented in Sect.\,\ref{sec:Obs}, we conclude that  LP\,714-47\,b is a transiting Neptune-like planet close to the lower edge of the hot Neptune desert, presented in Sect.\,\ref{sec:Analysis}. Adding a new planet with precise mass and radius determination is, therefore, helpful  in shedding more light on the origins of the hot Neptune desert, which may, in turn, place constraints on planet formation scenarios, as discussed in Sect.\,\ref{sec:Discussion}.

\section{{\em TESS} photometry}

\begin{figure}[t]
    \centering
    \includegraphics[width=0.48\textwidth]{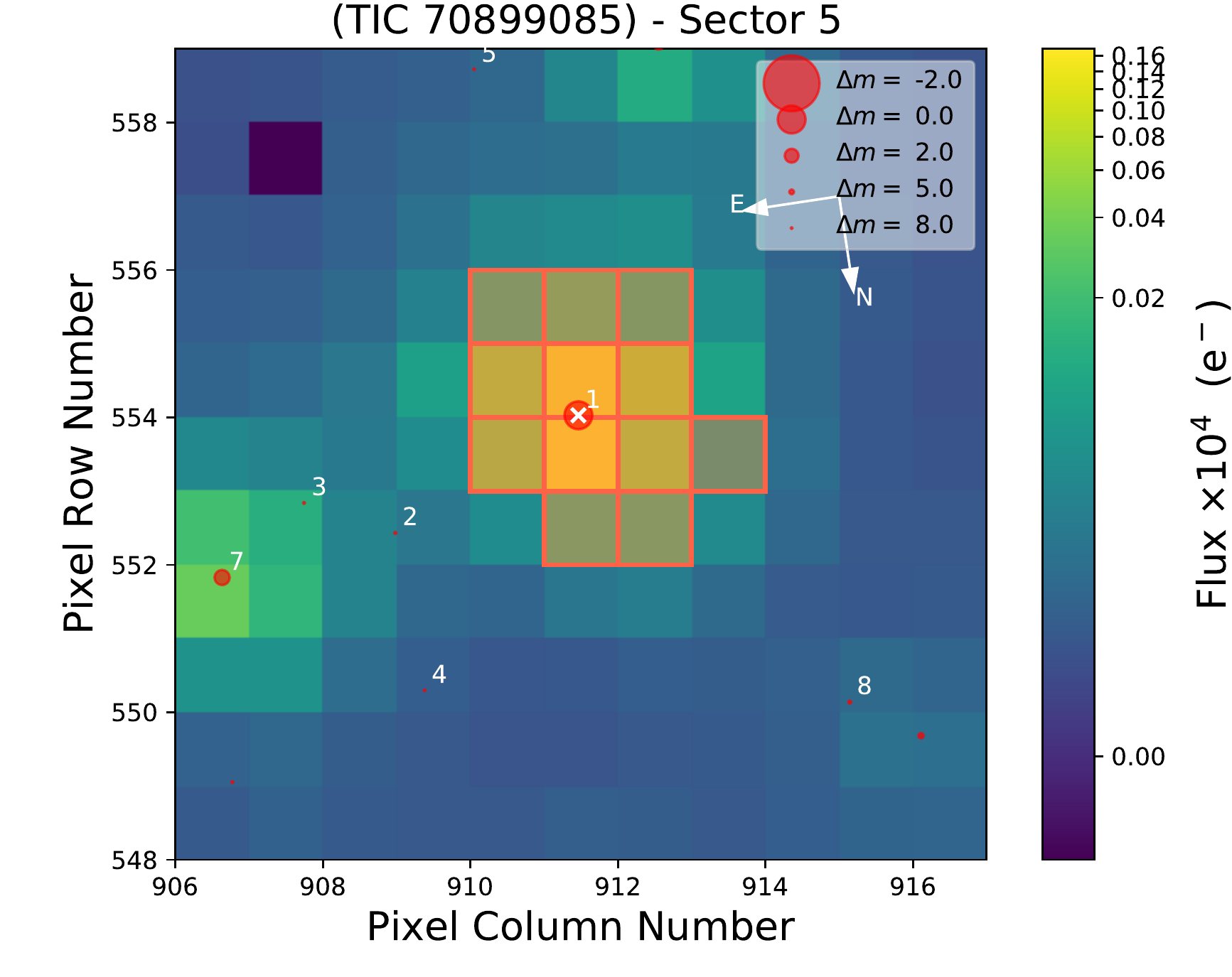}
    \caption{{\em TESS} Target Pixel File (TPF) of TOI\,442 (created with \texttt{tpfplotter}\protect\footnotemark, \citealt{aller20}). The pipeline aperture mask is shown as a red shaded region. Objects in the {\em Gaia} DR2 catalogue are overplotted with red circles (size depending on brightness as indicated in the legend). TOI~442 is marked with a white cross.}
    \label{fig:FFI}
\end{figure}
\begin{figure}[t]
    \centering
    \includegraphics[width=0.49\textwidth]{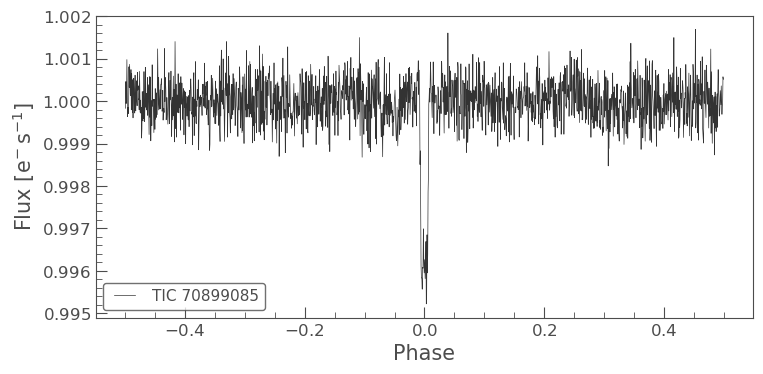}    \caption{Phase-folded {\em TESS} lightcurve of LP\,714-47 with a period of 4.0520\,d, generated with {\tt lightkurve}\protect\footnotemark{} using a window length of 1001 and a sigma clipping threshold of 4.}
    \label{fig:TESS_lc}
\end{figure}

LP\,714-47 was observed in {\em TESS} sector 5 in two-minute cadence mode between 15 November and 11 December 2018 (see Figs.\,\ref{fig:FFI}\footnotetext{\url{https://github.com/jlillo/tpfplotter}} and \ref{fig:TESS_lc}\footnotetext{\url{https://github.com/KeplerGO/lightkurve}}).
The planet candidate TOI442.01 was announced on 31 January 2019. 
The {\em TESS} Data Validation diagnostic (DV)
\citep{Twicken:DVdiagnostics2018,Li:DVmodelFit2019} revealed a planet
candidate at a period of 4.05\,d (TOI~442.01) with depths of $5685\pm
134$\,ppm (S/N = 45.3), the planet radius of the candidate was
determined as $4.7\pm 0.7$\,R$_\oplus$, slightly different values are
reported on the ExoFOP web
page\footnote{\url{https://exofop.ipac.caltech.edu/tess/target.php?id=70899085}}. The
DV difference image centroid offsets for TOI\,442.01 indicate that the
source of the transit signature is within 2 arcsec (0.1 {\em TESS}
pixels). The {\em TESS} lightcurves produced by the Science Process
Operation Centre \citep[SPOC,][]{Jenkins2016SPIE.9913E..3EJ} are
available at the Mikulski Archive for Space
Telescopes\footnote{\url{https://mast.stsci.edu/portal/Mashup/Clients/Mast/Portal.html}}. For
our analysis, we used the de-trended pre-search data conditioning
simple aperture photometry (PDC-SAP) lightcurve
\citep{2012PASP..124..985S,2012PASP..124.1000S,2014PASP..126..100S}. The
total time-span of the observation covered seven transits, although
one of them occurs during a gap in the data caused by the re-orientation of the spacecraft needed for the data downlink.

We independently searched for a transit signal in the light curve using the transit-least-squares method {with a signal detection efficiency of 19.8} \citep[TLS,][]{Hippke2019A&A...623A..39H} and the box-least-squares algorithm \citep[BLS,][]{Kovacs2002A&A...391..369K}. We recovered the 4.05\,d signal in both cases with a depth of $4.8\pm0.3$\,ppt.
After removing the transits, the strongest peak in the periodogram of the residuals was a modulation with a period of 9.2\,d and an amplitude of 230\,ppm.

The announcement of TOI442.01 as a potential planet host star initiated a series of follow-up observations: space and ground based transit photometry, high-resolution and low-resolution spectroscopy, and high-contrast imaging. Between the various teams, we decided to analyse all available data jointly.

\section{\label{sec:Obs}Follow-up observations}

\subsection{{\em Spitzer} photometry}

We observed a transit of the system with the {\em Spitzer Space Telescope}'s IRAC instrument \citep{fazio:2005}, as part of {\em Spitzer} GO~14084 (PI Crossfield).  The observations on 21 May 2019 used the channel 2 (4.5\,$\mu$m) broadband filter and we acquired 228 sets of 64 sub-array frames with integration times of 2\,s. The observations used the standard {\em Spitzer} approach to observing, with an initial peak-up observation to place the target near the well-calibrated {sweet spot} of the IRAC2 detector. Because our target is fainter than the recommended brightness level for peak-up observations, we used the brighter, nearby star HD\,27112 for peak-up. In total, the {\em Spitzer} observations spanned 8.3\,h and covered a full transit (Sect.~\ref{sec:Analysis}).

In order to account for the intra-pixel variations, we used the pixel-level de-correlation method developed by \cite{Deming2015}. This method reconstructs the observed transit signal from a linear combination of the varying contributions from the nine pixels covered by the instrument point spread function (PSF) and a transit model. The linear fit of the correlation coefficients is part of the optimisation procedure.

\subsection{Ground-based photometry}
\label{sect:GB-photo}

We acquired ground-based, time-series follow-up photometry of TOI\,442 as part of the {\em TESS} Follow-up Observing Program\footnote{\url{https://tess.mit.edu/followup/}} to attempt to: 
($i$) rule out nearby eclipsing binaries as potential sources of the {\em TESS} detection,
($ii$) detect the transit-like event on target to confirm the event depth and, thus, the {\em TESS} photometric deblending factor, 
($iii$) refine the {\em TESS} ephemeris,
($iv$) provide additional epochs of transit center time measurements to supplement the transit timing variation (TTV) analysis, 
and ($v$) place constraints on transit depth differences across optical filter bands. 
We used the {\em TESS} {\tt Transit Finder}, which is a customised version of the {\tt Tapir} software package \citep{Jensen:2013}, to schedule our transit observations. Unless otherwise noted, the transit follow-up data were extracted using the {\tt AstroImageJ} ({\tt AIJ}) software package \citep{Collins:2017}. We observed full transits from the MONET/South telescope (MOnitoring NEtwork of Telescopes) at the South African Astronomical Observatory, MuSCAT2 \citep{narita_2019} at the 1.5\,m Telescopio Carlos S\'anchez at the Observatorio del Teide, the George Mason University campus 0.8\,m telescope, the El Sauce private observatory, the Las Cumbres Observatory Global Telescope (LCOGT) telescope network \citep{LCO}, T\"UB\.{I}TAK National Observatory (TUG)\footnote{\url{http://tug.tubitak.gov.tr/en}}, TRAPPIST-South \citep{jehin2011}, and University of Louisville Manner Telescope (ULMT). The transit light curves were linearly de-correlated using information of airmass and centroid position, the de-correlation coefficients were then kept fixed during the analysis. Overall, we covered seven transit epochs with 15 data sets in various filters summarised in Table\,\ref{tab:photometry}. The instruments are further described in Appendix~\ref{sec:phot_inst}.

In an attempt to determine the stellar rotation period, we also compiled photometric long-term time series from public surveys including the All-Sky Automated Survey \citep[ASAS,][]{Pojmanski2002AcA....52..397P}, the Northern Sky Variability Survey \citep[NSVS,][]{Wozniak2004AJ....127.2436W}, and the WASP survey \citep{2006PASP..118.1407P}. The time series covered nine years (ASAS), seven months (NSVS), and seven years (WASP). From the WASP dataset, the most useful data come from four months of monitoring by WASP-South in 2008--2009.


\subsection{\label{sec:spec_inst}High resolution spectroscopy}
As a summary of the various instruments used for radial velocity follow-up of TOI\,442.01, we list the basic properties of the radial-velocity data and the number of observations from each instrument in Table\,\ref{tab:spec_fac} (see also Fig.\,\ref{fig:allRVdata}), followed by a more detailed description below. 
With CARMENES, we started the RV follow-up after the candidate alert in February 2019 and re-started in October 2019. HIRES and PFS observations started in August 2019, and those with iSHELL in November 2019.

\begin{table}[t]
    \centering
    \caption{Summary of radial-velocity data sets.}
    \begin{tabular}{@{}lrlc@{}}
        \toprule
         Observatory & Spectral & Observing & $N$  \\
         ~~Instrument & resolution & season (2019)\\
        \midrule
         Calar Alto& \\
         ~~CARMENES VIS & 94\,600 & Spring+Autumn & 33\\
         ~~CARMENES NIR & 80\,400 & Autumn & 21\\
         Paranal   \\
         ~~ESPRESSO     & 140\,000& September &19\\
         Keck      \\
         ~~HIRES        &  60\,000& Autumn & 14\\
         Las Campanas\\
         ~~PFS        & 127\,000&  Autumn & 6\\
         IRTF      \\
         ~~iSHELL$^{\mathrm (a)}$      &  75\,000&  Dec 2019 -- Jan. 2020 & 9\\
        \bottomrule
    \end{tabular}
    \tablefoot{\tablefoottext{a}{For iSHELL, we list the number of epochs, which consist of multiple co-added exposures.}} \label{tab:spec_fac}
\end{table}

\subsubsection{CARMENES spectra}
\label{sect:carmenes}

The {CARMENES} spectrograph consists of a red-optical (VIS) channel and a near-infrared (NIR) channel, covering a wide wavelength range of 550--960\,nm 
and 960--1700\,nm, and achieving spectral resolutions of 94\,600 and 80\,400, respectively \citep{Quirrenbach2014SPIE.9147E..1FQ,Quirrenbach2018SPIE10702E..0WQ}.  
The average sampling per resolution element is 2.8\,px.  
The spectra were processed by the standard pipeline {\tt caracal} \citep[CARMENES reduction and calibration software;][]{Zechmeister2014A&A...561A..59Z} and went through the guaranteed observation time data flow \citep{Caballero2016SPIE.9910E..0EC}.  

Triggered by the {\em TESS} alert, we monitored LP\,714-47 with {CARMENES} from February 2019 to November 2019.  
We collected 33 spectra, each with an exposure time of 30\,min (Table~\ref{tab:spec_fac}).  
The median typical signal-to-noise ratio (S/N) per pixel of the observations was $\sim$63 at 746\,nm. 
The RVs were computed with {\tt serval}\footnote{\url{https://github.com/mzechmeister/serval}} \citep[Spectrum radial-velocity analyzer][]{Zechmeister2018A&A...609A..12Z}; the uncertainties range from about 2 to 4\,m/s, with a median of 2.6\,m/s.

The python code {\tt serval} also provides spectroscopic activity indicators and the equivalent widths of key diagnostic photospheric and chromospheric lines, such as the differential line width, dLW, and the chromatic index, CRX \citep{Zechmeister2018A&A...609A..12Z}. Due to an apparently systematic offset, the CARMENES NIR data from the first season were excluded from the fit.

\subsubsection{ESPRESSO spectra}
ESPRESSO is a new high-resolution spectrograph at the ESO Very Large Telescope \citep{Pepe2010SPIE.7735E..0FP}. It covers the wavelength range 380--788\,nm with a resolution of 140\,000 and sampling of 4.5\,px per resolution element.

We obtained 19 spectra of LP\,714-47 in September 2019 in the single-telescope high-resolution mode with 2$\times$1 binning (HR21). The exposure times were at least 10\,min, and the S/N was about 30 per pixel at 573\,nm. The spectra were processed with the standard ESO pipeline. The RVs were calculated with {\tt serval}.

\subsubsection{HIRES spectra}
\label{sect:HIRES}
We acquired 14 spectra with the {HIRES} spectrograph with the
standard CPS setup \citep{howard:2010}: the C2 decker,
the usual {HIRES} iodine cell \citep[used to measure precise RVs;][]{marcy:1992,butler:1996}, and exposures of 10--30\,min (depending on
seeing conditions). The {HIRES} observations began on 14~Aug 2019 and
ended on 15~Nov 2019. As part of the HIRES analysis, we also acquired
an iodine-free observation to serve as a template spectrum on 31~Oct 2019 using the B3 decker and an exposure time of 45\,min.

\subsubsection{PFS spectra}
LP\,714-47 was observed on six nights with the {Carnegie Planet Finder Spectrograph} \citep[PFS;][]{crane:2006, crane:2008,crane:2010} mounted on the Magellan~2 (Clay) 6.5\,m telescope at Las Campanas Observatory in Chile.  Two consecutive exposures were taken each night.  The individual exposure times ranged from 15 to 20\,min, depending on the conditions.  Due to the faintness of this star, the CCD was 3$\times$3 binned to minimise readout noise.  The spectral resolution of all observations was 127\,000
and the typical S/N of the individual observations was 25.  
Target stars are routinely observed through a custom built iodine cell, which was scanned by the NIST Fourier transform spectrometer at a resolution of $10^6$ \citep{nave:2017} and provides a rich embedded reference from 5000 to 6200\,\AA.  As with the HIRES spectra, the velocity reduction used the approach described by \cite{butler:1996}. 

\subsubsection{iSHELL spectra}
We obtained 198 spectra of LP~714-47 on nine nights, spanning a 53\,d interval with the iSHELL spectrometer on the NASA Infrared Telescope Facility \citep[IRTF,][]{2016SPIE.9908E..84R}. The exposure times were 5\,min, repeated 17--22 times over a single a night to reach a cumulative photon S/N per spectral pixel at about $2.2\,\mu$m (at the approximate centre of the blaze for the middle order) varying from 107 to 171 to achieve a per-night RV precision of 3--10\,m\,s$^{-1}$ (median 6\,m\,s$^{-1}$). Spectra were reduced and RVs extracted using the methods outlined in \citet{Cale2019AJ....158..170C}. 


\subsection{Low-resolution spectroscopy}
\label{sec:low-res}
On 03 December 2019 (UT), we collected low-resolution optical spectra of LP\,714-47 using the Alhambra Faint Object Spectrograph and Camera (ALFOSC) mounted on the Cassegrain focus of the 2.56\,m Nordic Optical Telescope (NOT), located at the Observatorio del Roque de los Muchachos on La Palma island. 
ALFOSC is equipped with a 2k\,$\times$\,2k pixel e2v detector providing a plate scale of 0.2138\,arcsec\,pixel$^{-1}$ on the sky. 
We used the grism number 5 and a long slit with a width of 1.0\,arcsec, which delivered spectral data with a nominal dispersion of 3.55\,\AA\,pixel$^{-1}$ (resolution of 17\,\AA; resolving power of $R \approx 420$ at 700\,nm) and a wavelength coverage of 500--1,000\,nm. A total of four exposures of 300\,s each were obtained at parallactic angle and at airmass of 1.33.
The observations were hampered by thin cirri and a poor seeing of about 3.0\,arcsec.

The raw data were reduced following standard procedures for long-slit spectra within the {\tt IRAF} environment, including debiasing and flat-fielding. The spectra were optimally extracted and wavelength calibrated using Th-Ar arc lamp exposures obtained at the beginning of the night. The dispersion of the wavelength calibration was 1.3\,\AA. The spectra were corrected for instrumental response using the spectrophotometric standard star G~191--B2B (white dwarf), which was observed on the same night and with an identical instrumental configuration as our target, but with a different airmass. This and the poor weather conditions prevented us from removing the telluric contribution from the ALFOSC spectra. All four individual spectra were combined to produce a final spectrum (cosmic rays were removed in the process). The ALFOSC spectrum of LP\,714-47 is shown in Fig.~\ref{alfosc_fig}. It has a S/N higher than 200.
This spectrum is used in Sect.\,\ref{sec:HostStarParams} to derive the stellar parameters. A comparison of this spectrum with stellar templates as well as model atmospheres is presented in Appendix\,\ref{alfosc_data}.

\subsection{High-contrast imaging}
\label{subsec:AO}
\begin{figure}[t]
    \centering
    \includegraphics[width=0.49\textwidth]{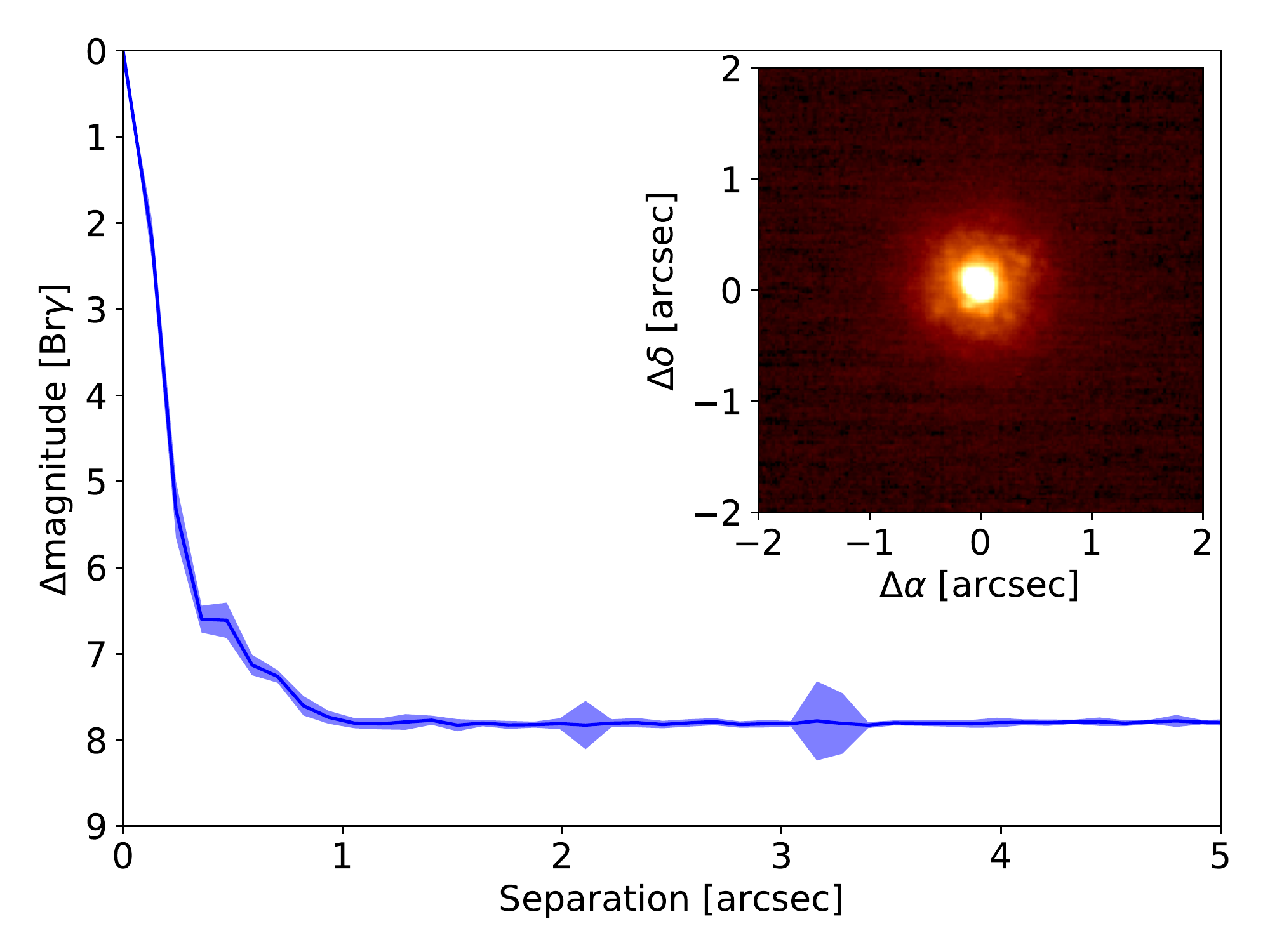}
    \caption{Sensitivity to visual companions in the Gemini/NIRI data. \textit{Inset:} Thumbnail image of the target. No stellar companions were detected anywhere in the field of view.}
    \label{fig:gemini-niri}
\end{figure}

We obtained diffraction-limited infrared images of LP\,714-47 on 18 March 2019 with the {NIRI} instrument at the Gemini Observatory \citep[][Program GN-2019A-LP-101, PI Crossfield]{hodapp2003}.  We used the {NIRI} camera behind the {ALTAIR} facility adaptive optics (AO) system.  Nine dithered exposures of 3\,s each were acquired, using the Br-$\gamma$ filter to approximate the $K$ band while avoiding saturation. A sky background image was constructed from the dithered science frames. The data were reduced using a custom code that corrects for bad pixels, flatfields, subtracts the sky background, aligns the star between images, and coadds data. Sensitivity to faint companions was calculated by injecting fake PSFs and scaling them to the magnitude at which they are detected at 5\,$\sigma$. No companions were seen anywhere in the field of view, which extends at least 13\,arcsec from the target in all directions, and we were sensitive to candidates 7\,mag fainter than the star beyond a few tenth of arcseconds. Our sensitivity to companions is shown in Fig.~\ref{fig:gemini-niri} along with a thumbnail image of the target. The candidate planet is, therefore, highly unlikely to be a false positive from a blended eclipsing binary.

Additional high-contrast imaging (Fig.\,\ref{fig:Speckle}) was obtained on 14 January 2020 using the Zorro speckle instrument on Gemini-South\footnote{ \url{https://www.gemini.edu/sciops/instruments/alopeke-zorro/}}. The results are reported in Appendix\,\ref{sec:Speckle}. Both direct imaging observations with high contrast and high spacial resolution exclude a companion which could either lead to a false positive transit detection as background binary or would lead to a third light contribution for the transit light curve.

\section{Host star}
\label{sec:HostStar}
\subsection{Basic stellar parameters}
\label{sec:HostStarParams}
\begin{table}[t]
    \caption{\label{tab:stellarparam}Basic properties of host star LP\,714-47.}
    \centering
    \begin{tabular}{@{}lcl@{}}
        \toprule
        Parameter & Value & Reference$^{a}$ \\
        \midrule
Name & LP\,714-47 & Luy63 \\
Name & G\,160-62 & Gic78 \\
TOI & 442 & {\em TESS} Alerts \\
TIC & 70899085 & Sta19 \\ 
Karmn & J04167--120 & Cab16 \\
$\alpha$ (J2000) & 04:16:45.65 & {\em Gaia} DR2$^c$ \\
$\delta$ (J2000) & --12:05:05.5 & {\em Gaia} DR2$^c$ \\
$d$ [pc] & $52.70\pm0.11$ & {\em Gaia} DR2 \\
$G$ [mag] &$11.7282\pm0.0004$ & {\em Gaia} DR2 \\
TESS [mag] & $10.733\pm0.007$ & Sta19\\
$J$ [mag]&$9.493\pm0.024$& 2MASS \\
Sp. Type & M0.0\,V & This work \\
$T_{\rm eff}$ [K]&3950$\pm$51& This work \\
$\log{g}$ [cgs] & 4.64$\pm$0.04& This work \\
$L_\star$ [$L_{\odot}$] & 0.075$\pm$0.001 & This work \\
$R_\star$ [$R_{\odot}$] & 0.584$\pm$0.016 & This work \\
$M_\star$ [$M_{\odot}$] & 0.59$\pm$0.02& This work \\
$[{\rm Fe/H}]^b$ & $0.41\pm 0.16$ & This work \\
pEW (H$\alpha$) [\AA] & $<$0.3 &  This work \\
$v\sin{i}$ [km\,s$^{-1}$] & $<$2 & This work \\
$P_{\rm rot}$ [d] & 33$\pm$3 & This work \\
        \bottomrule
    \end{tabular}
    \tablefoot{\tablefoottext{a}{
        2MASS: \cite{Skrutskie2006AJ....131.1163S};
        Cab16: Caballero et al. 2016;
        {\em Gaia}: \cite{Gaia2018A&A...616A...1G};
        Gic78: \cite{Giclas1978LowOB...8...89G};
        Luy63: \cite{1963BPM...C......0L};
        Sta19: \cite{StassunTIcv8}. 
        \tablefoottext{b}{Metallicity is not well constrained. See Appendix~\ref{alfosc_data}.}
        \tablefoottext{c}{Note that the {\em Gaia} DR2 coordinates are for equinox J2000 and at epoch J2015.5.}
    }}
\end{table}

\begin{figure}
    \centering
    \includegraphics[width=\linewidth]{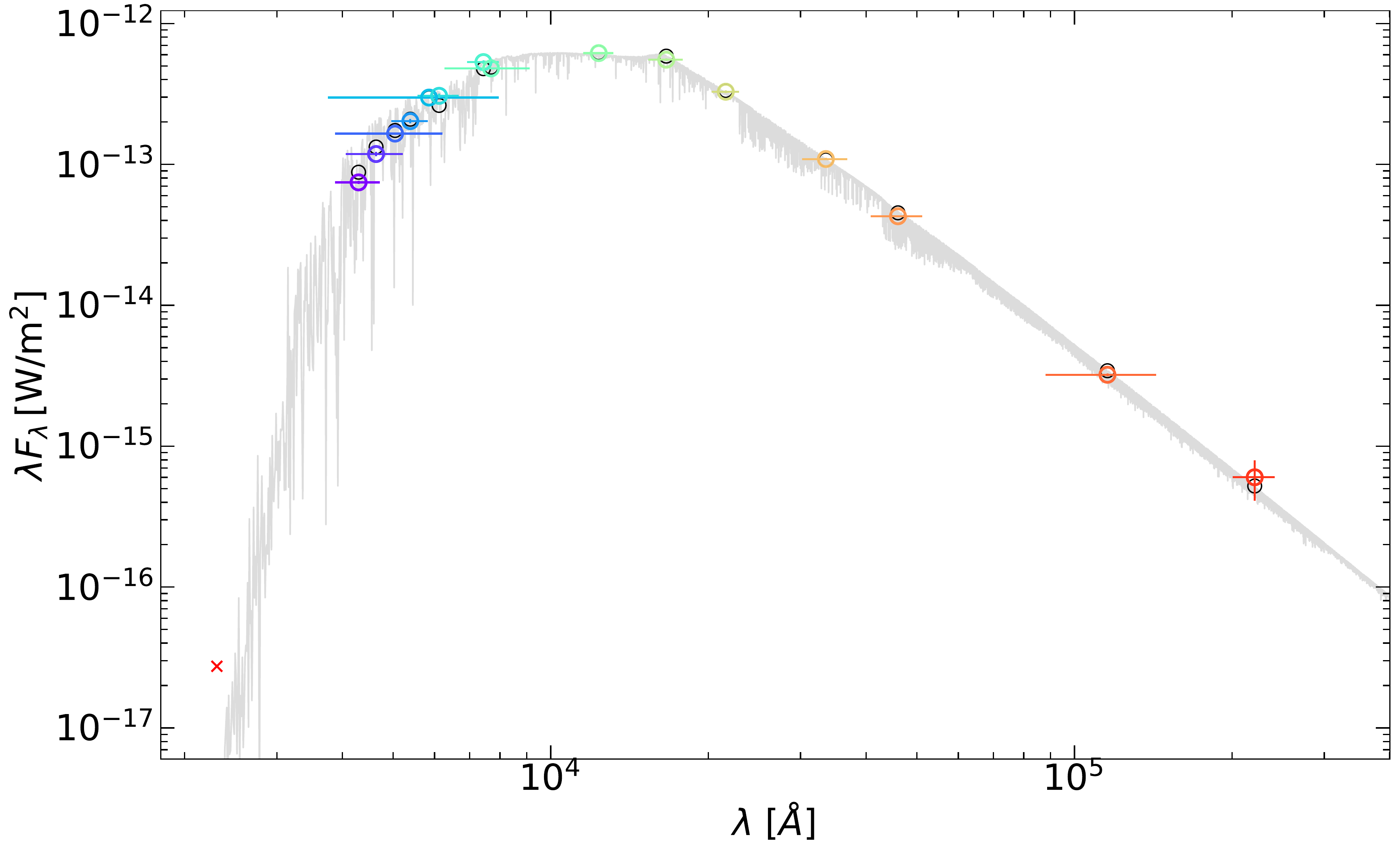}
    \caption{Spectral energy distribution (SED). Coloured circles
    represent the photometry, where the horizontal bars represent the effective width of the passbands. 
    Black circles are predicted magnitudes within the observed bandpasses integrated from the best-fit BT-Settl CIFIST atmosphere model \citep{2015A&A...577A..42B} (light grey curve) available to VOSA \citep{2008A&A...492..277B}.
    \label{fig:sed}}
\end{figure}

The host star LP\,714-47 was a poorly investigated, late-type, high-proper-motion star tabulated first by \citet{1963BPM...C......0L} and \citet{1971lpms.book.....G}, and later by \citet{2003ApJ...582.1011S}, \citet{2011AJ....142..138L}, \citet{2013MNRAS.435.2161F}, and \citet{2016ApJS..224...36K}.
The TESS Input Catalog Version\,8 \citep{StassunTIcv8} lists TOI\,442=TIC\,70899085 as a star with an effective temperature of $3779\pm157$\,K, a mass of 0.59\,$\pm0.02$\,M$_\odot$, and a radius of 0.61\,$\pm0.02$\,R$_\odot$.
The entry in the ExoFOP database 
also lists user-uploaded analyses by Jason Eastman and Abderahmane Soubkiou (A.S.). 
Both used {\tt exofastv2} \citep{ExoFAST2019}, which fits planetary and stellar parameters simultaneously using MIST isochrones \citep{MIST2016,Choi2016}, the spectral energy distribution (Fig.\,\ref{fig:sed}) and the {\em Gaia} DR2 distance (Table\,\ref{tab:stellarparam}). A.S. used Gaussian priors on the effective temperature and metallicity from the CARMENES spectral analysis as well as Gaussian priors on the {\em Gaia} DR2 parallax (adding 82\,$\mu$as to the reported value and adding 33\,$\mu$as in quadrature to the reported error, following the recommendation of \citet{StassunTorres:2018}). An upper limit on the extinction of $A_V = 0.15$ from the \citet{2011ApJ...737..103S} dust maps was enforced.  The {\tt exofastv2} fit without priors results in a slightly higher effective temperature (4200$\pm 160$\,K) and lower mass (0.58\,$\pm0.03$\,M$_\odot$), the metallicity being slightly sub-solar. The one with priors is in very good agreement with the values reported in Table\,\ref{tab:stellarparam}. 

We re-determined the stellar parameters from {CARMENES} spectra of LP\,714-47 taken during the radial-velocity follow-up (Sect.\,\ref{sect:carmenes}). 
As explained by \citet{passegger_2018,passegger_2019}, we used a $\chi^2$ method together with a downhill simplex to obtain $T_{\rm eff}$, $\log g$, and [Fe/H] by fitting the most recent PHOENIX spectra \citep{Hauschildt1992,Hauschildt1993,passegger_2019} to the co-added CARMENES spectra in both VIS and NIR channels. 
The luminosity, radius, and mass were determined following \citet{Schweitzer2019A&A...625A..68S}; that is, together with the {\em Gaia} DR2 parallactic distance, we obtained the luminosity, $L_\star,$ by integrating the broad-band spectral energy distribution from the blue optical to the mid-infrared, using the Virtual Observatory Spectral Energy Distribution Analyser \citep[VOSA,][Fig.\,\ref{fig:sed}]{2008A&A...492..277B}.
After applying Stefan-Boltzmann's law to obtain the radius, $R_\star$, we used the linear mass-radius relation from \citet{Schweitzer2019A&A...625A..68S} to arrive at a mass, $M_\star$.
The rotational velocity upper limit was determined as in \citet{Reiners2018}, and the spectral type was estimated from a comparison with other early-type M dwarfs in their catalogue. 
All these results, as well as other basic parameters compiled from the literature, are listed in Table~\ref{tab:stellarparam}. 

Using the HIRES data (Sect.\,\ref{sect:HIRES}), we also applied the {\tt SpecMatch-Empirical} spectral characterisation tool \citep{yee:2017} to the template spectrum, which yielded $T_\mathrm{eff}=3869\pm70$\,K, [Fe/H]$=+0.38\pm0.09$\,dex, and $R_\star=0.60 \pm 0.10$\,R$_\odot$. This is in agreement with the results from the CARMENES data.

In the spectroscopic analysis, the CARMENES-derived effective temperature is found to be in between the TIC\,v8 and {\tt exofastv2} results. 
However, the mass of 0.59\,$\pm0.02$\,M$_\odot$ and radius of 0.58\,$\pm0.02$\,R$_\odot$ match the previous results. 
As an independent check on the derived stellar parameters, we also performed a very similar analysis of the spectral energy distribution following the procedures described in \citet{Stassun2018AJ....155...22S}, which resulted in $R = 0.590 \pm 0.015$\,R$_\odot$ and $M = 0.57 \pm 0.03$\,M$_\odot$.
The latter mass determination is also consistent with the one that is based on the spectroscopic $\log g$ and the photometry-based radius, $M = 0.55 \pm 0.06$\,M$_\odot$. Finally, our mass also agrees with the results from the mass-magnitude relation by \citet{Mann2019ApJ...871...63M}, which yields $M = 0.595 \pm 0.015$\,M$_\odot$ when applying the metallicity-independent version.

The CARMENES and HIRES spectral analyses result in a super-solar metallicity.  A note of caution on such a super-solar metallicity should be included, as a low-resolution spectrum of LP\,714-47 with ALFOSC at the 2.56\,m Nordic Optical Telescope points to a slightly sub-solar metallicity (see Appendix~\ref{alfosc_data} for further details).  The most accurate way to determine the metallicity of LP\,714-47 is thus an unsettled issue. This is not surprising, given the intensive discussion in the literature about the best way to determine M-star metallicities \citep[][and references therein]{2005A&A...442..635B,2006PASP..118..218W,2010ApJ...720L.113R,2013A&A...551A..36N,2014ApJ...791...54G,2015ApJ...804...64M,Alonso-Floriano2015A&A...577A.128A,2018MNRAS.479.1332M}.

For the following analysis of the radial velocity and transit light curve data, we use the CARMENES spectroscopically derived values. 
In particular, the stellar mass and radius, which are the most important parameters for exoplanet characterisation, are in good agreement with the various analysis results.

\subsection{Stellar activity and rotation period}
\label{sect:StellarActivity}

We investigated the stellar activity of LP\,714-47 using the {CARMENES} activity indicators (Sect.\,\ref{sect:carmenes}) presented by \cite{Schoefer2019A&A...623A..44S}. 
The variability across a variety of potential stellar activity indicators such as H$\alpha$, He~D3, Na~{\sc i}~D, Ca~{\sc ii} IRT, He~{\sc i} $\lambda$10830\,{\AA}, Pa-$\beta$, TiO bands, and the differential line width (dLW, Sect.\,\ref{sect:carmenes}), all show that LP\,714-47 is a relatively inactive star. 
In particular, the measured pseudo-equivalent width of the H$\alpha$ line is less than +0.3\,{\AA}; LP\,714-47 is thus considered to be an inactive star.  
This is consistent with the results of \cite{Jeffers2018A&A...614A..76J} and many other authors, who found that only about 10\,\% of early-M stars are H$\alpha$-active. 
There is an indication of variability in the TiO bands, the chromatic index (CRX), and the dLW, which would indicate a very low level of spot coverage.  

\begin{figure*}
    \centering
    \includegraphics[width=1.\textwidth]{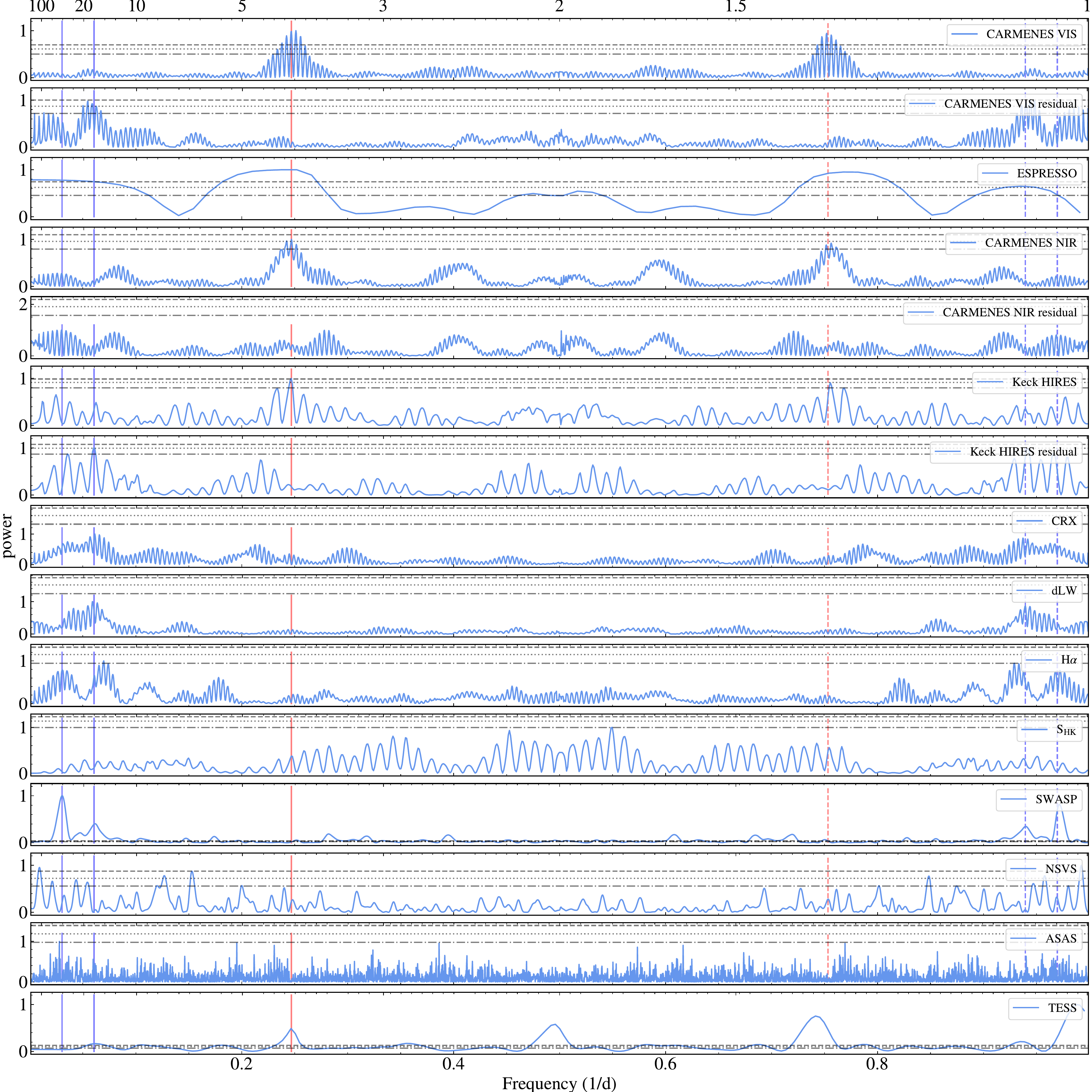}
    \caption{Periodogram of radial-velocity data with sufficiently long coverage, spectral activity indicators, long-term photometry, and {\em TESS} light curve. The red vertical line indicates the planet period, the blue vertical lines indicate the potential rotation period at 33\,d and its first harmonic, while the corresponding dashed lines mark the 1-day alias in the ground-based data sets. For CARMENES VIS and NIR, as well as for Keck/HIRES, we also show the periodogram of the residuals after subtracting the planet signal. Horizontal lines indicate the false alarm probablility of 10\%, 1\%, and 0.1 \%, respectively}
    \label{fig:GLS_alldata}
\end{figure*}

Using the generalised Lomb-Scargle (GLS) periodogram\footnote{\url{https://github.com/mzechmeister/GLS}} and the corresponding false-alarm probabilities \citep{Zechmeister2009A&A...496..577Z}, we investigated the available long-term ground-based photometry (see Fig.\,\ref{fig:GLS_alldata}). While the ASAS and NSVS data do not show significant periods, the WASP-South data show a peak at 33\,d and likely the first harmonic at 16\,d. We interpret the 33\,d signal as the stellar rotation period, as discussed further in Sect.\,\ref{subsec:rvActivity}. Finally, the very low stellar activity of LP\,714-47 is consistent with the
upper limit of 2~km\,s$^{-1}$ on its  projected rotational velocity and
with the relatively long rotation period of 33\,d.


\section{\label{sec:Analysis}Analysis}

The analysis of the various data sets was performed in a sequence of steps, with the modelling details described below. At first, we analyse the radial-velocity data and the transit photometry data separately. In the former, we focus on the planetary origin of the signals; in the latter, we focus on possible transit time variations. As a last step, we performed a combined analysis.

\subsection{\label{subsec:Method} Method and modelling details}

The analysis was done based on a collection of {\tt python} scripts for fitting Keplerian orbits for the radial velocity and transit data together with Gaussian process regression (GP) accounting for correlated noise, either due to stellar activity or due to coming from an instrumental source or caused by the Earth's atmosphere. The GP model is therefore used as a parametric noise model to account for residuals between observations and the planetary model. For the GP modelling, we used {\tt celerite} \citep{celerite2017}, which offers a fast and reliable implementation of GP regression, and two sets of {\tt celerite} kernels describing the correlation between each pair of data points, namely {\em REAL} and {\em SHO}. The former represents an exponential decay at a characteristic time scale $\tau$, while the latter is a stochastically-driven damped harmonic oscillator with an oscillator period $P$ and a damping time scale $\tau$, and can describe a quasi-periodic behaviour. Mathematical details are given in Appendix\,\ref{sec:ModDet}.

For the analysis of the radial-velocity data, we assume that the noise
is (mostly) due to stellar activity. Sufficiently long-living active
regions would cause correlations on time scale of their lifetimes and a (quasi) periodicity on the stellar rotation period or its harmonics, represented by the {\em SHO} kernel. Active regions with life times sufficiently below the rotation period or small-scale regions would cause correlated noise, which shows short-term correlations better represented by the {\em REAL} kernel. In the model, it is required for the same GP model to fit all radial-velocity data simultaneously. In order to constrain the GP regression, the fit of the GP parameters can simultaneously take the spectral activity indicators into account. We did not add a GP model for the transit data sets.

Each data set was analysed with an individual offset and with the option of a jitter, quadratically added to the measurement uncertainties, for each data set separately. We checked for a linear trend within the radial-velocity data, but found it to be consistent with zero. Therefore, we did not include a linear term in the RV fitting.

The Keplerian model had the following physical parameters: using the radial velocity we fitted the orbital period, $P$, the eccentricity, $e$, the longitude of periastron, $\omega$, and the time of periastron, $t_{\rm peri}$, as well as a semi-amplitude of the radial velocity variation, $K$. For the analytic transit light curve model \citep{MandelAgol2002}, we used the orbital period, the time of periastron, the orbital inclination, $i$, the planet-to-star radius ratio, $R_{p}/R_\star$, as well as the semi-major axis in units of the stellar radius, $a/R_\star$, as fit parameters. 

In the combined fit of radial-velocity data and transit photometry, we used the stellar mass and its uncertainty from Table\,\ref{tab:stellarparam} as input values and derived $a$ and $R_\star$ independently. Additionally, the planetary mass, radius, density, and equilibrium temperature could then be derived.

When investigating transit timing variations, the transit time of each transit was an additional free parameter. Depending on the limb darkening law applied, additional free parameters were needed. In particular, we used a quadratic limb darkening adding two parameters. Here, we only fitted the limb darkening parameters of the {\em TESS} light curve, but kept the others fixed using values from stellar atmosphere models for the ground-based \citep{Husser2013} and {\em Spitzer} data \citep{Claret2013} to limit the number of free parameters. Since the second coefficient of the quadratic limb darkening law was unconstrained by the {\em TESS} data, we also fixed to the value derived from the stellar atmosphere models. 

The fitting procedure started with an initial guess of the planetary parameters guided by a periodogram analysis. These parameters were optimised using {\tt scipy.optimize.minimize}. We then ran the Markov Chain Monte Carlo (MCMC) procedure {\tt emcee} \citep{emcee2013} with 400 walkers and 10\,000 steps after already 10\,000 burn-in steps, initialised with a Gaussian distribution around the previous best fit and a standard deviation 10 times larger than that of the first optimisation. We made sure that the initial distribution was far broader than the final one from the MCMC posteriors. Boundaries for the parameters were only set to guarantee positive definite values for the amplitude $K$, $R_{p}/R_\star$, and $R_\star$. The eccentricity is limited to bound orbits, the absolute value of the eccentricity was used when calculating the Keplerian models, as discussed in \citet[][Appendix~D]{Eastman2013PASP..125...83E}. We use uniform priors for all parameters within their bounds.

\subsection{\label{subsec:toi442b}Confirmation of LP 714-47 b from radial velocity variations}

For our initial values, we calculated the GLS periodogram and the corresponding false-alarm probabilities \citep{Zechmeister2009A&A...496..577Z} for all radial-velocity data and the spectroscopic and photometric activity indicators (Fig.\,\ref{fig:GLS_alldata}).
The {CARMENES} VIS and the {CARMENES} NIR and HIRES data sets all show a signal at the expected period of the transiting planet at 4.05\,d, with daily aliasing. The structure of the peaks show the mean separation of the two runs (1/220\,d$^{-1}$ and 1/75\,d$^{-1}$ for CARMENES and HIRES, respectively) and the run length of about 20\,d as width of the alias pattern. The signal is above the 0.1\,\% false-alarm probability (FAP) in the VIS and HIRES data, and at 1\,\% in the NIR data. 
The ESPRESSO data set is by itself too limited to show the 4.05\,d signal, but a broad peak in that period range is present in the periodogram of the data. Likewise, the sampling was also insufficient to detect the 4.05\,d signal in the PFS data set by itself; the 4.05\,d signal could not be detected in the periodogram of the iSHELL data alone.

After subtracting the signal at 4.05\,d, additional power is present in the periodograms in the range of 14--16\,d for VIS and HIRES, but not for NIR. The inspection of the periodograms of the spectroscopic activity indicators CRX, dLW, and H$\alpha$ revealed power in the 14--16\,d (dLW and H$\alpha$) and 25--40\,d (CRX) ranges, respectively. While the former period range overlaps with the second highest peak in the periodogram of the radial-velocity data, the latter is about twice that period. As mentioned in Sect.\,\ref{sect:StellarActivity}, the long-term photometry of WASP-South shows a clear peak at 33\,d, which we interpret as the stellar rotation period. 
Therefore, we used the activity indicators in order to discriminate between the signals of a true companion and of stellar activity in the radial-velocity data as discussed in the following.

For the detailed analysis, we ran a sequence of models with increasing complexity for the radial-velocity data (Table\,\ref{tab:RV_results}). For the model selection, we used the Bayesian Information Criterion (BIC) defined as $-2\ln \mathcal{L}_{\rm max} + k\ln N$ with the maximum likelihood $\mathcal{L}_{\rm max}$, the number of free parameters $k$, and the number of data points $N$. Following \citet{KassRaftery1995},  we  convert the difference of the BIC value of two models $\mathcal{M}_1$ and $\mathcal{M}_2$ into the  Bayes Factor (BF) by assuming a single  Gaussian  posterior  distribution, $\ln {\rm BF}_{21}=1/2 ({\rm BIC}_{\mathcal{M}_1}-{\rm BIC}_{\mathcal{M}_2})$. An appropriate threshold for the selection is $\ln BF_{21}>5$ according to \citet{KassRaftery1995}.

\begin{table}[t]
    \centering
    \caption{Sequence of models for the analysis of the radial-velocity data with $n=0\ldots2$ planets in circular (C) or Keplerian (K) orbits, with or without GP kernels or jitter terms with their corresponding BIC values. The BF is calculated relative to the model with the lowest BIC value. The selected model is in boldface.}
    \begin{tabular}{@{}cccrrrr@{}}
        \hline\hline
        $n$ & orbit &  GP & \multicolumn{2}{c}{BIC} & \multicolumn{2}{c}{BF} \\
                  &       &     & no jitter & jitter & no jitter & jitter \\
        \noalign{\smallskip}
        \hline
        \noalign{\smallskip}
                0 & $-$ &  $-$ &  8192 & 889&3755  & 103.5\\
                1 & C &  $-$   & 1360 & 752 & 339  & 35\\
                1 & K &  $-$   & 1344 & 761 & 331  & 39.5\\
                {\bf 1} & {\bf K} &  {\bf REAL}  & {\bf 693} & 720  & {\bf 4.5}  & 18\\
                1 & K &  SHO   & 696 & 720  & 6    & 18 \\
                2 & K &  $-$   & 709 & 705  & 12.5 & 10.5\\
                2 & K &  REAL  & 684 & 709  & 0    & 12.5\\
                2 & K &  SHO   & 690 & 719  & 3    & 17.5\\
      \noalign{\smallskip}
                  \hline
    \end{tabular}
    \label{tab:RV_results}
\end{table}

\begin{figure}
    \centering
    \includegraphics[width=0.49\textwidth]{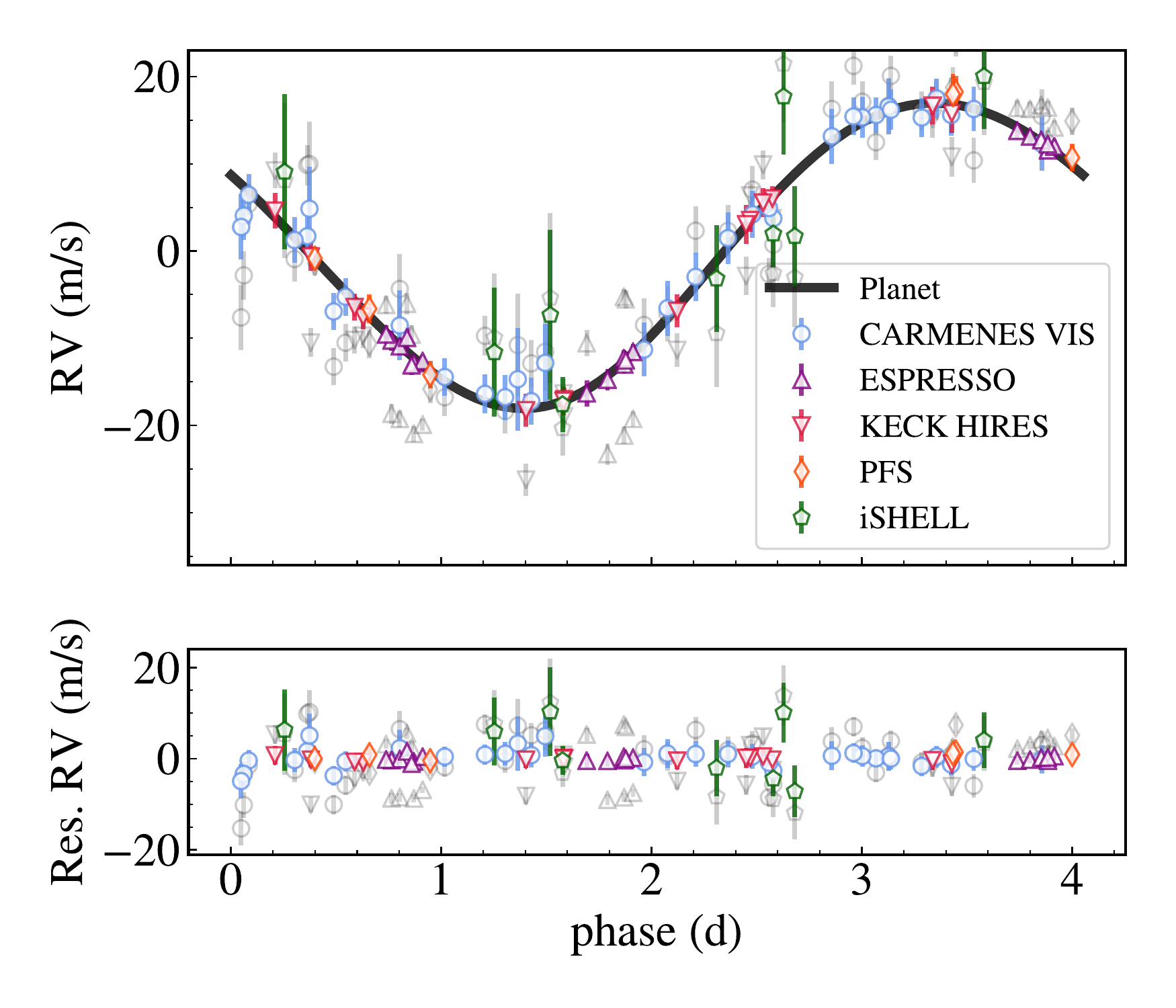}
    \caption{Radial-velocity data (grey symbols) and corrected for correlated noise by subtracting the contribution of the GP model (coloured symbols), and phase-folded to the planet period $P=4.052$\,d. CARMENES NIR RVs are not shown because of their larger uncertainties.}
    \label{fig:phase}
\end{figure}

Our base model was the no-planet model that allowed for instrumental offsets for the individual data sets. We then added one Keplerian signal with a period of 4.05\,d (see Table\,\ref{tab:orbit}). The eccentricity was small and compatible with zero, as expected for a close-in planet. We also checked the match between the mid transit time predicted from the fit of all radial-velocity data only with those from all transit data. From the former, the $1\,\sigma$ uncertainty is 0.095\,d compared to 0.0003\,d from the latter. From the radial-velocity data, the transit time of the first transit in the {\em TESS} data is predicted within $2\,\sigma$.
The improvement in the BIC value constitutes a clear detection of the transit signal period in the radial-velocity data (Table\,\ref{tab:RV_results}, Fig.\,\ref{fig:phase}). This improvement was complemented by the analysis of the high-spatial resolution imaging (Sect.\,\ref{subsec:AO}), which excluded any background eclipsing binary. Further confirmation was obtained from the comparison of the analysis of the visual ({CARMENES VIS, HIRES, PFS}) versus infrared ({CARMENES NIR, iSHELL}) data. The radial-velocity semi-amplitudes determined from both data sets agree within the uncertainties of our preferred model listed in Table\,\ref{tab:orbit}, showing that the strength of the signal is wavelength independent, as expected for a planetary signal.

\subsection{\label{subsec:rvActivity} \label{subsec:AdditionalPlanets} Correlated noise, stellar rotation, and search for additional planets}

\begin{figure}
    \centering
    \includegraphics[width=0.495\textwidth,trim=20 40 45 40,clip]{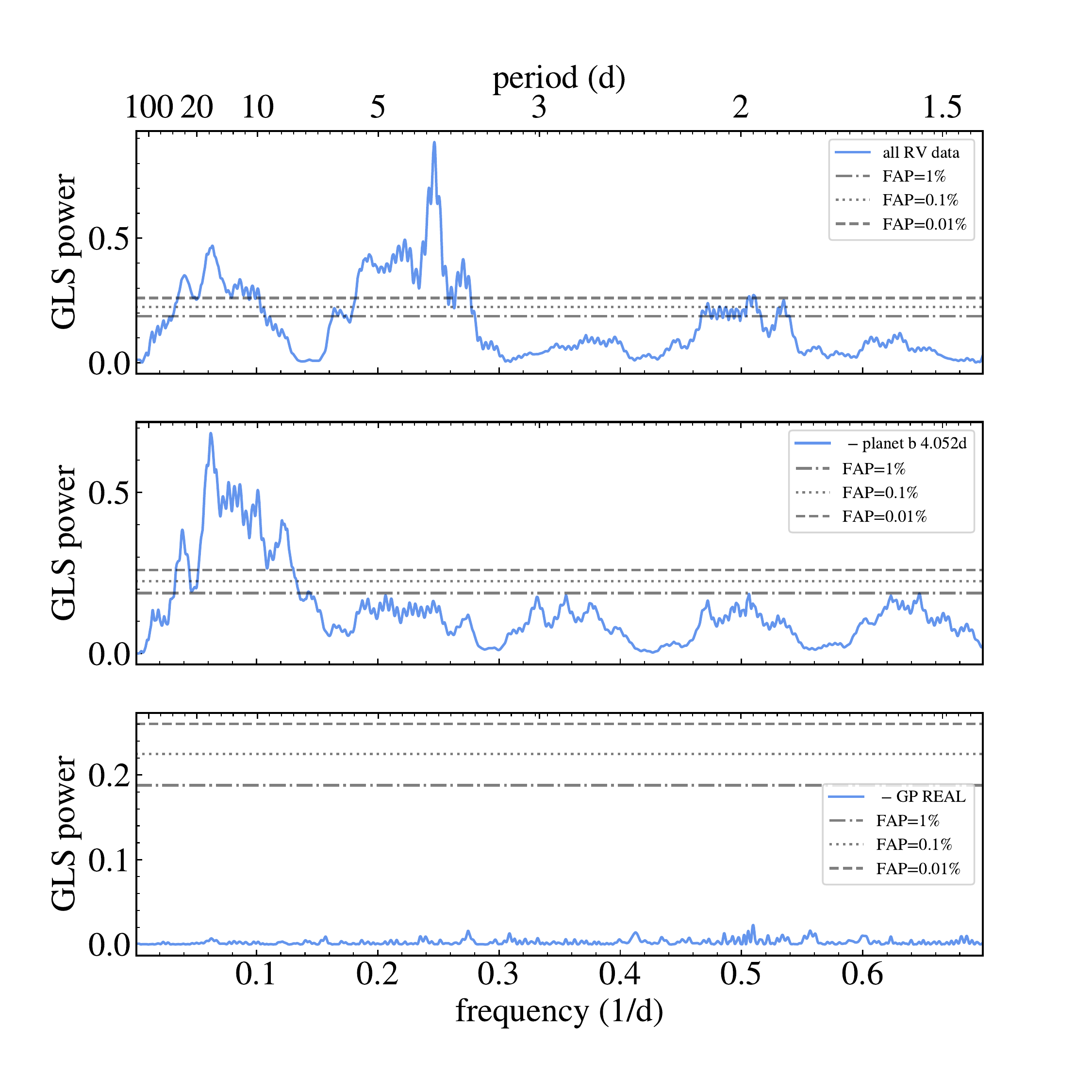}
    \caption{ GLS periodograms of the combined radial-velocity data corrected for individual offsets: Subsequent pre-whitening of the radial-velocity data ({\em top}) with the signal of LP\,714-47\,b ({\em middle}) and the GP modelling the correlated noise ({\em bottom}). Horizontal lines indicate the false alarm probablility of 1\%, 0.11\%, and 0.01 \%, respectively.}
    \label{fig:prewhite}
\end{figure}

\begin{figure}
    \centering
    \includegraphics[width=0.495\textwidth,trim=20 40 45 40,clip]{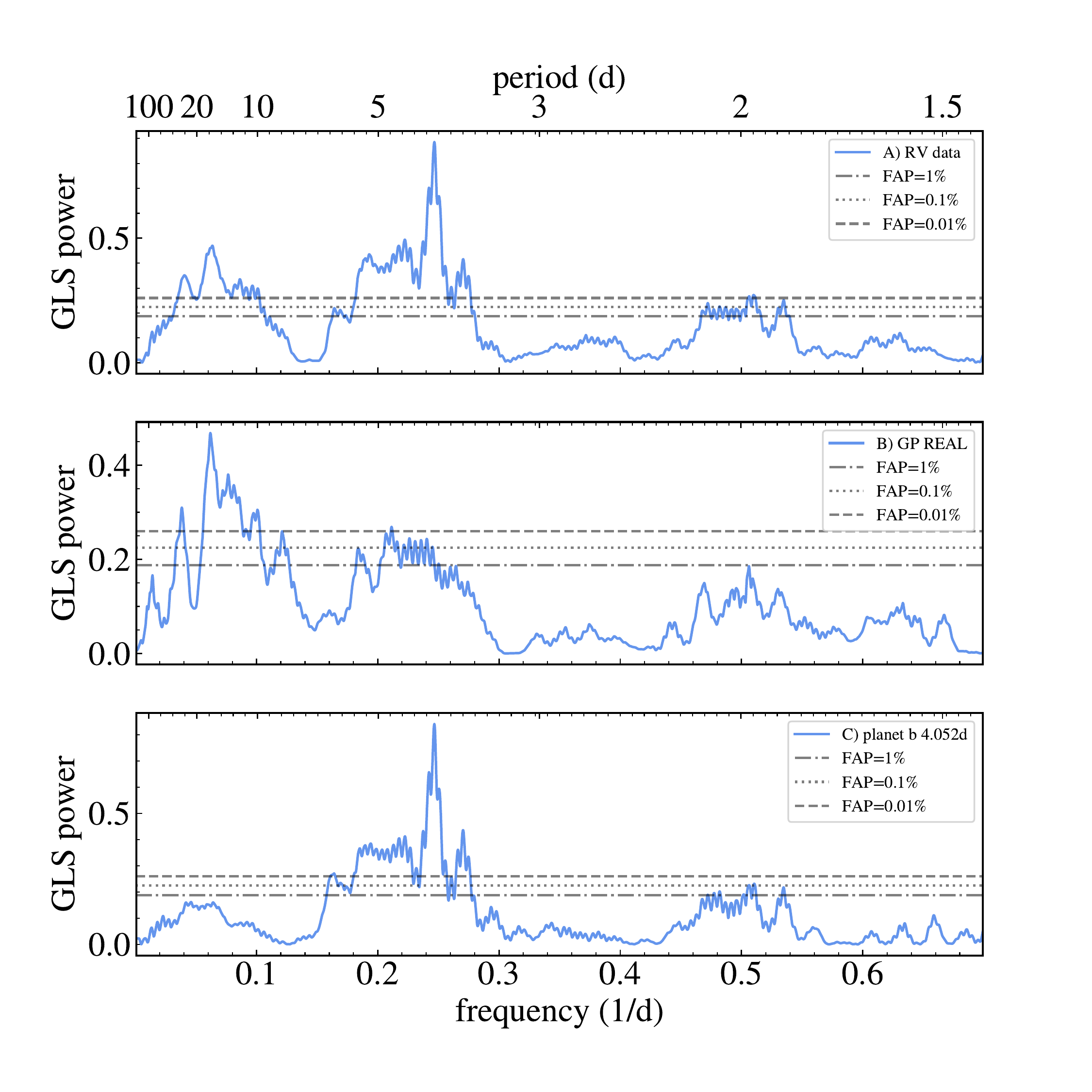}
    \caption{Periodogams of the two model components of the best-fit model: the radial-velocity data ({\em top}), the GP modelling the correlated noise ({\em middle}), and the planet ({\em bottom}). Horizontal lines indicate the false alarm probablility of 1\%, 0.11\%, and 0.01 \%, respectively.}
    \label{fig:signal}
\end{figure}

After subtracting the Keplerian signal, a second signal at about 16\,d (0.06\,d$^{-1}$) is detected in the periodograms of the residuals of individual data sets   (Fig.\,\ref{fig:GLS_alldata}, panels 2, 5, and 7 from the top) as well as in the combined radial velocity data (Fig.\,\ref{fig:prewhite}).  
The GLS-priodogram in this figure was calculated from all radial velocity data, corrected for individual offsets taken from the best-fit model (Table\,\ref{tab:orbit}).

As mentioned in Sect.\,\ref{sect:StellarActivity}, this period is close to half of the stellar rotation period of 33\,d (\ref{fig:GLS_alldata}) and this signal could be due to stellar activity rather than to a second planet. Modelling this signal as a second planet or as correlated noise shows a significant improvement in the BIC value (bottom five rows in Table\,\ref{tab:RV_results}).

A fit with a second Keplerian signal results in a good fit for a Neptune-mass planet (Table\,\ref{tab:orbit}, Figs.\,\ref{fig:allRVdata_2planets}, and \ref{fig:phase_planet2}). In combination with the {\em REAL} kernel, the two-planet model has the lowest BIC value. The eccentricity of the second planet would be rather high ($e=0.26$, without GP). Nonetheless, we checked the dynamical stability of this hypothetical two-planet system, which would be close to a 4:1 mean motion resonance, using the Hill stability criterion in the angular momentum deficit (AMD) framework introduced by \citet{Laskar2018} implemented in the {\tt Exo-Striker} package \citep{ExoStriker}. A second (co-planar) planet at a 4:1 period commensurability would be stable up to an eccentricity of about 0.4. The {\em TESS} data did not show indications of another transit. However, transits corresponding to most of the viable orbital solutions of this potential planet would have gone undetected due to the gap in the {\em TESS} data during the time needed for the data downlink.

Alternatively, we fit the second signal assuming it to be caused by correlated noise due to stellar activity with {\em REAL} and {\em SHO} kernels. The oscillator period of the SHO kernel representing the stellar rotation periods was constrained by the simultaneous fit of the chromatic index with a wide uniform prior between 5\,d and 200\,d. Despite the simultaneous fit of the activity indicator, the fit with the {\em SHO} kernel resulted in a strongly damped oscillator equivalent to an exponential decay, which was better modelled using the {\em REAL} kernel. This is also indicated by the better BIC value of the fits with the {\em REAL} kernel (Table\,\ref{tab:RV_results}). In Fig.\,\ref{fig:prewhite}, the middle and lower panels show the subsequent removal of the two model components, that is, the one-planet and the GP model (REAL kernel).  The models are sampled at the times of observation. In Fig.\,\ref{fig:signal}, the periodograms of these two model components are shown in comparison to the full dataset.

The two-planet model with GP has a lower BIC value compared to the one-planet model that includes a GP model, however, the Bayes factor differs slightly less than the threshold of five. This weak detection could argue for using this as our final best model, but there is currently more evidence for an interpretation of the 16\,d signal as correlated noise due to stellar activity. First, the activity indicators showed periods at 33\,d (H$\alpha$) and 16\,d (dLW and CRX), as seen in Fig.\,\ref{fig:GLS_alldata}. The photometric monitoring with {WASP-South} showed a peak at 33\,d. Secondly, a given GP kernel might not adequately represent the effect of stellar activity. Therefore, we interpret the 33\,d as the stellar rotation period and the 16\,d signal is potentially the second harmonic. 

While none of these arguments rule out a second planet, its confirmation would need a longer time base to check the coherence of this signal. We show the comparison of the two-planet model with the GP in more detail in Appendix~\ref{sect:2-planet} (in particular Figs.\,\ref{fig:allRVdata_2planets}, \ref{fig:phase_planet2}, \ref{fig:prewhite_2planets}, and~\ref{fig:signal_2planets}). 

All the variants of the modelling do not have an impact on the parameters derived for the planet\,b, as demonstrated by the comparison between the parameters for the one planet+GP, the two planets, and the two planets+GP models shown in Table\,\ref{tab:orbit}. 

\begin{table*}
   \centering
   \caption{\label{tab:orbit} Model and reference statistical parameters from the simultaneous fit of radial velocities and transit photometry\tablefootmark{a}. Uncertainties are given as 68\% intervals.}
   \begin{tabular}{@{}lccccc@{}}
      \hline 
      \hline 
      \noalign{\smallskip}
      Parameter & one planet+GP\tablefootmark{b} & \multicolumn{2}{c}{two planets} & \multicolumn{2}{c}{two planets+GP} \\
      planet & b & b & c & b & c \\
      \noalign{\smallskip}
      \hline 
      \noalign{\smallskip}
      $P$ [d]        &  $4.052037\pm 0.000004$& $4.052039\pm 0.000005$ & $16.04\pm 0.04$ & $4.052037\pm 0.000005$ & 16.06$^{+0.10}_{-0.09}$\\[0.3em]
      $K$ [m/s]      &  $17.6\pm 0.8$         & $17.5^{+0.3}_{-0.2}$   & $6.9\pm 0.3$&17.4 $^{+0.5}_{-0.6}$&6.3$^{+0.4}_{-0.7}$\\[0.3em]
      $e$            &  $0.04\pm 0.02$        & $0.030^{+0.008}_{-0.009}$&$0.26\pm 0.06$& $0.04\pm 0.02$&0.12$^{+0.16}_{-0.09}$\\[0.3em]
      $\omega$ [deg] & $219\pm 19$            & $191^{+12}_{-9}$       & $297\pm 11$2& $21\pm 7$&92$^{+44}_{-31}$\\[0.3em]
      $t_{\rm peri}$ [d]\tablefootmark{c}
                     &  $1.80\pm 0.23$        & $1.48^{+ 0.15}_{-0.10}$& $7.3\pm 0.9$& $1.82\pm 0.09$&14.0$^{+0.5}_{-0.4}$\\[0.3em]
      $i$ [deg]      & $87.3\pm 0.2$          & $87.2^{+0.2}_{-0.2}$ & ... & $87.3\pm 0.2$ & ...  \\[0.3em]
      $R_{\rm p}/R_\star$
                     &  $0.0751\pm 0.0009$    & $0.0754^{+0.0010}_{- 0.0009}$& ... & $0.0750^{+0.0010}_{- 0.0009}$& ... \\[0.3em]
      $R_\star$ [$R_\odot$]
                     &  $0.57\pm0.02$         & $0.58\pm0.02$ & ... & $0.57\pm0.02$ & ... \\[0.3em]
      $u_1$          &  $0.48\pm 0.08$        & $0.5\pm 0.1$ & ... &$0.47\pm 0.10$& ... \\[0.3em]               
      $u_2$\tablefootmark{d}& $0.245$                & $0.245$ & ... & $0.245$ & ... \\[0.3em]               
      \hline 
      \noalign{\smallskip}
      \multicolumn{5}{@{}l}{Derived parameters}\\
      \noalign{\smallskip}
      \hline 
      \noalign{\smallskip}
      $a$ [au]        &  $0.0417\pm 0.0005$   & $0.0417\pm 0.0005$   & $0.104\pm 0.001$& $0.0417\pm 0.0005$ & $0.104\pm 0.001$\\[0.3em]
      $\lambda$ [deg]\tablefootmark{c,e} 
                      &  $336.1\pm 0.8$       & $337.0\pm 0.9$& $135\pm 12$& 336.8$^{+1.9}_{-1.8}$&$134\pm 27$\\[0.3em]
      $t_{\rm c}$ [d]\tablefootmark{c,f}
                      &0.38421 $^{+ 0.00025}_{- 0.00024}$ &0.38401 $^{+0.00041}_{-0.0004}$&13.22$^{+ 0.92}_{-0.97}$& 0.38419 $^{+0.00027}_{-0.00023}$ & 13.51 $^{+1.51}_{-1.69}$\\[0.3em]
      $m_{\mathrm p}$  [M$_\oplus$] 
                      &  $30.8\pm 1.5$ &  $30.7\pm 0.8$ & $18.4\pm 1.0$& $30.5\pm 1.2$ &17.2$^{+1.4}_{-2.2}$\\[0.3em]
      $a/R_{\star}$ & 15.9 $^{+1.0}_{-0.7}$& 15.8$^{+1.2}_{-0.8}$ & 39.5 $^{+3.1}_{-2.0}$& 15.7 $^{+ 1.0}_{- 0.8}$ & 39.4 $^{+2.5}_{-2.0}$\\[0.3em]
      $R_{\mathrm p}$ [R$_\oplus$]
                      &  $4.7\pm 0.3$  & $4.8\pm 0.3$  & ... & $4.7\pm 0.3$ & ... \\[0.3em]
      $\rho_{\mathrm p}$ [g/cm$^{3}$]
                      &  $1.7\pm 0.3$  & $1.5^{+0.3}_{-0.2}$& ... & $1.7^{+0.4}_{-0.3}$& ... \\[0.3em]
      $T_{\rm eq}$ [K]\tablefootmark{g} & 700 $^{+19}_{-24}$ & 703 $^{+21}_{-29}$& 444 $^{+13}_{-18}$& 706 $^{+21}_{-25}$& 445 $^{+13}_{-16}$\\[0.3em]
      \hline 
      \noalign{\smallskip}
      \multicolumn{5}{@{}l}{Instrumental parameters offset and weighted {\em rms}}\\
      \noalign{\smallskip}
      \hline 
      \noalign{\smallskip}
      CARMENES VIS [m/s]      &  --2.5 $^{+ 1.3}_{- 1.5}$ \,\,\,\,1.64 & \multicolumn{2}{c}{--1.0 $^{+0.5}_{-0.4}$\,\,\,\,4.28} &\multicolumn{2}{c}{--2.4$^{+0.8}_{-0.7}$ \,\,\,\,1.85} \\[0.3em]
      CARMENES NIR [m/s]      &   0.0 $^{+ 2.7}_{- 2.6}$ \,\,\,\,12.3& \multicolumn{2}{c}{5.6 $^{+0.5}_{-0.4}$\,\,\,\,11.8} &\multicolumn{2}{c}{--2.8$^{+0.9}_{-0.8}$\,\,\,\,12.3}\\[0.3em]
      ESPRESSO [m/s] &   3.6 $^{+ 3.2}_{- 3.1}$ \,\,\,\,0.50& \multicolumn{2}{c}{3.1 $^{+0.2}_{-0.3}$\,\,\,\,1.26} &\multicolumn{2}{c}{4.8$^{+0.6}_{-0.7}$ \,\,\,\,0.52}\\[0.3em]
      HIRES [m/s]    &   2.3 $^{+ 1.9}_{- 1.7}$ \,\,\,\,0.44& \multicolumn{2}{c}{1.5 $^{+0.4}_{-0.5}$\,\,\,\,2.80} &\multicolumn{2}{c}{3.5$^{+0.7}_{-1.0}$ \,\,\,\,0.42}\\[0.3em]
      PFS [m/s]      &  --3.6 $^{+ 2.4}_{- 2.5}$ \,\,\,\,0.65& \multicolumn{2}{c}{--0.6 $^{+0.5}_{-0.3}$\,\,\,\,2.49} &\multicolumn{2}{c}{--4.6$^{+1.0}_{-0.7}$ \,\,\,\,0.90}\\[0.3em]
      iSHELL [m/s]   &  --0.3 $^{+ 2.8}_{- 3.0}$ \,\,\,\,5.03& \multicolumn{2}{c}{0.02 $^{+0.2}_{-0.1}$\,\,\,\,7.61} &\multicolumn{2}{c}{0.0$^{+0.8}_{-1.0}$ \,\,\,\,5.49}\\[0.3em]
\noalign{\smallskip}
      \hline 
      \noalign{\smallskip}
       \multicolumn{5}{@{}l}{GP hyper parameters}\\
      \hline 
      \noalign{\smallskip}
       variance [m$^2$/s$^2$]   &  27 $^{+ 8}_{- 6}$     & \multicolumn{2}{c}{...}&\multicolumn{2}{c}{16.9$^{+5.5}_{-4.6}$}\\[0.3em]
      $\tau_{\rm d}$ [d] &  3.2 $^{+ 1.5}_{- 0.9}$& \multicolumn{2}{c}{...}&\multicolumn{2}{c}{2.2$^{+1.3}_{-0.7}$}\\[0.3em]
      \noalign{\smallskip}
      \hline 
      \noalign{\smallskip}
      $-\ln L$ &  316 &  \multicolumn{2}{c}{336} &  \multicolumn{2}{c}{303} \\[0.3em]
      red. $\chi^2$ & 0.81 & \multicolumn{2}{c}{2.64} &  \multicolumn{2}{c}{1.04} \\[0.3em]
      \noalign{\smallskip}
      \hline 
    \end{tabular}
    \tablefoot{
        \tablefoottext{a}{The planetary parameters of the transiting planet from the different models are within 1-sigma of each other, proving the consistency of this signal. Posterior distributions of the parameters are displayed in Fig.\,\ref{fig:MCMC_planet}. The derived values for semi-major axis and planetary mass take the stellar mass uncertainty listed in Table\,\ref{tab:stellarparam} into account. The likelihood $-\ln L$ and the reduced $\chi^2$ is given for the RV fit.}
        \tablefoottext{b}{We found the best model to be the one\,planet+GP model.}
        \tablefoottext{c}{Reference time is BJD=2458438.}
        \tablefoottext{d}{Fixed.}
        \tablefoottext{e}{Mean longitude.}
        \tablefoottext{f}{Time of inferior conjunction.}
        \tablefoottext{g}{Assuming zero albedo and full heat redistribution.}
    }
\end{table*}

\subsection{\label{subsec:transits}Analysis of the transit light curves}

\begin{figure}
    \centering
    \includegraphics[width=0.49\textwidth]{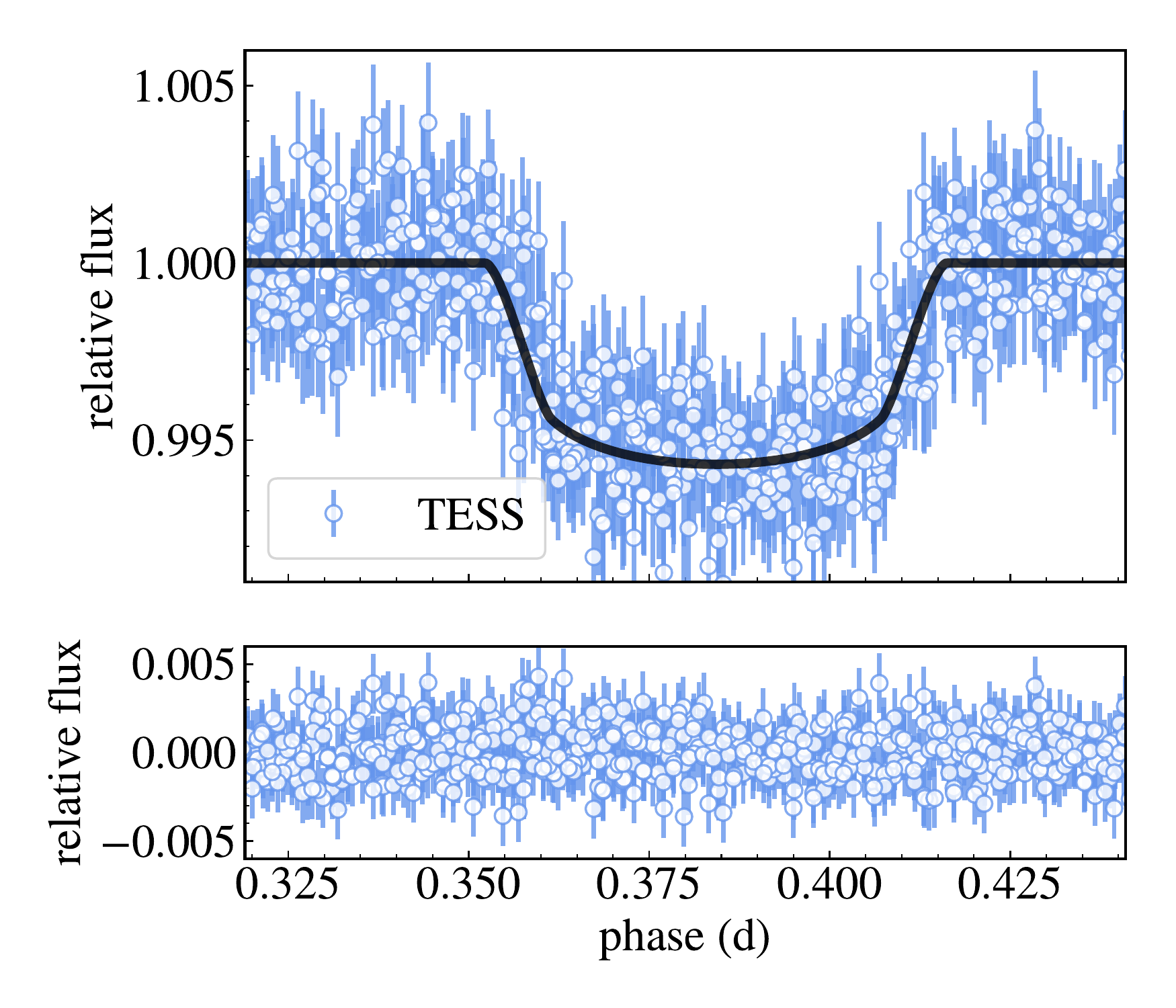}
    \includegraphics[width=0.49\textwidth]{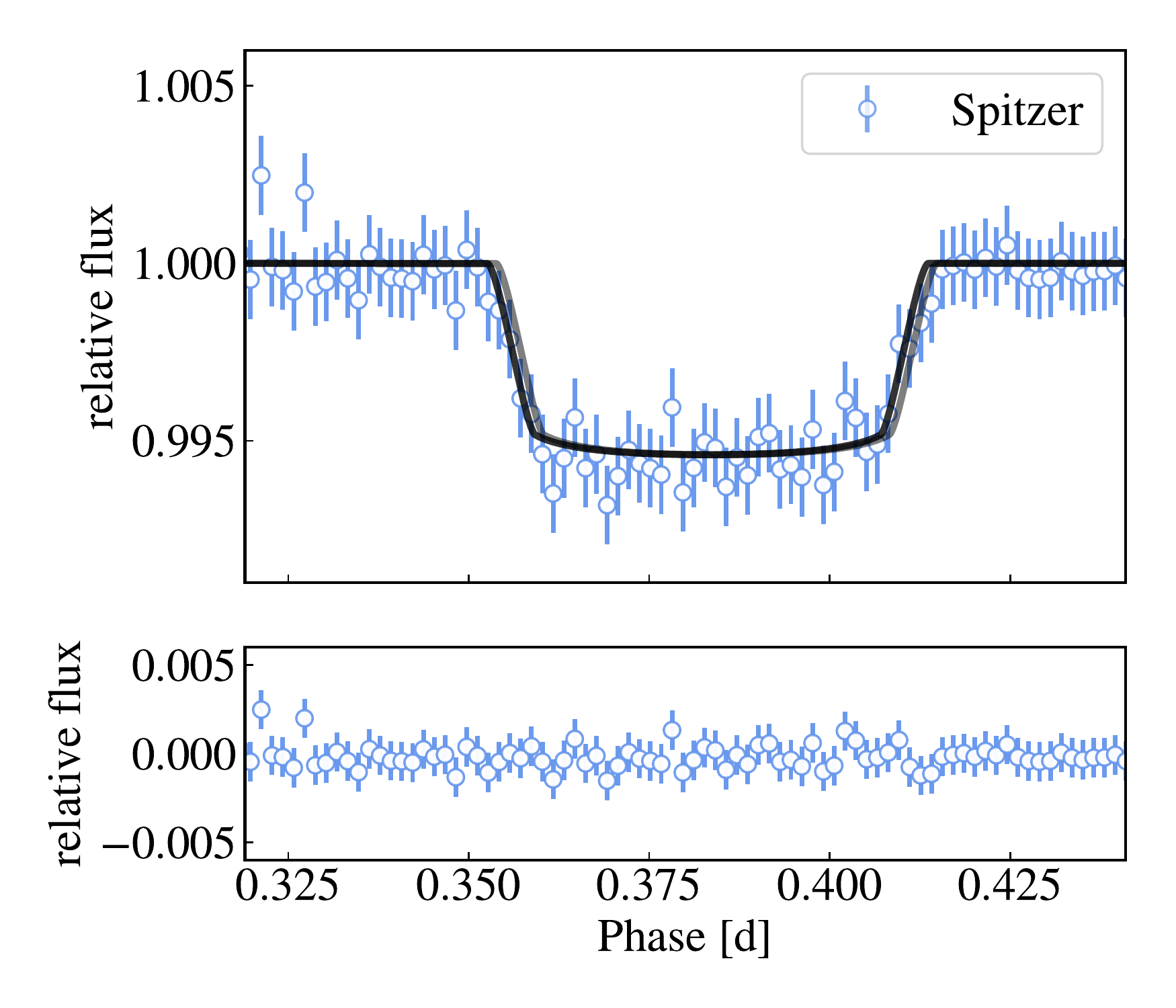}
    \caption{Six transits observed by {\em TESS} ({\it top}) and one transit from {\em Spitzer} ({\it bottom}) phase-folded to the transiting planet period of 4.052\,d. The grey model in the lower panel corresponds to the best joint model fit, whereas the black line allowed for individual mid transit time shifts (Sect.\,\ref{subsec:transits}). The shift is of the order of 2\,min.}
    \label{fig:phaseTR}
\end{figure}

\begin{figure}
    \centering
    \includegraphics[width=0.49\textwidth]{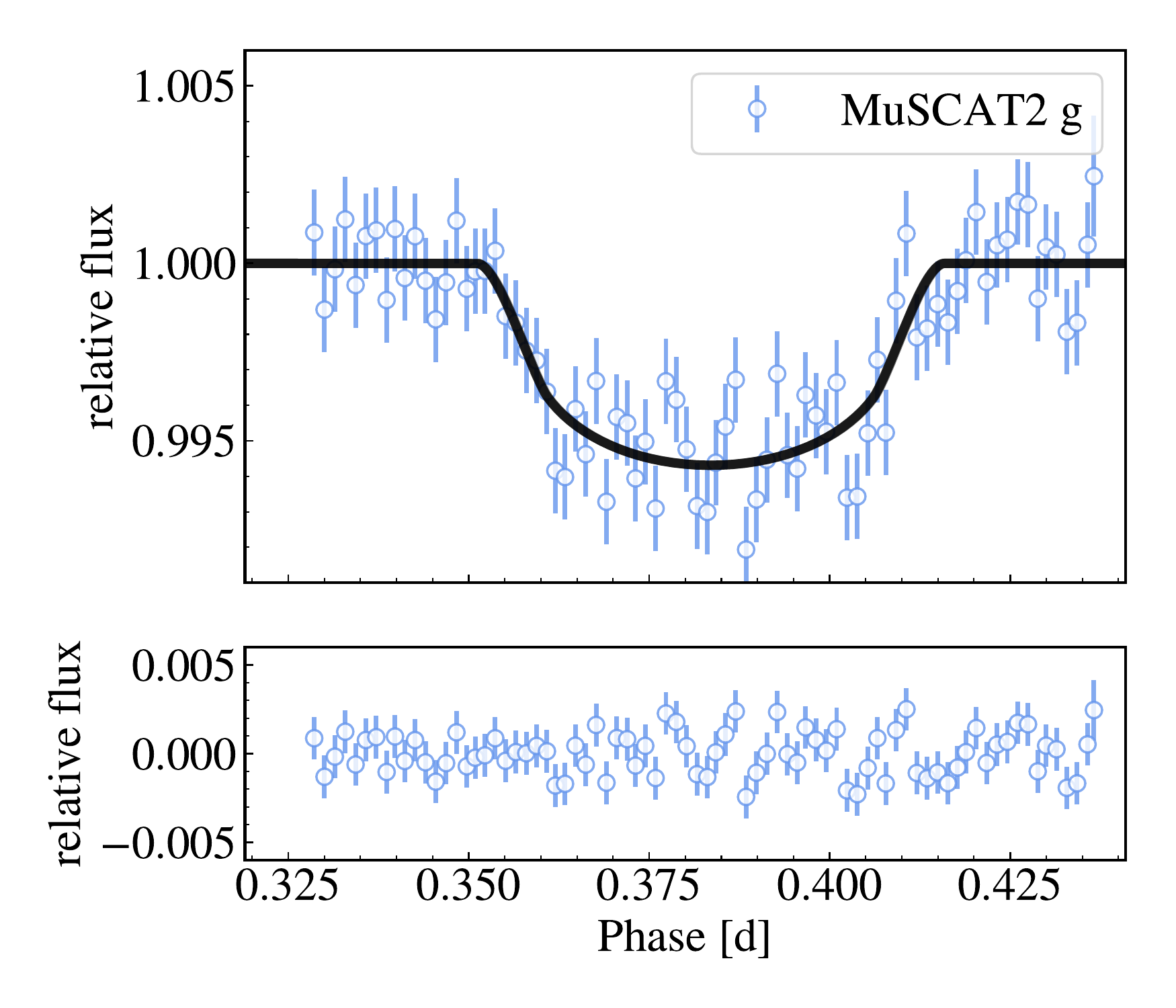}
    \includegraphics[width=0.49\textwidth]{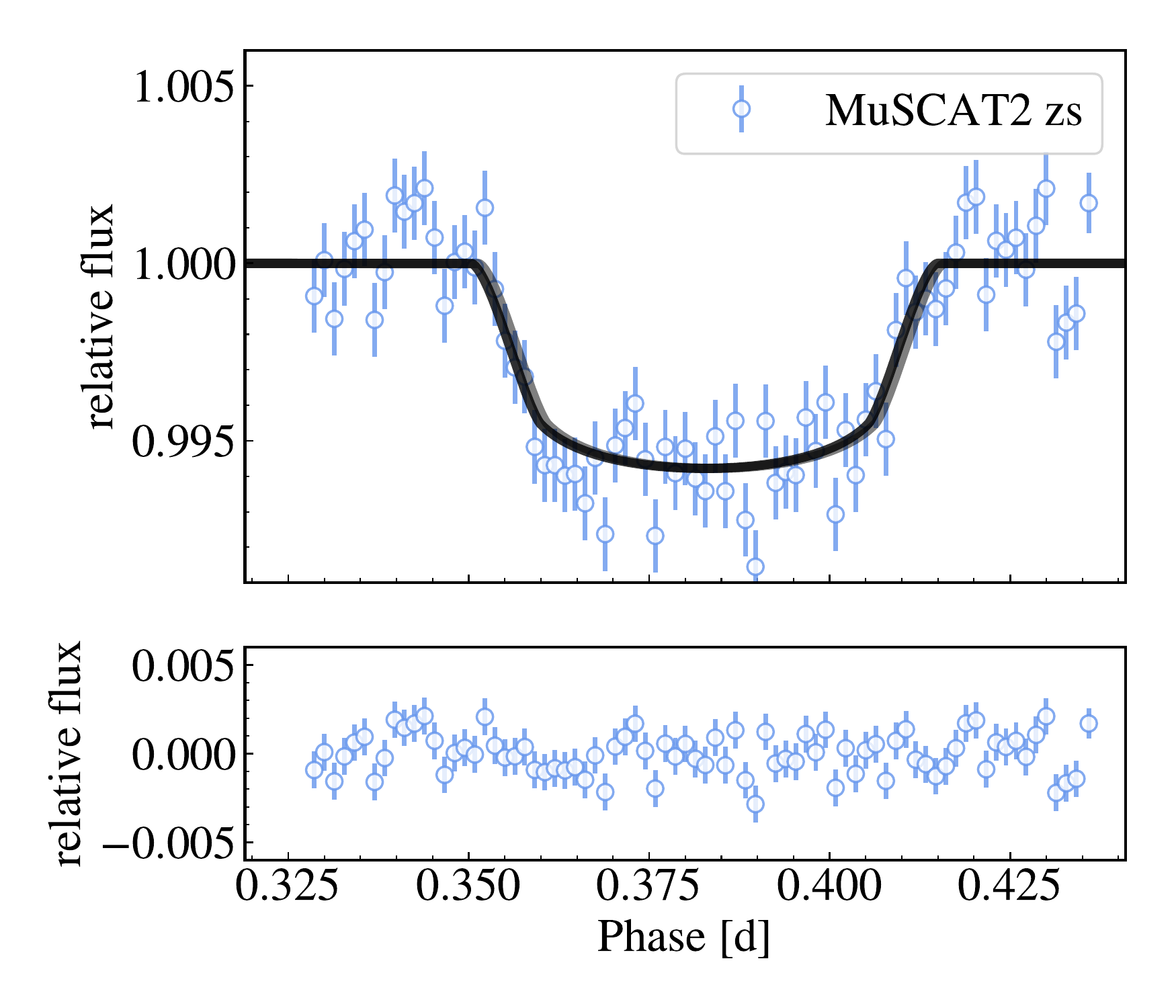}
    \caption{MuSCAT2 $g$ ({\it top}) and $z_s$ ({\it bottom}) data phase-folded to the planet period. The grey model allowed in the lower panel corresponds to the best joint model fit, whereas the black line allowed for individual mid transit time shifts (Sect.\,\ref{subsec:transits}).The other ground-based transit light curves are shown in Fig.\,\ref{fig:ttvs}. We note the effect of the different limb darkening in the various filters.}
    \label{fig:phaseTR_Muscat}
\end{figure}

As the next step in our analysis, we performed a simultaneous fit of the transit light curves (Figs.\,\ref{fig:phaseTR}, \ref{fig:phaseTR_Muscat}, \ref{fig:transits}) to obtain the additional planet parameters, namely the planet-to-star radius ratio and a refined orbital period (see Sect.\,\ref{subsec:Method}). We first used the linear ephemeris to calculate all mid-transit times during the fit procedure.
The good fit of the ground-based transit follow-up light curves additionally confirms the planetary origin of the 4.05\,d signal. 

We then allowed for small shifts in the mid-transit times for each transit.
The very good match of all the transits with a Keplerian model indicates very small or absent TTVs and, hence, potential planet-planet interaction below the current detection threshold. The possible indication of a small shift in mid transit time of the TUG, GMU, and MONET data (Fig.\,\ref{fig:ttvs}), on the order of 3\,min, can be attributed to correlated noise. A small transit midpoint shift of about 2\,min ($1 \sigma$ deviation) is also visible in the {\em Spitzer} light curve (Fig.\,\ref{fig:phaseTR}). By fitting the five {\em TESS} transits individually, we also see a scatter in transit time by that amount. The transit times were also checked and confirmed independently within the team using {\tt juliet} and {\tt exofast}. 

This marginal transit time variation in the {\em Spitzer} light curve may indicate that a less massive planet could be hidden in the data. As a result, we explored this possibility using {\tt TTVFaster} \citep{TTVFaster}. At a 2:1 orbit commensurability or mean motion resonance to planet\,b, a low-mass planet would produce TTVs of the order of minutes, strongly depending on the period and mass ratio, while its radial velocity signal could be sufficiently low to escape detection in the current data. 
A more in-depth investigation of the possibility of an additional planet is beyond the scope of this paper and requires more data, in particular, more transit-time measurements for planet\,b.

\subsection{Final model for \texorpdfstring{LP\,714-47\,b}{LP~714-47~b}}

\begin{figure*}
    \centering
    \includegraphics[width=0.99\textwidth]{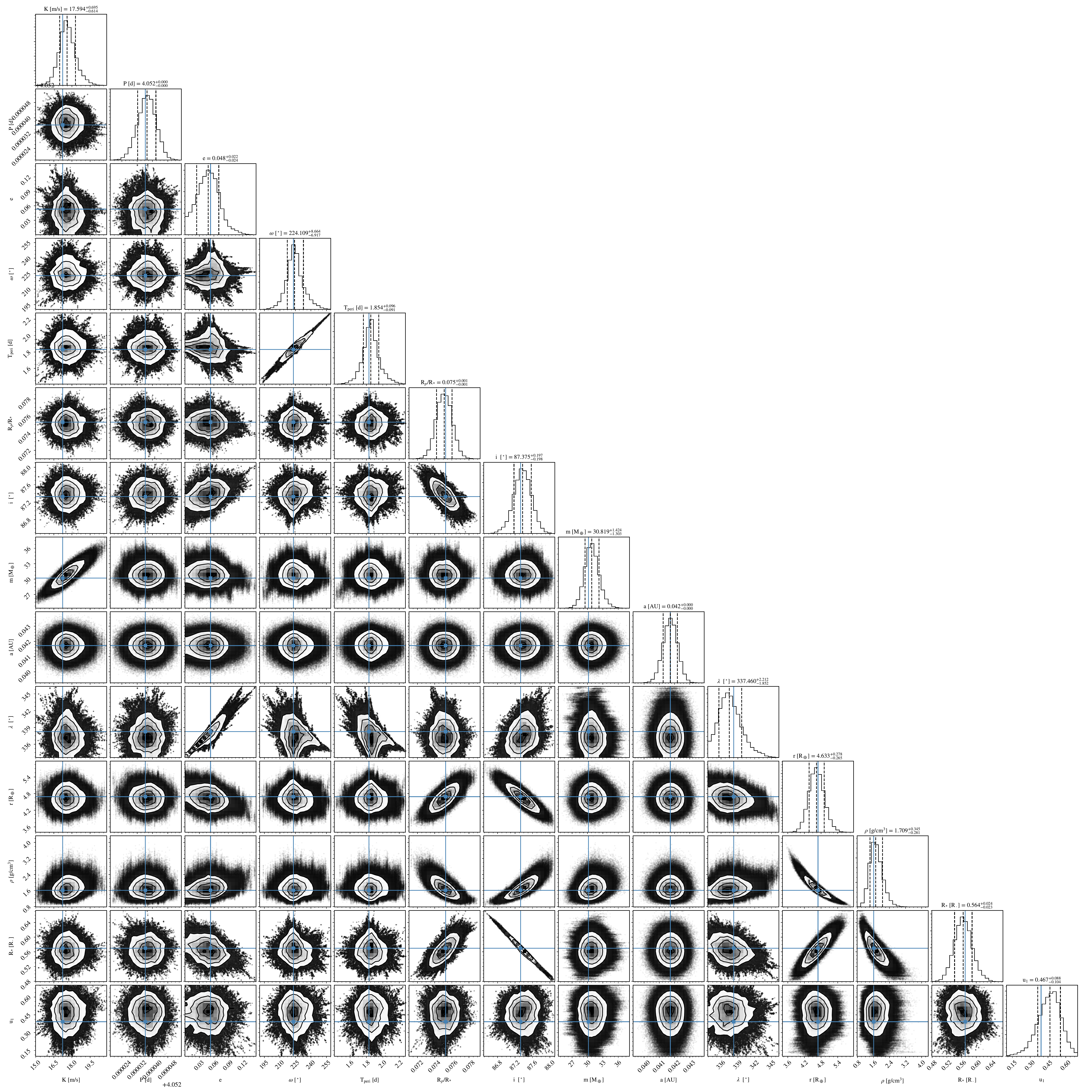}
    \caption{Posterior distribution of planetary parameters from our best-fit model.}
    \label{fig:MCMC_planet}
\end{figure*}

In a final analysis step, we simultaneously fit the radial-velocity data together with the transit light curves. The simultaneous fit slightly improves the parameters. The results are listed in Table\,\ref{tab:orbit}. In summary, we firmly detect a transiting planet with about twice the mass of Neptune ($m_{\mathrm b}=30.8\pm1.5$ M$_\oplus$) orbiting the early M star LP\,714-47 at the period of the signal reported for TOI\,442.01. The radius of $r_{\rm b}=4.7\pm0.3$ R$_\oplus$ makes LP\,714-47\,b a Neptune-like planet with a mean density of $\rho=1.7\pm 0.3$\,g\,cm$^{-3}$. We obtain the stellar radius from the stellar mass prior and the transit parameters, which results in $R_\star=0.57\pm0.02$ R$_\odot$. This is in excellent agreement with the spectroscopic radius reported in Table\,\ref{tab:stellarparam}. Additional radial velocity variations are most likely due to stellar activity. The components of our preferred model (one planet+GP) are shown in Fig.\,\ref{fig:signal}. The GLS-periodogram on the top panel is computed from the combined radial-velocity data, corrected for individual offsets (Table\,\ref{tab:orbit}). The middle and bottom panels show the two model components, that is, the GP model and the planetary model. The comparison to the radial-velocity data is shown in Fig.\,\ref{fig:allRVdata}, as well as to the phase-folded data in Fig.\,\ref{fig:phase}. In the latter, we omit CARMENES NIR data due to larger uncertainties as they would not contribute any additional information. A comparison of our final model to the transit photometric data sets is displayed in Figs.\,\ref{fig:phaseTR}, \ref{fig:phaseTR_Muscat}, and \ref{fig:ttvs}. The posterior parameter distribution is presented as a corner plot in Fig.\,\ref{fig:MCMC_planet}.

\section{\label{sec:Discussion}Discussion and conclusions}

\begin{figure}
    \centering
    \includegraphics[width=\linewidth]{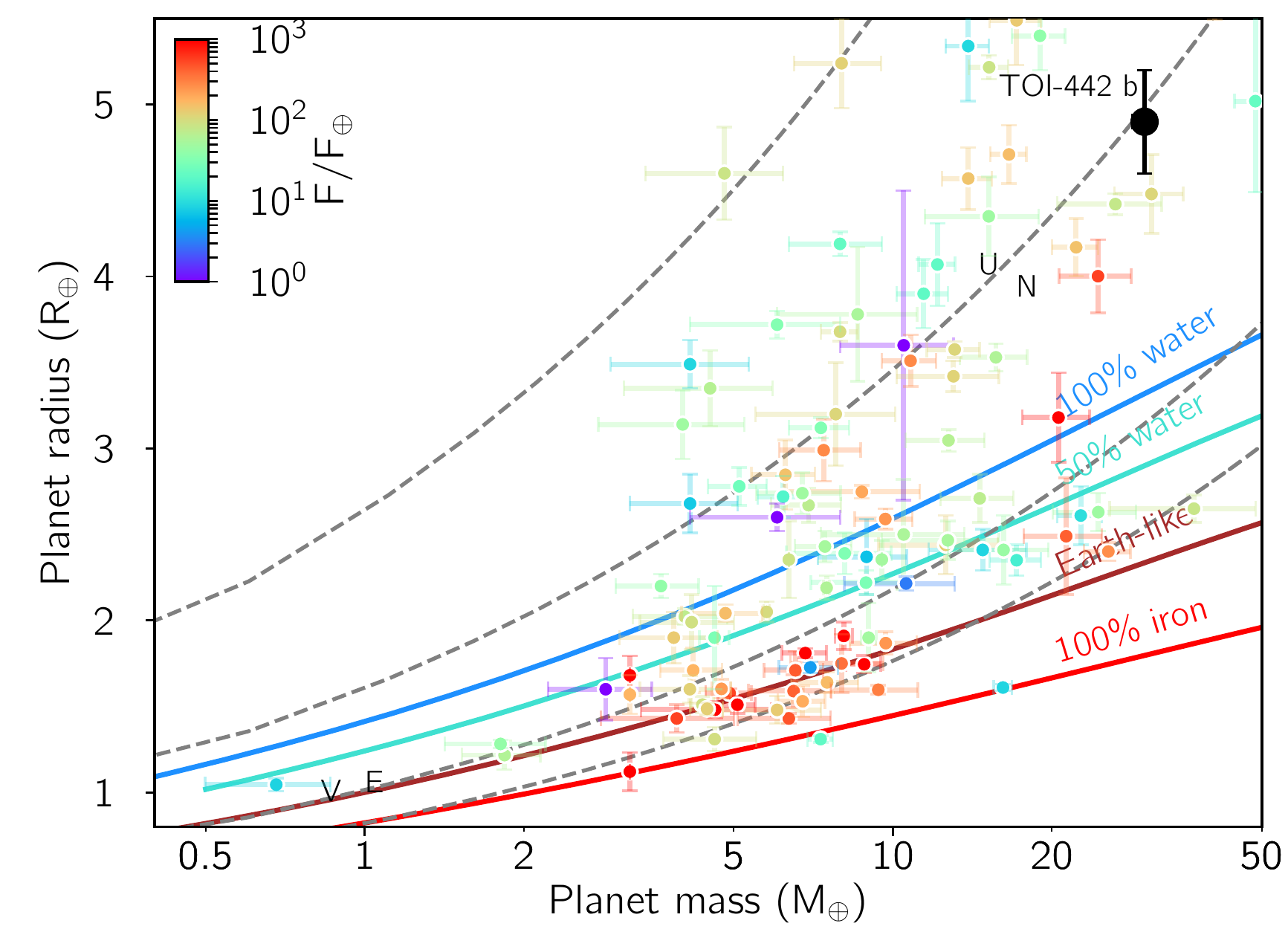}    
    \caption{\label{fig:massradius} Mass-radius diagram for low-mass planets. Mass-radius relations for various compositions are overlayed. Coloured lines are planetary models by \citet{Zeng2016ApJ...819..127Z} for idealised pure compositions. Dashed lines indicate constant mean densities. The incoming stellar radiation in present terrestrial units is colour-coded.}
\end{figure}

We used radial-velocity data from CARMENES, ESPRESSO, HIRES, PFS, and iSHELL, and light curves from {\em TESS}, {\em Spitzer}, and ground-based photometry, as well as high-resolution AO imaging using Gemini/NIRI to confirm the planetary nature of LP\,714-47\,b and determine the parameters of the planetary system with high accuracy. The simultaneous fit of radial velocities and transits allowed a determination of the planetary mass and radius within 5\,\% and 6\,\%, respectively, while the planet-to-star radius ratio was determined with an even smaller fractional uncertainty of 1.5\,\%. The limiting factor was the uncertainty in the host star mass. Comparing these uncertainties to those listed in the NASA Exoplanet Archive\footnote{\url{https://exoplanetarchive.ipac.caltech.edu/}} LP\,714-47\,b is among the well characterised planets, allowing for a meaningful comparison with planet structure models (Fig.\,\ref{fig:massradius}).

As discussed in Sect.~\ref{sec:Analysis}, the signal at 16\,d is significant. From the existing data, it is, however, not possible to identify this signal in the radial-velocity data as a second planet due to the proximity of the period to half of the stellar rotation period at 33\,d. We therefore do not claim the detection of a second planet in a 16\,d orbit based on the current data set. The transit time of the {\em Spitzer} light curve may indicate a weak TTV, which could be due to a rocky planet close to 2:1 mean motion resonance. A longer monitoring of the radial velocity variations to check the coherence of the 16\,d signal, as well as the re-visit of {\em TESS} in its extended mission to check for changes in transit times and inclination, will shed new light on this issue. 

\begin{figure*}
    \centering
    \includegraphics[width=\linewidth]{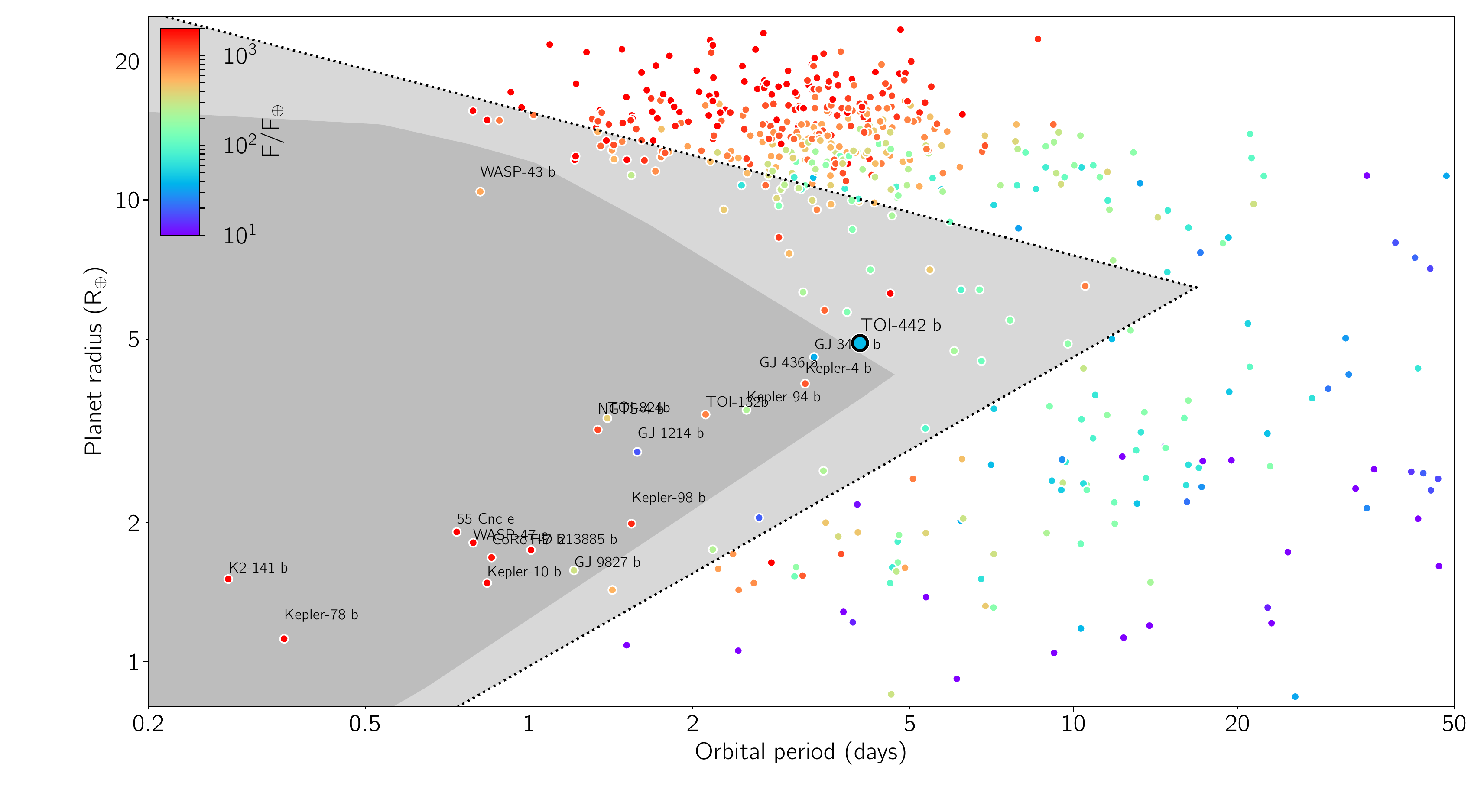}
    \caption{\label{fig:desert} Period-radius diagrams for planets with radius precision better than 10\,\%. The light grey shaded region represents the mean boundaries of the Neptune desert derived by \cite{Mazeh2016A&A...589A..75M} and the dark grey shaded region corresponds to the interior envelope of the 1$\sigma$ boundaries. The incoming stellar radiation is colour coded.} 
\end{figure*}

As shown in Fig.\,\ref{fig:massradius}, LP\,714-47\,b has the same bulk density as Neptune at twice the Neptune mass and higher equilibrium temperature. Together with its orbital period, this places LP\,714-47\,b  as an apparently typical warm Neptune-like planet at the edge of the Neptune desert.
In the orbital period range of tens of days or less, and for a given planet size, the planet occurrence density $\mathrm{d}n/\mathrm{d}\log{P}$ rises steeply at a specific period before flattening out to a roughly constant value.  The position of this rise shifts to larger orbits with rising planet mass and radius (see Fig.\,\ref{fig:desert}), creating a diagonal boundary between a rather densely populated zone of super-Earths and an area of very low occurrence  represented in grey in Fig.\,\ref{fig:desert} \citep[e.g.][]{Szabo2011, Mazeh2016A&A...589A..75M}.
 The occurrence of planets such as LP\,714-47\,b within the Neptune desert provides constraints for the possible scenarios to explain this feature in the planetary system architectures.


Multiple origins of this Neptune desert have been considered:
\begin{itemize}

    \item Photoevaporation of the planetary atmosphere once a gas-rich planet arrives close to its central star, explaining the lower boundary of the desert \citep{OweWu2013,LopezFortney2014,Jinetal2014}. Either additionally or alternatively, the subsequently released primordial energy from the formation may (further) clean up the Neptune desert, since this energy is comparable to the atmospheric binding energy. This core-powered mass-loss scenario was discussed later by \citet{2016ApJ...825...29G}, \citet{2019MNRAS.487...24G}, and \citet{2020MNRAS.tmp..303G}.
   
    \item Interplay of gas accretion and planet migration in the core accretion scenario. Planets of low and intermediate masses undergo fast, inward type-I migration and arrive at inner orbits where they cannot accrete efficiently anymore \citep[e.g.][]{Cimerman2017MNRAS, LambrechtsLega2017}. These planets presumably shape the lower border of the triangular desert region. On the other hand, planets that enter runaway gas accretion in the outer regions can open a gap in the disc and enter the less efficient type-II migration regime. These emerging gas giants move diagonally in the radius-period diagram and arrive at the upper border of the desert. In addition, scattering events could move gas giants that remained in the outer disc close to their host star, populating that region as well \citep[e.g.][]{Dawson2018ARAA}. The resulting desert has been observed in several independent planet population syntheses \citep[e.g.][]{IdaLin2008,Mordasinietal2009,BitschJohansen2016,Ndugu2018}.

    \item Migration of low-mass planets of several Earth masses to the inner regions of the disc \citep[e.g.][]{Flock2019}.
    At these close orbits, further gas accretion is unlikely due to the high temperatures \citep{LambrechtsLega2017,Cimerman2017MNRAS} and the planets are stuck at low masses.

    \item High-eccentricity migration followed by circularisation \citep{2016ApJ...820L...8M}. When circularised close to the host star, planets in the intermediate mass range are preferentially subject to tidal disruption. Predictions related to the Neptune desert limits are in good agreement with the observations.
   
\end{itemize}

LP\,714-47\,b is among the few objects that populate this desert, residing at its lower boundary.  The low density of the planet (see mass-radius diagram Fig.\,\ref{fig:massradius}) suggests that it hosts an atmosphere, although its composition is not yet known. This low bulk density indicates that at some point during its evolution, LP\,714-47\,b accreted a sizeable gaseous envelope and retained parts of this atmosphere even if it was subject to photoevaporation, which is relatively low due to the late stellar type (see the colour code in Fig.\,\ref{fig:desert}). This scenario is supported by the mass of the planet, which is sufficiently high to prevent significant erosion through this mechanism \citep[e.g.][]{OweWu2013}.  If, however, a future characterisation of the atmosphere of LP\,714-47\,b showed a preferential loss of light elements, this would indicate that it experienced photoevaporative loss \citep{Benneke2019}.

The eccentricity of LP\,714-47\,b is very low. If it did indeed experience a high-eccentricity migration in the past, it obviously did survive the event. This is in agreement with its position in Figure\,1 of \citet{2016ApJ...820L...8M}, where it is inside the stable region. 

The close orbit of LP\,714-47\,b implies a warm environment with equilibrium temperatures of about 700\,K.
At such temperatures and at its current location, the contraction of an early atmosphere is hindered due to recycling flows that penetrate the planetary Hill sphere, preventing the planet from growing into a gas giant \citep[e.g.][]{Cimerman2017MNRAS}.
This leads to the tentative conclusion that the planet most likely formed in colder regions further out, possibly beyond the water ice line, before migrating inwards while its atmosphere contracted.
In summary, LP\,714-47\,b adds to planets in or close to the Neptune desert and, therefore, it contributes to the building up of a sufficiently large sample that can provide constraints for the planet formation scenarios discussed above.


\begin{acknowledgements}
  CARMENES is an instrument for the Centro Astron\'omico Hispano-Alem\'an de
  Calar Alto (CAHA, Almer\'{\i}a, Spain). 
  CARMENES is funded by the German Max-Planck-Gesellschaft (MPG), 
  the Spanish Consejo Superior de Investigaciones Cient\'{\i}ficas (CSIC),
  the European Union through FEDER/ERF FICTS-2011-02 funds, 
  and the members of the CARMENES Consortium 
  (Max-Planck-Institut f\"ur Astronomie,
  Instituto de Astrof\'{\i}sica de Andaluc\'{\i}a,
  Landessternwarte K\"onigstuhl,
  Institut de Ci\`encies de l'Espai,
  Institut f\"ur Astrophysik G\"ottingen,
  Universidad Complutense de Madrid,
  Th\"uringer Landessternwarte Tautenburg,
  Instituto de Astrof\'{\i}sica de Canarias,
  Hamburger Sternwarte,
  Centro de Astrobiolog\'{\i}a and
  Centro Astron\'omico Hispano-Alem\'an), 
  with additional contributions by the Spanish Ministry of Economy, 
  the German Science Foundation through the Major Research Instrumentation 
    Programme and DFG Research Unit FOR2544 ``Blue Planets around Red Stars'', 
  the Klaus Tschira Stiftung, 
  the states of Baden-W\"urttemberg and Niedersachsen, 
  and by the Junta de Andaluc\'{\i}a.
  Based on data from the CARMENES data archive at CAB (INTA-CSIC).
    
We acknowledge the use of public TESS Alert data from pipelines at the TESS Science Office and at the TESS Science Processing Operations Center.
This research has made use of the Exoplanet Follow-up Observation Program website, which is operated by the California Institute of Technology, under contract with the National Aeronautics and Space Administration under the Exoplanet Exploration Program. Resources supporting this work were provided by the NASA High-End Computing (HEC) Program through the NASA Advanced Supercomputing (NAS) Division at Ames Research Center for the production of the SPOC data products.

Some of the observations in the paper made use of the High-Resolution Imaging instrument Zorro. Zorro was funded by the NASA Exoplanet Exploration Program and built at the NASA Ames Research Center by Steve B. Howell, Nic Scott, Elliott P. Horch, and Emmett Quigley. Zorro was mounted on the Gemini South telescope of the international Gemini Observatory, a program of NSF’s OIR Lab, which is managed by the Association of Universities for Research in Astronomy (AURA) under a cooperative agreement with the National Science Foundation. on behalf of the Gemini partnership: the National Science Foundation (United States), National Research Council (Canada), Agencia Nacional de Investigación y Desarrollo (Chile), Ministerio de Ciencia, Tecnología e Innovación (Argentina), Ministério da Ciência, Tecnologia, Inovações e Comunicações (Brazil), and Korea Astronomy and Space Science Institute (Republic of Korea).

This work makes use of observations from the LCOGT network. This article is based on observations made with the MuSCAT2 instrument, developed by ABC, at Telescopio Carlos S\'anchez operated on the island of Tenerife by the IAC in the Spanish Observatorio del Teide. 
This paper is also based on observations made in the Observatorios de Canarias del IAC with the Nordic Optical Telescope operated on the island of La Palma by NOTSA in the Observatorio del Roque de los Muchachos.
Data were partly obtained with the MONET/South telescope of the MOnitoring NEtwork of Telescopes, funded by the Alfried Krupp von Bohlen und Halbach Foundation, Essen, and operated by the Georg-August-Universit\"at G\"ottingen, the McDonald Observatory of the University of Texas at Austin, and the South African Astronomical Observatory.

This research has made use of the NASA Exoplanet Archive, which is operated by the California Institute of Technology, under contract with the National Aeronautics and Space Administration under the Exoplanet Exploration Program.

This research made use of Lightkurve, a Python package for Kepler and TESS data analysis \citep{2018ascl.soft12013L}.

  We acknowledge financial support from the Spanish Agencia Estatal de Investigaci\'on
  of the Ministerio de Ciencia, Innovaci\'on y Universidades and the European
  FEDER/ERF funds through projects 
  AYA2015-69350-C3-2-P,     
  PGC2018-098153-B-C31/C33,   
  AYA2016-79425-C3-1/2/3-P, 
  AYA2018-84089,            
  BES-2017-080769,          
  ESP2016-80435-C2-1/2-R,   
  ESP2017-87676-C5-1/2/5-R, 
  Instituto de Astrof\'isica de Andaluc\'ia (SEV-2017-0709), 
the Generalitat de Catalunya through CERCA programme'',
the UK Science and Technology Facilities Council through grant ST/P000592/1,
the JSPS KAKENHI through grants 17H04574, JP18H01265, and 18H05439, 
and the JST PRESTO through grant JPMJPR1775,
the “la Caixa” INPhINIT Fellowship Grant LCF/BQ/IN17/11620033 for Doctoral studies at Spanish Research Centres of Excellence,  
V.M.P. acknowledges support from NASA Grant NNX17AG24G, 

Support for this work was provided to J.K.T. by NASA through Hubble Fellowship grant HST-HF2-51399.001 awarded by the Space Telescope Science Institute, which is operated by the Association of Universities for Research in Astronomy, Inc., for NASA, under contract NAS5-26555.

The research leading to these results has received funding from the ARC grant for Concerted
Research Actions, financed by the Wallonia-Brussels Federation. TRAPPIST is funded by the Belgian Fund for Scientific Research (Fond National de la Recherche Scientifique, FNRS) under the grant FRFC 2.5.594.09.F, with the participation of the Swiss National Science Fundation (SNF). M.G. and E.J. are FNRS Senior Research Associates.

S.D. acknowledges support from the Deutsche Forschungsgemeinschaft under Research Unit FOR2544 ``Blue Planets around Red Stars'', project no. RE 281/32-1. 
M.Z. acknowledges support from the Deutsche Forschungsgemeinschaft under DFG RE 1664/12-1 and Research Unit FOR2544 ``Blue Planets around Red Stars'', project no. RE 1664/14-1. 
M.S. was supported by the DFG Research Unit FOR2544 “Blue Planets around Red Stars”, project no. RE 2694/4-1.
I.J.M.C.\ and E.M.\ acknowledge support from the National Science Foundation through grant AST-1824644.
B.B. thanks the European Research Council (ERC Starting Grant 757448-PAMDORA) for their financial support.
D. D. acknowledges support from NASA through Caltech/JPL grant RSA-1006130 and through the TESS Guest Investigator Program Grant 80NSSC19K1727.
P.P. acknowledges support from NASA (16-APROBES16-0020 and the Exoplanet Exploration Program) and the National Science Foundation (Astronomy and Astrophysics grant 1716202), the Mt Cuba Astronomical Foundation, and George Mason University start-up and instructional equipment funds. The NASA Infrared Telescope Facility is operated by the University of Hawaii under contract NNH14CK55B with the National Aeronautics and Space Administration.
M.R.K is supported by the NSF Graduate Research Fellowship, grant No. DGE 1339067.
M.Y. and H.V.S. thank to T\"UB\.{I}TAK for a partial support in using T100 telescope with project number 19AT100-1474. 
J.N.W.\ thanks the Heising-Simons Foundation for support.
D.H.\ acknowledges support from the Alfred P. Sloan Foundation, the National Aeronautics and Space Administration (80NSSC18K1585, 80NSSC19K0379), and the National Science Foundation (AST-1717000).
T.D. acknowledges support from MIT’s Kavli Institute as a Kavli postdoctoral fellow.
This work made use of \texttt{tpfplotter} (developed by J. Lillo-Box, which also made use of the python packages \texttt{astropy}, \texttt{lightkurve}, \texttt{matplotlib} and \texttt{numpy}.

Part of this research was carried out at the Jet Propulsion Laboratory, California Institute of Technology, under a contract with the National Aeronautics and Space Administration (NASA).

The authors wish to recognise and acknowledge the very significant
cultural role and reverence that the summit of Mauna Kea has always
had within the indigenous Hawaiian community. We are most fortunate to
have the opportunity to conduct observations from this mountain.

Finally, we thank the referee for detailed and helpful comments.
\end{acknowledgements}

\interlinepenalty=10000

\bibliographystyle{aa}
\interlinepenalty=1000000
\bibliography{main}

\appendix
\section{\label{sec:phot_inst}Photometric facilities}
\begin{table*}
    \caption{\label{tab:photometry} Photometric facilities.}
    \centering
    \begin{tabular}{@{}ll@{}c@{}ccclc@{}c@{}}
        \toprule
        Instrument  & Transit date   & Diameter  & FOV           & CCD   & Scale                & Filter(s) & $N$ & $\Delta t$ \tabularnewline
        \multicolumn{2}{l}{Observatory}            & [m]   & [arcmin$^2$]  &       & [arcsec\,pix$^{-1}$] & & & [min]\tabularnewline
        \midrule
        TRAPPIST-South & 2019-02-17 &0.6 & 22\,$\times$\,22 & 2k\,$\times$\,2k & 0.65 & $z'$ & 506 & 190 \tabularnewline
        \multicolumn{2}{l}{Paranal Observatory}& \tabularnewline
        LCOGT &  2019-02-17 & 1.0  &  26$\times$26   &  4k\,$\times$\,4k    & 0.39 & $z'$ & 156 & 184 \tabularnewline
        \multicolumn{2}{l}{Las Cumbres Observatory Global Telescope}    \tabularnewline
        El Sauce  & 2019-02-17  & 0.36  & 18.8$\times$12.5 & 1.5k\,$\times$\,1k & 0.74 & $R$ & 147 & 196  \tabularnewline
        \multicolumn{2}{l}{El Sauce Observatory} &\tabularnewline
        MONET-S & 2019-09-27   & 1.2  &  12.6\,$\times$\,12.6     & 2k\,$\times$\,2k & 0.37 & $V$ & 236 & 130\tabularnewline
        \multicolumn{2}{l}{South African Astronomical Observatory}  &  \tabularnewline
        TUG & 2019-10-05 & 1.0 & 21.5\,$\times$\,21.5  & 4k\,$\times$\,4k & 0.31 & $R$ & 114 & 219 \tabularnewline
        \multicolumn{2}{l}{National Observatory, Turkey}\tabularnewline
        ULMT & 2019-10-30 & 0.6 & 26$\times$26 & 4k\,$\times$\,4k & 0.39 &  $g'$ & 88 & 266 \tabularnewline
        \multicolumn{2}{l}{Steward Observatory} & \tabularnewline
        MuSCAT2 & 2019-12-17 &1.52&7.4$\times$7.4&4$\times$ 1k\,$\times$\,1k&0.44& $g',r',i',z'$ & 4$\times 230$ & 229 \tabularnewline
                & 2019-12-21 &&&&&& 4$\times 157$ & 156 \tabularnewline
        \multicolumn{2}{l}{Teide Observatory} & \tabularnewline
        GMU     & 2019-12-25 & 0.8 &23$\times$23&4k\,$\times$\,4k&0.34&$R_c$ & 186 & 257 \tabularnewline        
        \multicolumn{2}{l}{Georg Mason University} & \tabularnewline
        \bottomrule
    \end{tabular}
\end{table*}

\begin{figure*}
    \centering
    \includegraphics[width=0.32\textwidth]{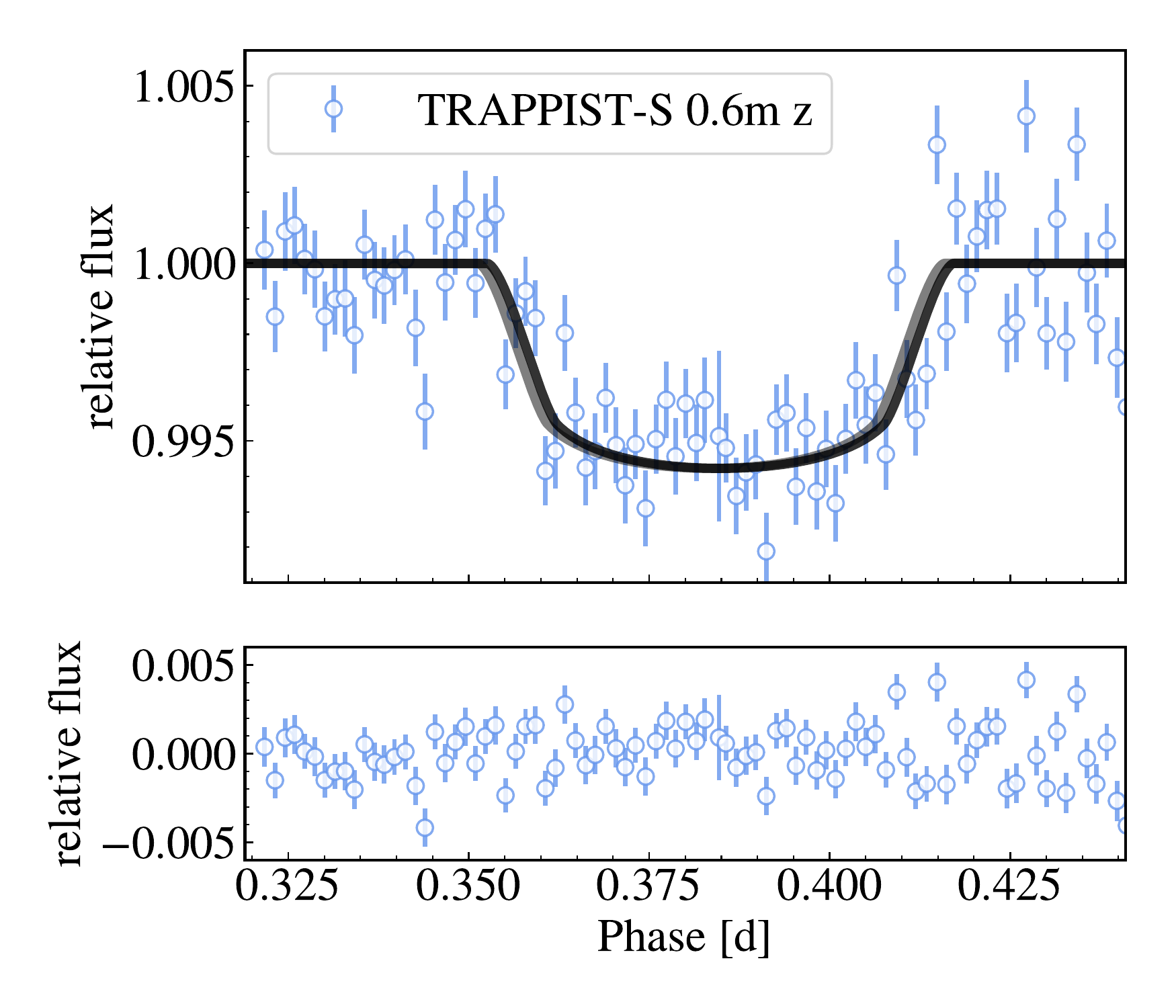}
    \includegraphics[width=0.32\textwidth]{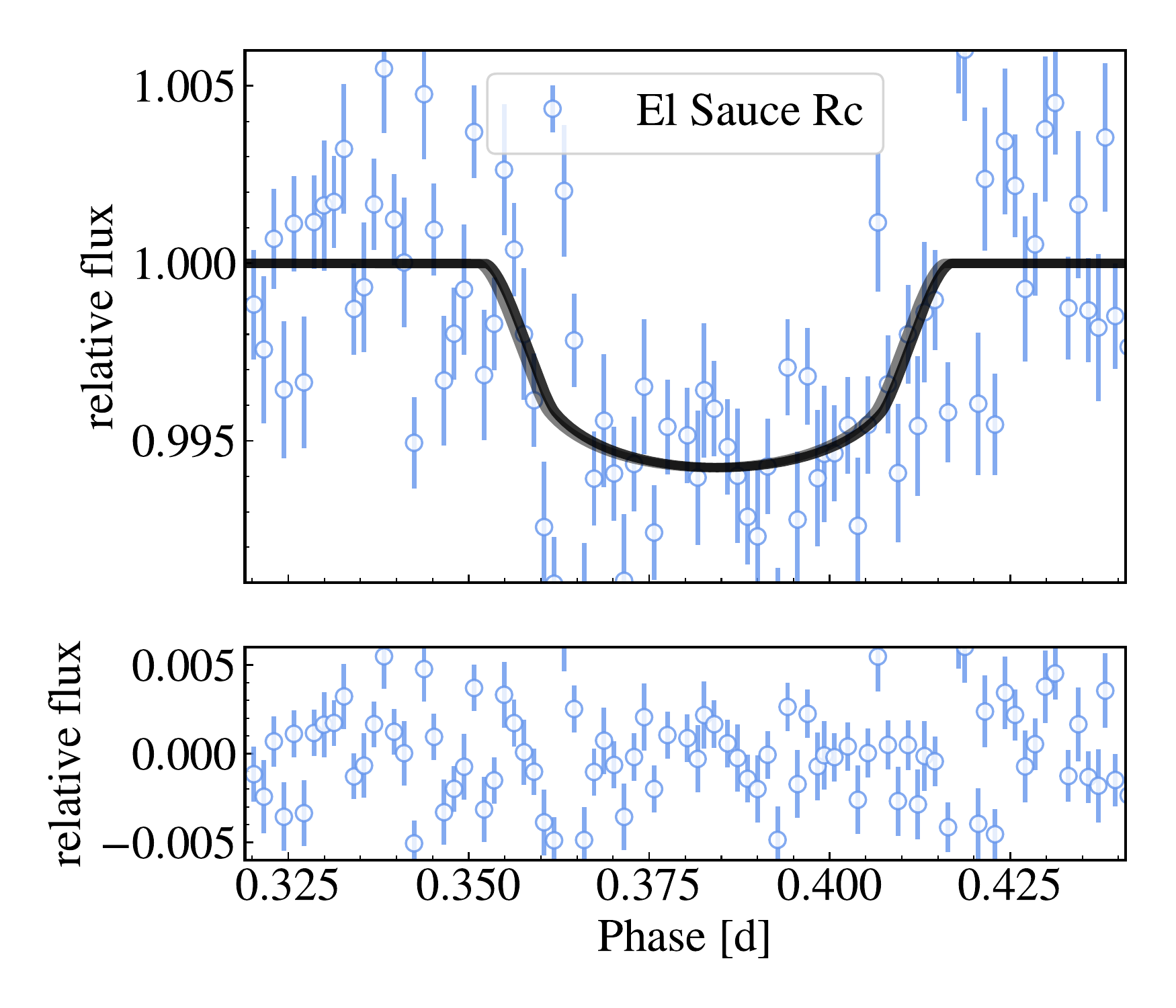}
    \includegraphics[width=0.32\textwidth]{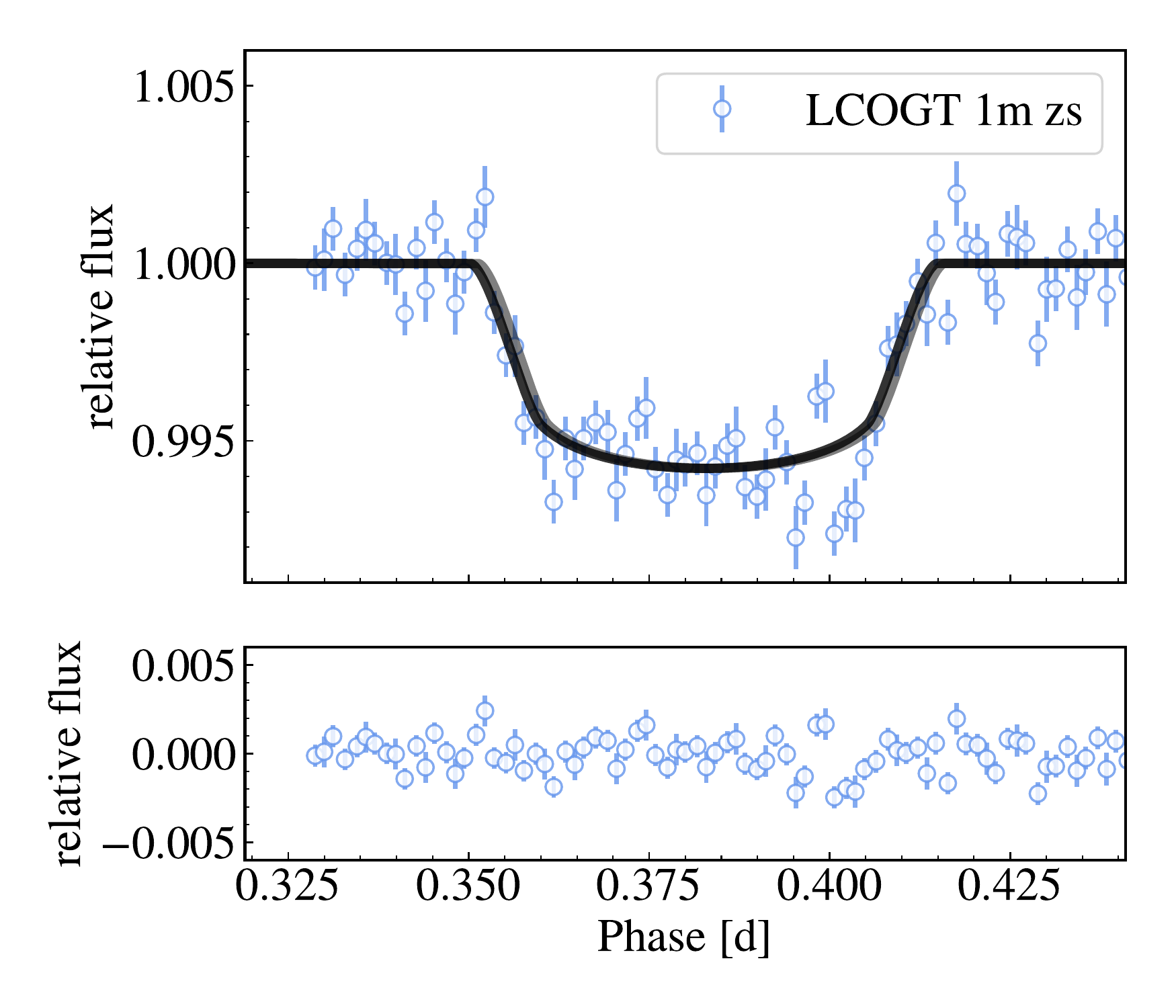}
    \includegraphics[width=0.32\textwidth]{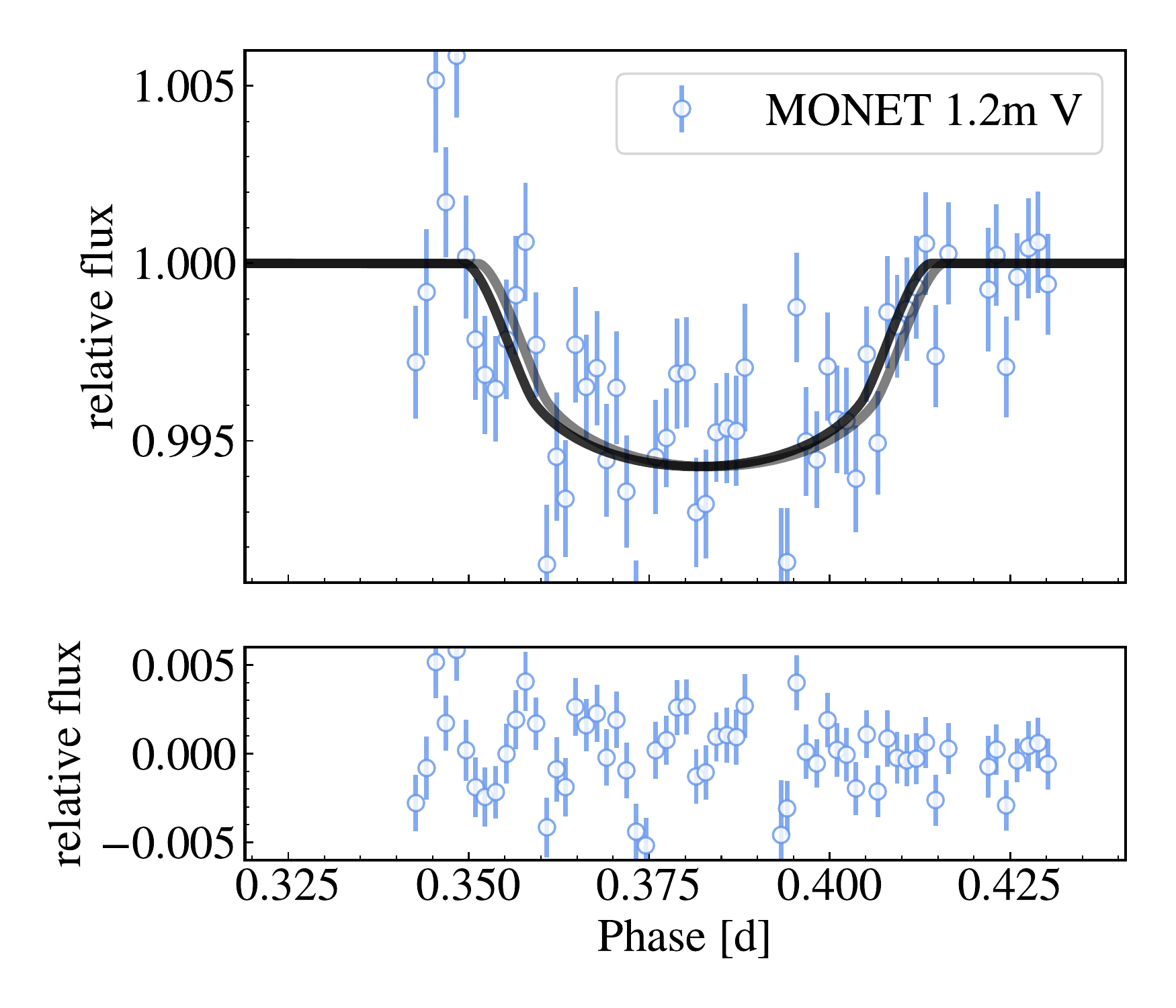}
    \includegraphics[width=0.32\textwidth]{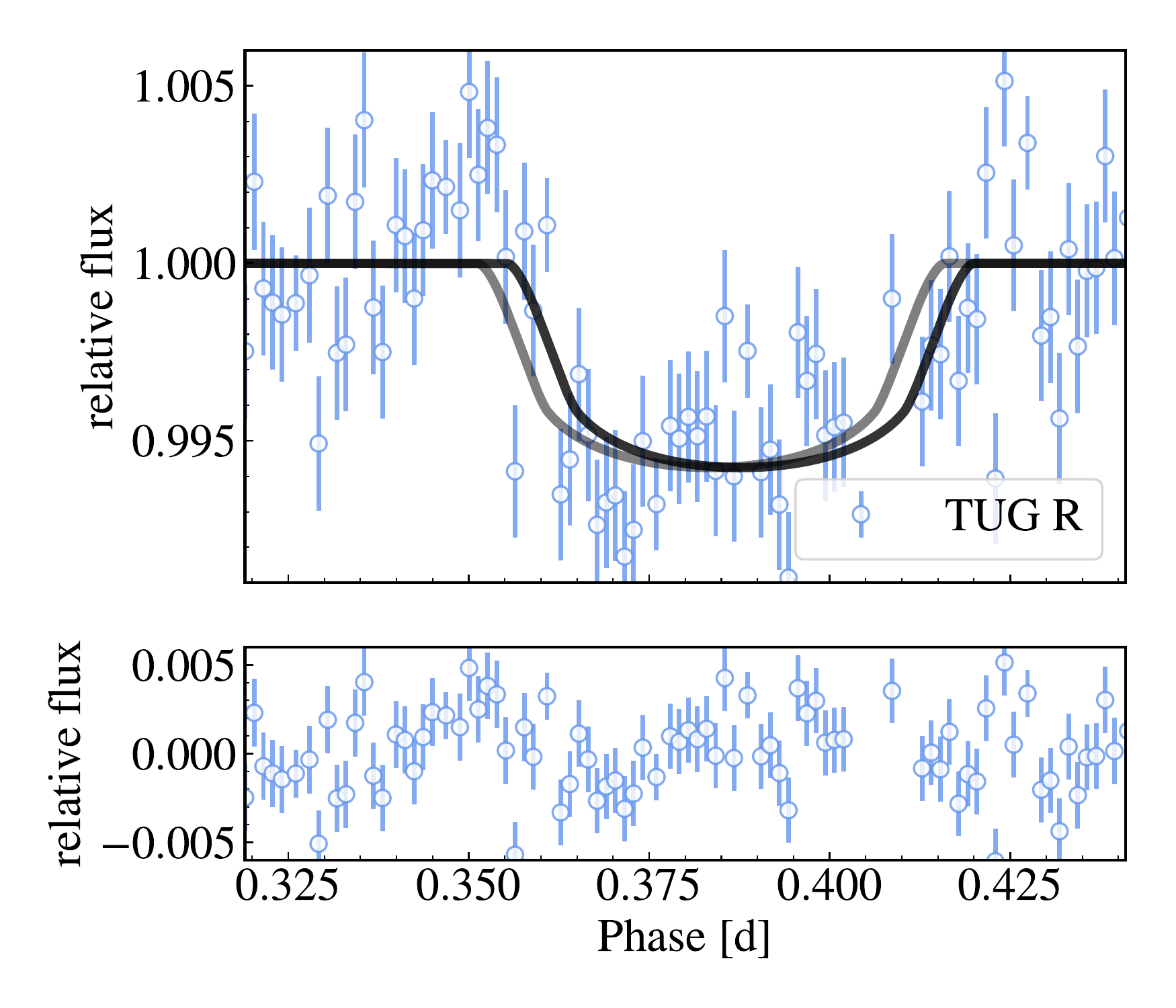}
    \includegraphics[width=0.32\textwidth]{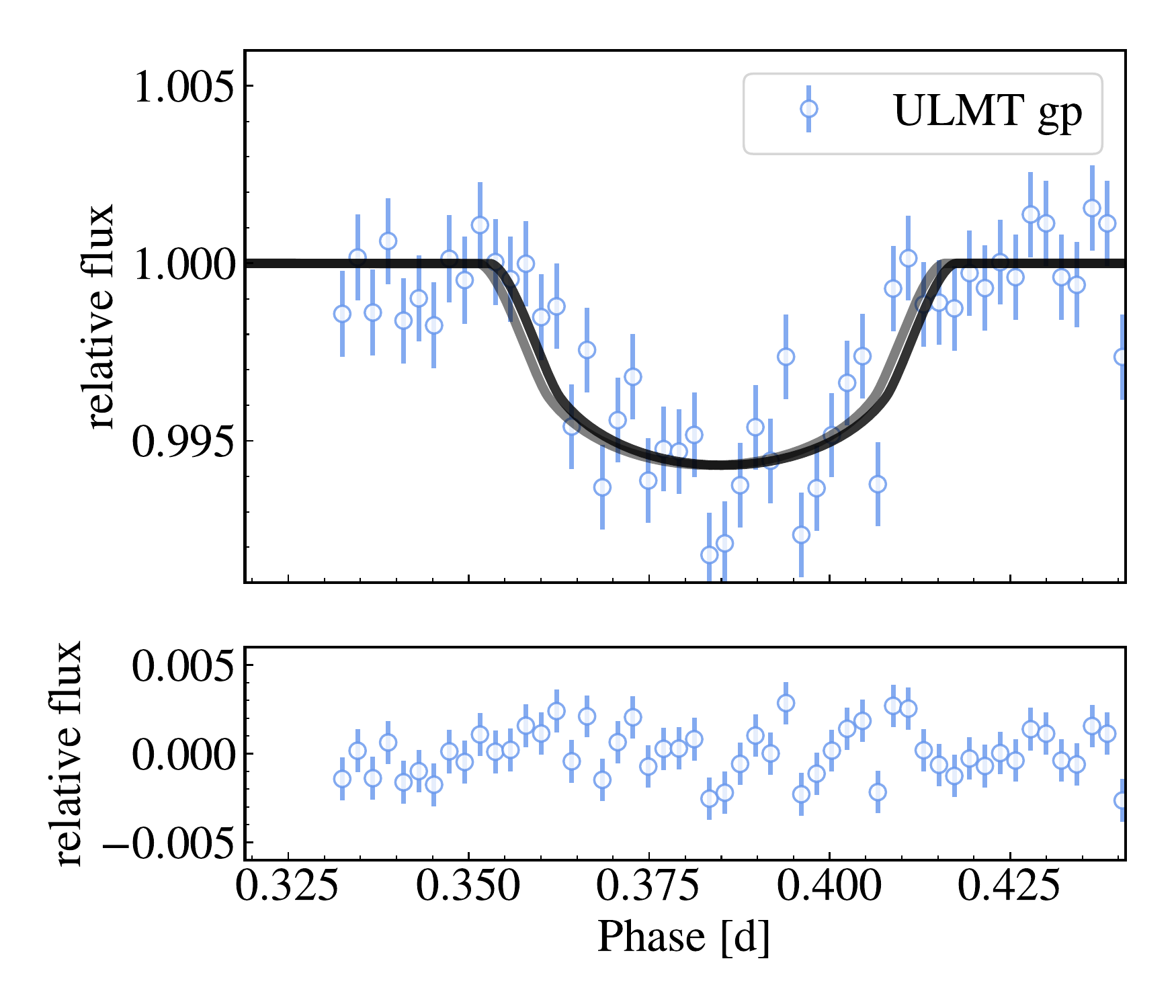}
    \includegraphics[width=0.32\textwidth]{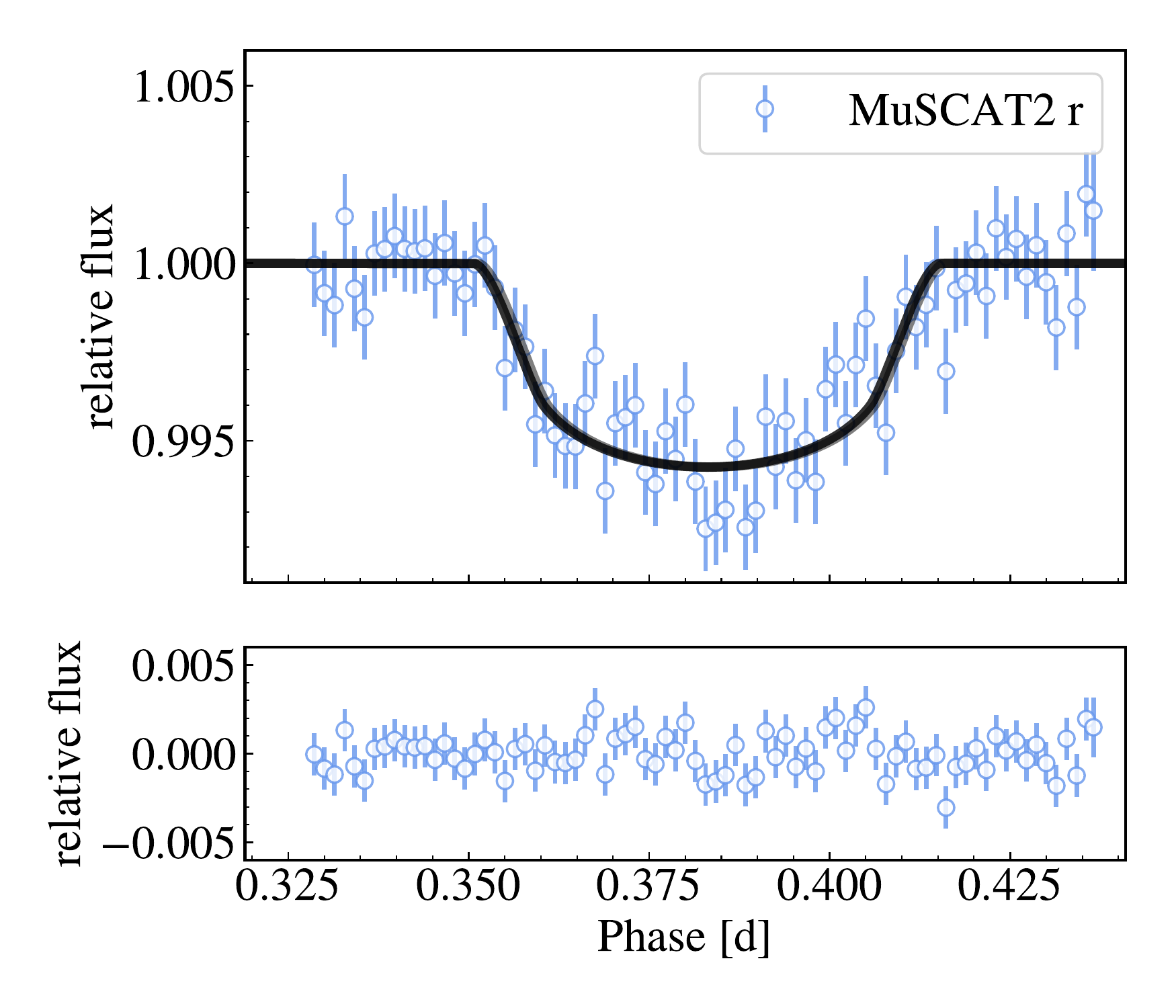}
    \includegraphics[width=0.32\textwidth]{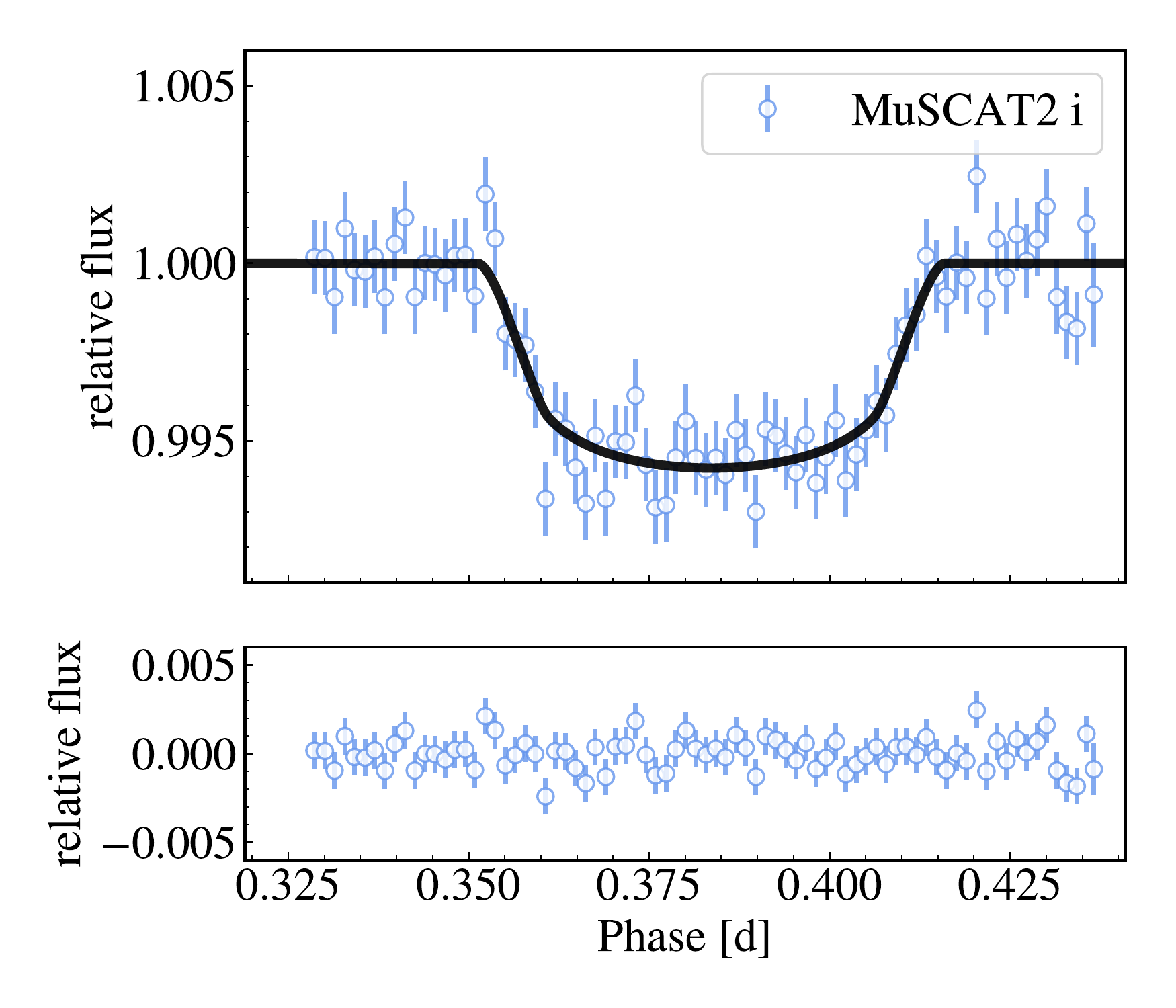}
    \includegraphics[width=0.32\textwidth]{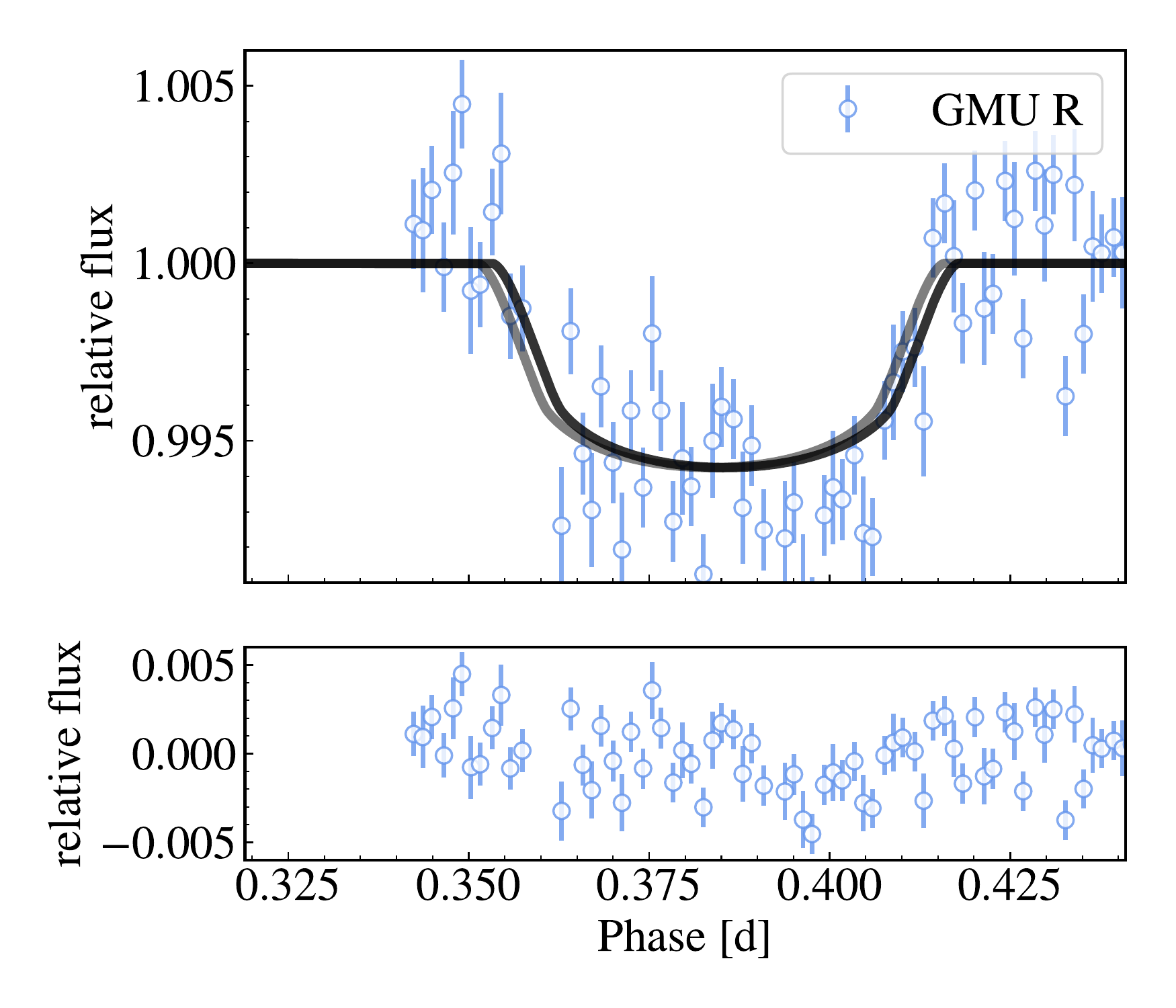}

    \caption{Ground-based transits (1 minute bins) compared to the best fit model (grey) and a model allowing small shifts in transit times (black). { TRAPPIST}, {LCOGT,} and {El Sauce} data were obtained at transit 24, {MONET} at 79, {TUG} at 81, {ULM} at 87, {MuSCAT2} at 99 and 100, and {GMU} at 101, relative to the first {\em TESS} transit. {MuSCAT2} transits in filter $g$ and $z$ are shown in the main text (Fig.\,\ref{fig:phaseTR_Muscat}).}
    \label{fig:transits}\label{fig:ttvs}
\end{figure*}

\paragraph{WASP-South.}
TOI~442 was monitored over 120 days in 2008/9 by WASP-South, the Southern station of the Wide Angle Search for Planets. WASP-South (located at SAAO, Sutherland, South Africa) was an array of 8 cameras using 200\,mm, f/1.8 lenses with a broadband filter spanning 400--700\,nm, equipped with $2048 \times 2048$ CCDs giving a plate scale of 13.7 arcsec per pixel.

\paragraph{LCOGT.}  We obtained a full transit on UTC 2019-02-17 from a 1\,m telescope of the LCOGT node at the Cerro Tololo Inter-American Observatory. A total of 155 frames covering 190\,min of the 2019-02-17 transit were obtained in $z_s$-band. The telescope is equipped with a Sinistro camera with a pixel scale of 0.389\,arcsec. The PSF is 1.75\,arcsec, sources are extracted with a 16 pixel aperture. The images were calibrated by the LCOGT BANZAI pipeline \citep{McCully:2018} and photometric data were extracted using {\tt AstroImageJ}.

\paragraph{MONET-South.}
The 1.2\,m MONET/South telescope (MOnitoring NEtwork of Telescopes) is located at the South African Astronomical Observatory (Northern Cape, South Africa). It is equipped with a Finger Lakes ProLine 2k$\times$2k e2v CCD and has a $12.6\times12.6$\,arcmin$^2$ field of view. We performed aperture photometry with {\tt AstroImageJ} using eight comparison stars. 236 images in $V$ have been obtained covering 130\,min of the 2019-09-27 transit.

\paragraph{MuSCAT2.}
The Multicolour Simultaneous Camera for studying Atmospheres of Transiting exoplanets 2 \citep[MuSCAT2;][]{narita_2019} is mounted at Telescopio Carlos S\'anchez in Teide observatory (Tenerife, Spain). MuSCAT2 observes simultaneously in the $g$, $r$, $i$, and $z_s$ bands using a set of dichroics to split the light between four separate cameras with a field of view of 7.4$\times$7.4\,arcmin$^2$ (0.44\,arcsec/pix). MuSCAT2 is designed to be especially efficient for science related to transiting exoplanets and objects varying on short timescales around cool stellar types. Aperture photometry is calculated using a Python-based pipeline especially developed for MuSCAT2 \citep[see][for details]{narita_2019}.

\paragraph{TRAPPIST-South.}
TRAPPIST-South at ESO's La Silla Observatory in Chile is a 60-cm Ritchey-Chr\'etien telescope, which
has a thermoelectrically cooled 2k$\times$2k FLI Proline CCD camera with a field-of-view of $22'\times22'$
and resolution of 0.65\,arcsec\,pix$^{-1}$. We carried out a full-transit observation of TOI~442 on 2019-02-17 with the $z'$ filter using an 
exposure time of 10~s. We took 505 images, and we made use of {\tt AstroImageJ} to perform aperture photometry, where the optimum aperture was 10~pixels (6.5~arcsec) and the PSF was 3.44~arcsec.
We cleared all the stars from eclipsing binaries within the 2.5~arcmin around the target star. 

\paragraph{ULMT.}
We obtained a full transit on UTC 2019-10-30 from The University Louisville Manner Telescope (ULMT), which is located at Steward Observatory on Mount Lemmon near Tucson, Arizona. The observations employed a 0.6\,m f/8 telescope equipped with an SBIG STX-16803 CCD which has a 4k$\times$4k array of 9$\,\mu$m pixels, yielding a 0.39\,arcsec pixel scale and a $26\times26$\,arcmin field  of view. A total of 89 $g'$-band exposures were obtained covering 266\,min. The images were calibrated and photometric data were extracted using {\tt AstroImageJ}.

\paragraph{El Sauce.}
We obtained a full transit on UTC 2019-02-17 from El Sauce private observatory in Coquimbo Province, Chile. The 0.36\,m telescope is equipped with an SBIG STT1603-3 camera with a pixel scale of 0.735\,arcsec resulting in an $18.8\times12.5$ arcmin field of view. In-camera binning was used at $2 \times 2$ giving an operating pixel scale of 1.47\, arcsec. A total of 146 exposures covering 196\,min were obtained in $R_c$-band.  The PSF is 5.7\,arcsec and sources were extracted with a 6 pixel aperture. The photometric data were extracted using {\tt AstroImageJ}.

\paragraph{TUG.}
The 1m Ritchey-Chr\'etien T100 telescope located at T\"UB\.{I}TAK (The Scientific and Technological Research Council of Turkey) National Observatory (TUG) is equipped with an SI 1100 4k$\times$4k CCD camera with $15\times15$ $\mu$m pixels and $21.5'\times21.5'$ FoV. The full-transit observations have been obtained in Bessell R filter using an exposure time of 60~s. We obtained 114 frames for a full-transit observation on 2019-10-05. During the data reduction, we performed aperture photometry using AstroImageJ with an aperture radius of 24 pixels (7.7 arc-seconds).

\paragraph{GMU.}
The 0.8 m Ritchey-Chr\'etien Optical Guidance Systems telescope is located at and operated by George Mason University. The observations employed an R$_c$ filter equipped with a SBIG16803 4k$\times$4k CCD with $9\times9$ $\mu$m pixels and $23'\times23'$ FoV. The images were calibrated and photometric data were extracted using AstroImageJ.

The main information of the ground-based photometry is also summarised in Table\,\ref{tab:photometry}. 
In the main text, we show the comparison of our final model to most of our data. Here, we add the comparison to the other ground-based transit data (Fig.\ref{fig:transits}).

\section{ALFOSC low-resolution spectrum \label{alfosc_data}}
\begin{figure}[ht]
    \centering
    \includegraphics[width=\linewidth]{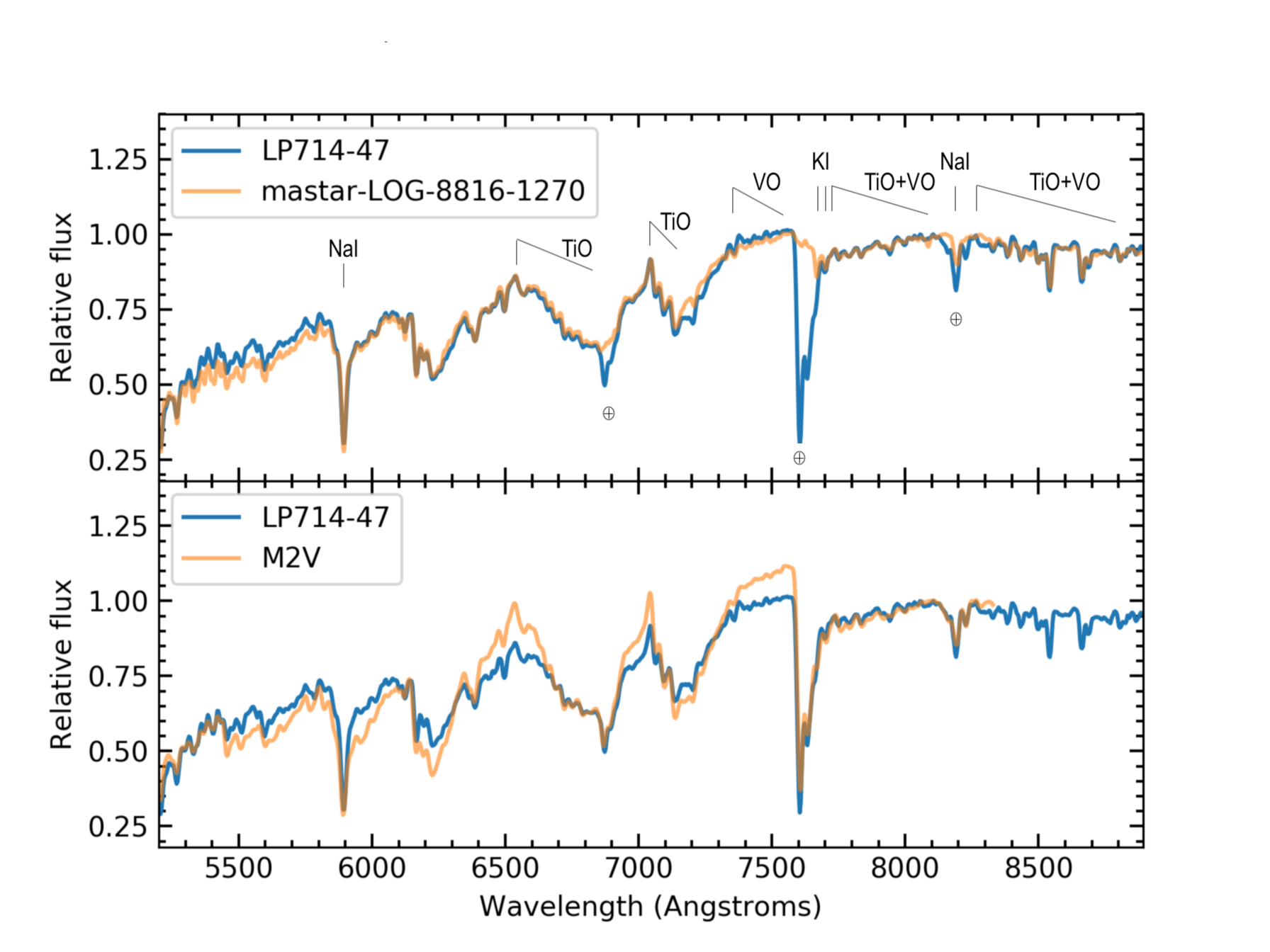}
    \caption{\label{alfosc_fig} {\it Top panel}: ALFOSC spectrum of LP\,714-47 (blue) and the MaStar spectrum (orange) of star mastar-LOG-8816-1270 (degraded to the spectral resolution of our data). The comparison star has a slightly sub-solar metallicity (Yan et al. 2019). The ALFOSC spectrum is not corrected for the telluric contribution, while the MaStar spectrum is telluric-free. Some atomic and molecular features are identified. {\it Bottom panel}: LP\,714-47 (blue) compared to an M2V standard star \citep[from][]{Alonso-Floriano2015A&A...577A.128A}.
}
\end{figure}

We used the spectroscopic catalogue of M dwarfs by \cite{Alonso-Floriano2015A&A...577A.128A} to determine the spectral type of LP\,714-47 to be M2\,V, with an error of half of a subtype. This classification is later than that determined by Lee (1984). However, as shown in the bottom panel of Fig.~\ref{alfosc_fig}, the match between our target and the M2V spectral standard is far from what is expected for a reliable classification, with significant deviations at wavelengths affected by the molecular absorption of TiO. This suggests that the metallicity of LP\,714-47 is likely non-solar. We used the NEXTGEN synthetic spectra computed for different stellar parameters \citep{1997ARA&A..35..137A, 1999ApJ...512..377H} with a resolution degraded to the ALFOSC data, and found that when all three parameters (temperature, surface gravity, and metallicity) are free, the solution is degenerate. Models with $T_{\rm eff}$ in the interval 3700--3900\,K, atmospheric gravity $\log{g}$ = 5.0--5.5 (cgs), and metallicity of [M/H] = $+$0.3 dex (metal-rich) and $-$0.5 dex (metal-depleted) yield acceptable fits to the ALFOSC observations. 

To conclude whether LP\,714-47 is a metal-rich or metal-poor star, we used data from the literature (see next), and the modelling of the CARMENES high-resolution spectra (see main text of this paper). The optical and near-infrared colours of LP\,714-47 appear to be consistent with the stars of the solar neighbourhood with similar spectral types, which implies a metallicity close to that of the solar vicinity. Using {\em Gaia} DR2 proper motion, parallax, and radial velocity, we obtained the space motions $U = 124.4 \pm 3.5$, $V = -253.2 \pm 2.8$, and $W = -67.6 \pm 2.0$ km\,s$^{-1}$. The equations of \citet{1987AJ.....93..864J} were employed, where $U$ is defined as positive away from the Galactic center. Following the criteria of \citet{1992ApJS...82..351L}, because of its very negative and high $VW$ velocities, LP\,714-47 may kinematically belong to a low-metallicity population of the Galaxy. From the extensive MaStar stellar library (described in \citealt{2019ApJ...883..175Y}), we found good matches to the ALFOSC spectrum using stars with $T_{\rm eff}$ = 3750\,K and [Fe/H] = $-$0.26 dex (see top panel of Fig.~\ref{alfosc_fig}). 

\section{Additional high-contrast imaging}
\label{sec:Speckle}

\begin{figure}
    \centering
    \includegraphics[width=0.49\textwidth]{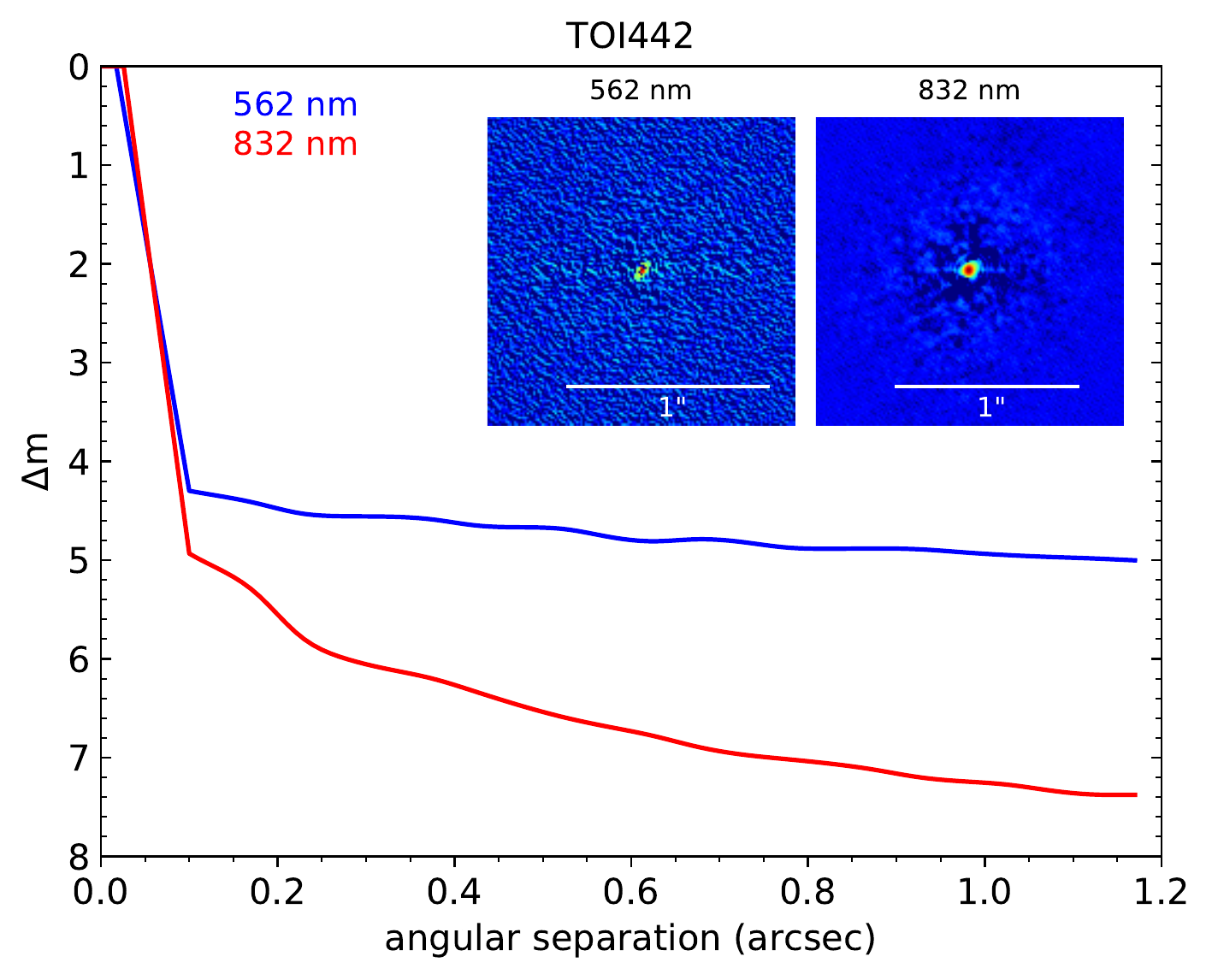}
    \caption{Direct imaging of LP\,714-47 with the Zorro speckle instrument on Gemini-South. The detection limits for the two filters are 5\,$\sigma$ limits. }
    \label{fig:Speckle}
\end{figure}

Direct high-resolution  imaging observations of LP\,714-47 were carried out on 14 January 2020 using the Zorro speckle instrument on Gemini-South\footnote{ \url{https://www.gemini.edu/sciops/instruments/alopeke-zorro/}}. Zorro simultaneously provides speckle imaging in two bands, 562\,nm and 832\,nm, with output data products including a reconstructed image, and robust limits on companion detections \citep{Howell2011}. The observations consisted of 5 sets of 1000, 0.06 sec observations each obtained during a night of good seeing (0.6\,arcsec). Fig.\,\ref{fig:Speckle} shows the speckle imaging contrast curves in both bands and the reconstructed high-resolution images for LP\,714-47. Based on these observations, we find that LP\,714-47 is a single star with no companion brighter than about 5-7.5 magnitudes detected within 1.2\,arcsec. These limits corresponds to a detection of no companion brighter than an M7-M9 main sequence star between the inner and outer working angle limits, for d=52\,pc, of 0.9\,au to 1.5\,au.

\section{Additional model comparison to data}

In addition to the phase-folded presentation of the radial-velocity data compared to our final model (Fig.\,\ref{fig:phase}), here we present the entire radial-velocity data set compared to the final model over the time base of the observations. In this plot, we also show the contribution of the Keplerian model only. The modification by the added GP model is especially visible in the regions of {\em ESPRESSO} data around BJD\,2458748 and {\em HIRES} data around BJD\,2458788.
\begin{figure*}
    \centering
    \includegraphics[width=\textwidth,trim=90 110 90 135,clip]{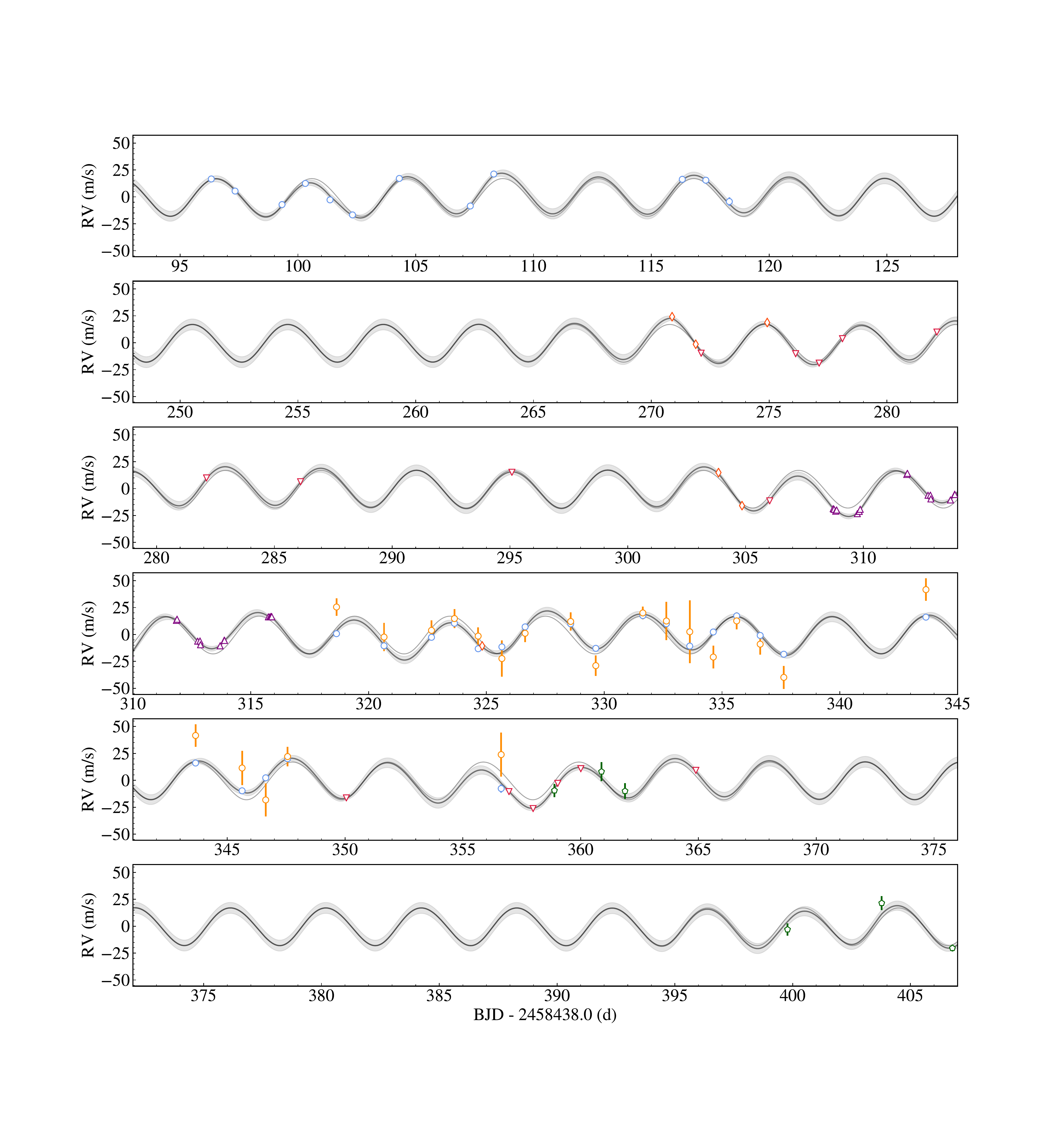}
    \caption{Radial-velocity data ({CARMENES VIS/NIR} in light blue and orange circles, respectively,  {ESPRESSO} in purple upper triangles, {HIRES} in crimson lower triangles, {PFS} in orange diamonds, and {iSHELL} in green pentagons) overlayed with our best fit model (dark grey, thick line) including the variation due to the GP (light grey). The one-planet model is shown in  darker grey and thin line.}
    \label{fig:allRVdata}
\end{figure*}

\section{Modelling details}
\label{sec:ModDet}

As mentioned in Sect.\,\ref{sec:Analysis}, we 
used {\tt celerite} \citep{celerite2017} as parametric noise model accounting for covariances between data points, which offers a fast and reliable implementation of GP regression. Covariances may be due to stellar, instrumental, or observational origin.

The likelihood function for $N$ data points $y_n$ at times $t_n$, with uncertainties $\sigma_n^2$, and model parameters $\theta$ (in our case the Keplerian orbital parameters) providing the residual vector $\boldsymbol{r}$ is:
\begin{equation*}
\ln p(\{y_n\}\,|\,\{t_n\},\,\{\sigma_n^2\},\,\theta) =
    -\frac{1}{2}\,\boldsymbol{r}^\mathrm{T}\,K^{-1}\,\boldsymbol{r}
    -\frac{1}{2}\,\ln\det K - \frac{N}{2}\,\ln 2\pi\ .
\end{equation*}    
The covariance matrix $K_{nm} = \sigma_n^2\,\delta_{nm} + k(\tau_{n,m})$ has two terms, the uncertainties on the diagonal, which may include a quadratically added jitter ($k(\tau_{n,m}) = \sigma^2\,\delta_{n,m}$), and the covariances or kernel as function of time differences $\tau_{n,m}=t_n-t_m$. For the latter, {\tt celerite} restricts the parametrisation to the following sum of complex exponential functions for $J$ kernels:
    
\begin{align}
k(\tau_{n,m}) &= \sum_{j=1}^J \frac{1}{2}\left[
    (a_j + i\,b_j)\,e^{-(c_j+i\,d_j)\,\tau_{n,m}} +
    (a_j - i\,b_j)\,e^{-(c_j-i\,d_j)\,\tau_{n,m}}
\right] \\
&= \sum_{j=1}^J \frac{1}{2}\left[
    \alpha_j\,e^{-\beta_j\,\tau} +
    {\alpha_j}^*\,e^{-{\beta_j}^*\,\tau_{n,m}}
\right]\ .
\end{align}

For b=0 and d=0, the kernel is called {\em REAL} and represents an exponential decay at a characteristic time scale $\tau$ and has two free parameters, the variance $a$ and $c=1/\tau$:

\begin{equation}
    k(\tau_{nm}) = a \, \exp \left (-c\,\tau_{n,m} \right )
.\end{equation}

A kernel representing damped oscillations driven by white noise is called {\em SHO}. It could represent solar-like oscillation or pseudo radial velocity variations from evolving spots on the rotating stellar surface. It has three parameters, 
the undamped oscillator period $P=2\pi/\omega_0$, a damping time scale $\tau$ related to the quality factor $Q$, and the variance $S_0$. For more details see Equations (19)-(24) in \citet{celerite2017}. It has the form:

\begin{align*}
k(\tau_{nm}) =& S_0 \omega_0 Q\exp{-\frac{\omega_0\tau_{n,m}}{2Q}} \\
 &\begin{cases}
    \cosh{(\eta\omega_0\tau_{nm})}+\frac{1}{2\eta Q} \sinh{(\eta\omega_0\tau_{nm})}& 0<Q<1/2\\
    2(1+\omega_0\tau_{n,m})& Q=1/2\\
    \cos{(\eta\omega_0\tau_{nm})}+\frac{1}{2\eta Q} \sin{(\eta\omega_0\tau_{nm})}& 1/2<Q\\
    \end{cases}
\end{align*}
with $\eta=\sqrt{\left|1-(4Q^2)^{-1}\right|}$.

\section{Two-planet model}
\label{sect:2-planet}

As discussed in Sect.\,\ref{sec:Analysis}, a second planet may be present. Its orbital period close to the first harmonic of the stellar rotation period requires a longer time base in order to check the coherence of that signal. We show the fit to the current data here. Additional ground-based data, especially longer coverage of a single instrument, would be necessary for discriminating between planetary origin and stellar activity.

\begin{figure*}
    \centering
    \includegraphics[width=\textwidth,trim=90 70 80 85,clip]{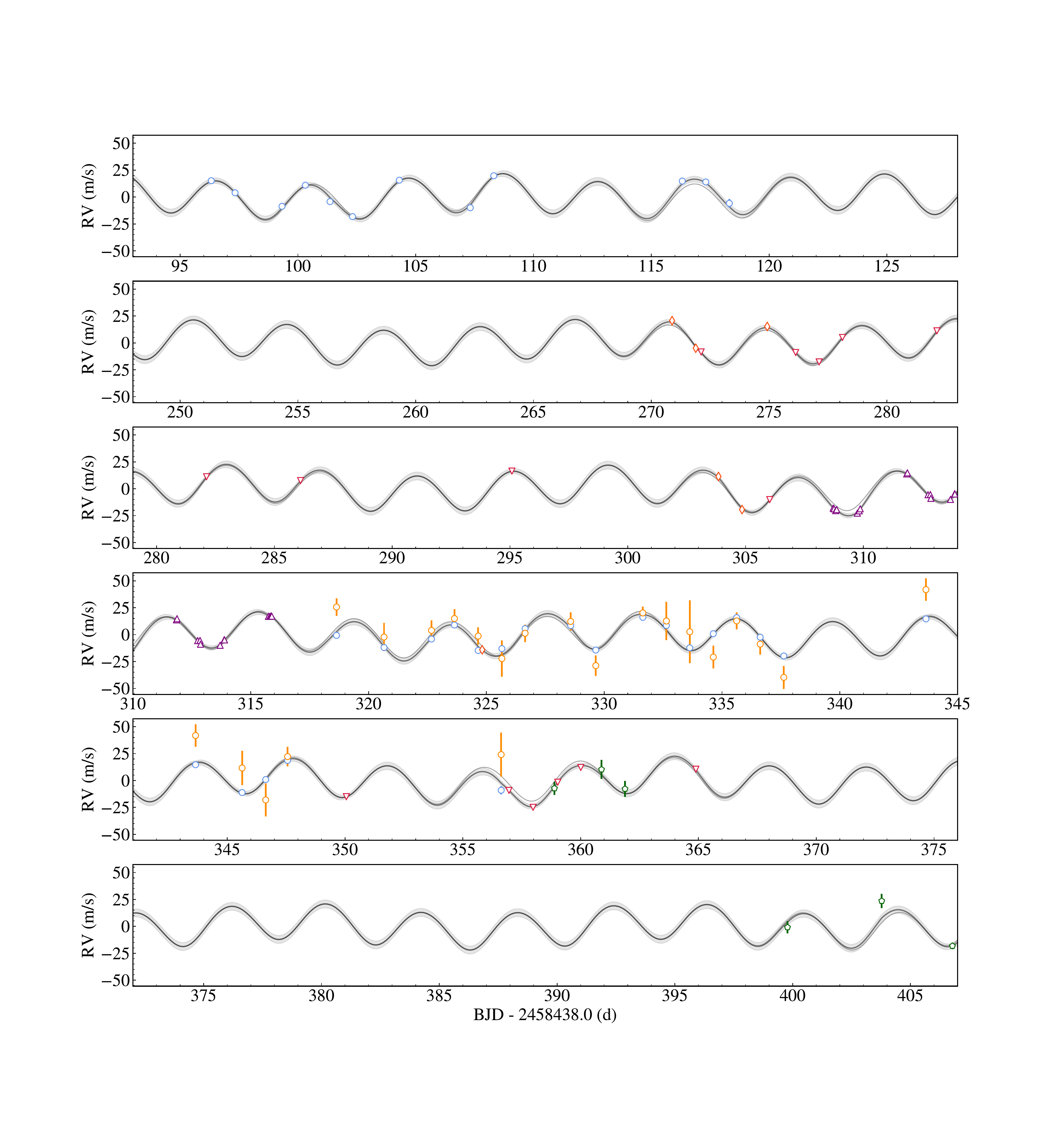}
    \caption{Radial-velocity data ({CARMENES VIS/NIR} in light blue and orange circles, respectively, {ESPRESSO} in purple upper triangles, {HIRES} in crimson lower triangles, {PFS} in orange diamonds, and {iSHELL} in green pentagons). The two-planet model is shown in dark grey (thick line) including the variation due to the GP (light grey).The two-planet model is shown in  darker grey and thin line.}
    \label{fig:allRVdata_2planets}
\end{figure*}

\begin{figure}
    \centering
    \includegraphics[width=0.49\textwidth]{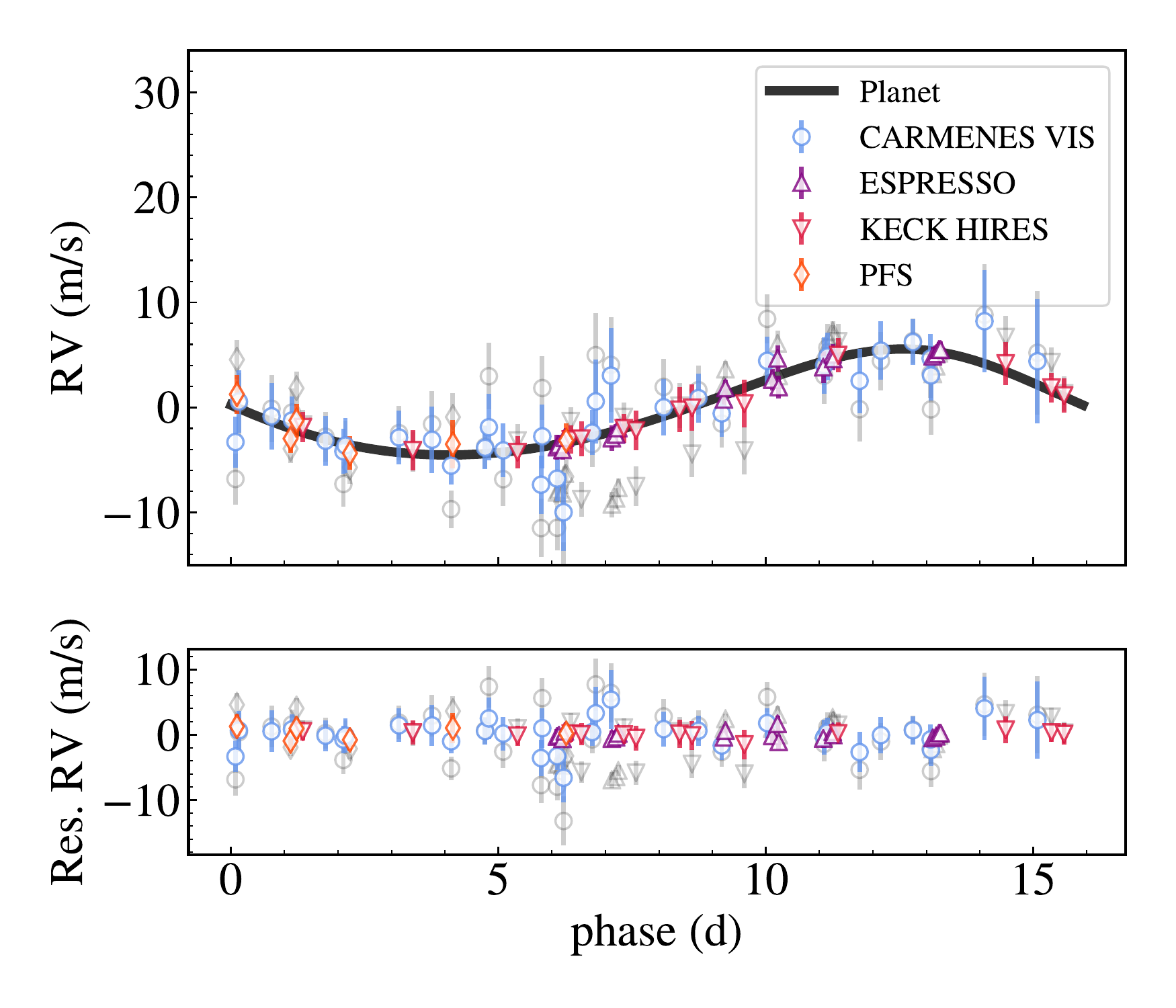}
    \caption{Radial velocity data (grey symbols) and corrected for correlated noise (coloured symbols) phase folded to the period of the putative second planet.}
    \label{fig:phase_planet2}
\end{figure}

\begin{figure}
    \centering
    \includegraphics[width=0.49\textwidth]{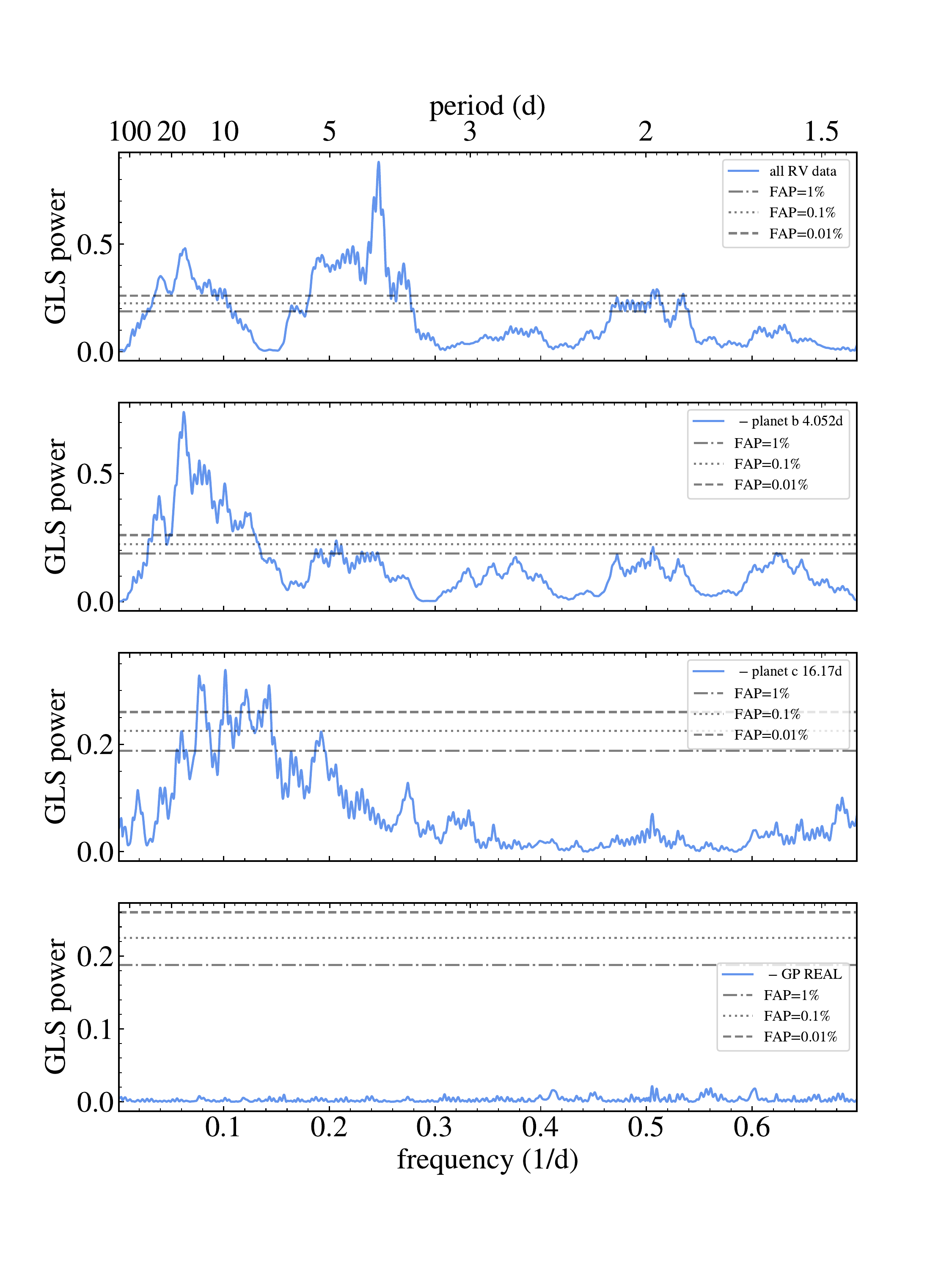}
    \caption{Subsequent pre-whitening of the radial velocity data ({\em top}) with the signals of TOI~442.01 and the potential second planet ({\em second panel}), the Gaussian Process modelling the correlated noise ({\em third panel}), and the residuals ({\em bottom}). Horizontal lines indicate the false alarm probablility of 1\%, 0.11\%, and 0.01 \%, respectively.}
    \label{fig:prewhite_2planets}
\end{figure}

\begin{figure}
    \centering
    \includegraphics[width=0.49\textwidth]{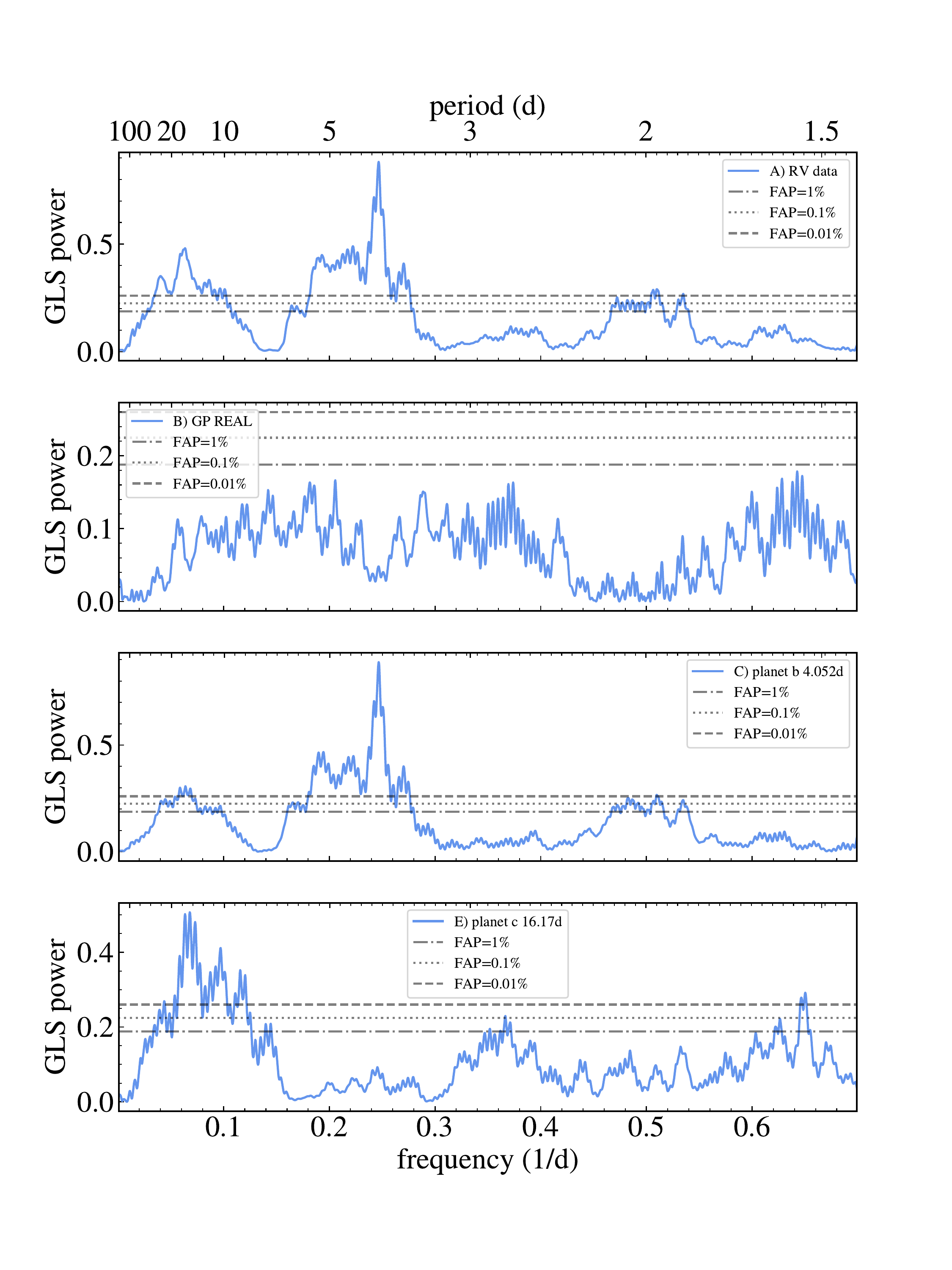}
    \caption{Periodogams of the model components: original radial-velocity data ({\em top}) the Gaussian process modelling the correlated noise (second panel), planet\,b (third panel), and the potential second planet ({\em bottom}). Horizontal lines indicate the false alarm probability of 1\%, 0.11\%, and 0.01 \%, respectively.}
    \label{fig:signal_2planets}
\end{figure}

Assuming the second signal to be of planetary origin, we show the fit to all radial velocity data in Fig.\,\ref{fig:allRVdata_2planets}, and the second phase-folded with the correlated noise and the first planet removed (Fig.\,\ref{fig:phase_planet2}). The decomposition of the radial velocity data into the two planetary signals and the correlated noise is shown in Figs.\,\ref{fig:prewhite_2planets} and \ref{fig:signal_2planets}.

\end{document}